\documentclass[useAMS,usenatbib]{mn2e}
\usepackage{psfig}

\usepackage{times}

\def \aap{A\&A}
\def \aaps{A\&AS}
\def \al{Astron. Lett.}
\def \aj{AJ}
\def \ap{Appl. Phys.}
\def \apj{ApJ}
\def \apjl{ApJ}
\def \apjs{ApJS}
\def \asp{Astron. Soc. Pac.}
\def \ca{Comments on Astrophys.}
\def \josb{J. Opt. Soc. Am. B}
\def \jpcrd{J. Phys. Chem. Ref. Data}
\def \jqsrt{J. Quant. Spectrosc. Radiat. Transfer}
\def \met{Metrologia}
\def \mnras{MNRAS}
\def \nat{Nat}
\def \npb{Nucl. Phys. B}
\def \nsrds{Natl. Stand. Rel. Data Ser.}
\def \pasp{PASP}
\def \plb{Phys. Lett. B}
\def \pr{Phys. Rev.}
\def \pra{Phys. Rev. A}
\def \prd{Phys. Rev. D}

\def \prl{Phys. Rev. Lett.}
\def \psc{Phys. Scr.}
\def \rmp{Rev. Mod. Phys.}
\def \zpa{Z. Phys. A}

\defcitealias{MurphyM_01a}{M01a}
\defcitealias{MurphyM_01b}{M01b}

\def \da{\Delta\alpha/\alpha}
\def \domg{\Delta\omega_{\rm g}/\omega_{\rm g}}
\def \dy{\Delta y/y}
\def \dx{\Delta x/x}
\def \dota{\dot{\alpha}/\alpha}
\def\bsp_small{\vspace{0.5cm}\small\noindent This paper
has been typeset from a \TeX / \LaTeX\ file prepared by the author.}

\title[Further evidence for variable $\alpha$ from QSO absorption
spectra]{Further evidence for a variable fine-structure constant from
Keck/HIRES QSO absorption spectra}

\author[M. T. Murphy, J. K. Webb \&
  V. V. Flambaum]{M. T. Murphy,$^{1,2}$\thanks{E-mail: mim@ast.cam.ac.uk
  (MTM); jkw@phys.unsw.edu.au (JKW)} J. K. Webb$^{2}$\footnotemark[1],
  V. V. Flambaum$^{2}$\\
$^{1}$Institute of Astronomy, University of Cambridge, Madingley Road,
  Cambridge, CB3 0HA, UK\\
$^{2}$School of Physics, University of New South Wales, UNSW Sydney
  N.S.W. 2052, Australia}

\begin{document}

\date{Accepted ---. Received ---; in original form ---}

\pagerange{\pageref{firstpage}--\pageref{lastpage}} \pubyear{2003}

\maketitle

\label{firstpage}

\begin{abstract}
We have previously presented evidence for a varying fine-structure
constant, $\alpha$, in two independent samples of Keck/HIRES QSO absorption
spectra. Here we present a detailed many-multiplet analysis of a third
Keck/HIRES sample containing 78 absorption systems. We also re-analyse the
previous samples, providing a total of 128 absorption systems over the
redshift range $0.2 < z_{\rm abs} < 3.7$. The results, with raw statistical
errors, indicate a smaller weighted mean $\alpha$ in the absorption clouds:
$\da = (-0.574 \pm 0.102) \times 10^{-5}$. All three samples separately
yield consistent and significant values of $\da$. The analyses of low-$z$
(i.e.~$z_{\rm abs} < 1.8$) and high-$z$ systems rely on different ions and
transitions with very different dependencies on $\alpha$, yet they also
give consistent results. We identify an additional source of random error
in 22 high-$z$ systems characterized by transitions with a large dynamic
range in apparent optical depth. Increasing the statistical errors on $\da$
for these systems gives our fiducial result, a weighted mean $\da = (-0.543
\pm 0.116) \times 10^{-5}$, representing 4.7\,$\sigma$ evidence for a
varying $\alpha$. Assuming that $\da = 0$ at $z_{\rm abs} = 0$, the data
marginally prefer a linear increase in $\alpha$ with time rather than a
constant offset from the laboratory value: $\dota = (6.40 \pm 1.35) \times
10^{-16}{\rm \,yr}^{-1}$. The two-point correlation function for $\alpha$
is consistent with zero over 0.2--13\,Gpc comoving scales and the angular
distribution of $\da$ shows no significant dipolar anisotropy. We therefore
have no evidence for spatial variations in $\da$.

We extend our previous searches for possible systematic errors, giving
detailed analyses of potential kinematic effects, line blending, wavelength
miscalibration, spectrograph temperature variations, atmospheric dispersion
and isotopic/hyperfine structure effects. The latter two are potentially
the most significant. However, overall, known systematic errors do not
explain the results. Future many-multiplet analyses of independent QSO
spectra from different telescopes and spectrographs will provide a now
crucial check on our Keck/HIRES results.
\end{abstract}


\section{Introduction}\label{sec:intro}

High resolution spectroscopy of QSO absorption systems provides a precise
probe of possible variations in several fundamental constants over
cosmological distances and timescales. \citet*{DzubaV_99b,DzubaV_99a} and
\citet{WebbJ_99a} introduced and applied the new and highly sensitive
many-multiplet (MM) method for constraining space-time variations of the
fine-structure constant, $\alpha \equiv e^2/\hbar c$. Using a MM analysis
of 49 absorption systems from two independent samples of Keck/HIRES
spectra, we recently reported statistical evidence for a variable $\alpha$
(\citealt[hereafter
\citetalias{MurphyM_01a}]{MurphyM_01a};
\citealt{WebbJ_01a}): $\da = (-0.72 \pm 0.18) \times 10^{-5}$, where $\da$
is defined as
\begin{equation}
\da = (\alpha_z - \alpha_0)/\alpha_0\,
\end{equation}
for $\alpha_z$ and $\alpha_0$ the values of $\alpha$ in the absorption
system(s) and in the laboratory
respectively\footnote{$\alpha_0^{-1}=137.03599958(52)$
\citep{MohrP_00a}.}. The absorption redshifts covered the range $0.5 <
z_{\rm abs} < 3.5$.

Here we significantly extend this work, presenting new measurements of
$\da$ in a third independent Keck/HIRES sample of 78 absorption systems
covering the range $0.2 < z_{\rm abs} < 3.7$. The present paper provides a
detailed description of our methods, results and analysis of potential
systematic effects. A companion paper, Webb et al. (in preparation),
presents a summary of the main results.

In the remainder of this section we outline the many multiplet technique
of obtaining constraints on $\da$ from QSO spectra. We describe the
different QSO and atomic datasets in Section \ref{sec:data} and our
analysis methods in Section \ref{sec:analysis}. The new results, together
with those from a re-analysis of the previous datasets, are presented in
Section \ref{sec:results}. We consider both temporal and spatial variations
in $\alpha$. Section \ref{sec:syserr} extends our search for systematic
errors, concentrating on those difficult to quantify in
\citet[hereafter \citetalias{MurphyM_01b}]{MurphyM_01b}. We summarize the
present work in Section \ref{sec:concs} and compare our new values of $\da$
with other constraints on $\alpha$-variation.

\subsection{QSO absorption line constraints on $\bmath{\da}$}\label{ssec:qso_da}

\subsubsection{The alkali-doublet (AD) method}\label{sssec:ad_method}

The relative wavelength separation between the two transitions of an alkali
doublet (AD) is proportional to $\alpha^2$
\citep[e.g.][]{BetheH_77a}. \citet{SavedoffM_56a} first analysed AD
separations seen in Seyfert galaxy emission spectra to obtain constraints
on $\alpha$-variation. Absorption lines from intervening clouds along the
line of sight to QSOs are substantially narrower than intrinsic emission
lines and therefore provide a more precise probe. \citet*{BahcallJ_67b}
first used AD absorption lines which seemed to be intrinsic to the
QSO. Narrower still are absorption lines arising from gas clouds at
significantly lower redshift than the background QSO. \citet*{WolfeA_76a}
first analysed Mg{\sc \,ii} doublets from a damped Lyman-$\alpha$ system at
$z_{\rm abs} \approx 0.524$. Since then, several authors
\citep[e.g.][]{CowieL_95a,VarshalovichD_96b,VarshalovichD_00a} have applied
the AD method to doublets of several different species (e.g.~C{\sc \,iv},
Si{\sc \,ii}, Si{\sc \,iv}, Mg{\sc \,ii} and Al{\sc \,iii}).

In \citet{MurphyM_01c} we obtained the strongest current AD constraint on
$\da$ by analysing 21 Si{\sc \,iv} doublets observed towards 8 QSOs with
Keck/HIRES. We found a weighted mean
\begin{equation}\label{eq:Si}
\da = (-0.5 \pm 1.3)\times 10^{-5}\,.
\end{equation}
The systematic error due to uncertainties in the \citet{GriesmannU_00a}
Si{\sc \,iv} laboratory wavelengths is $\pm 0.2\times 10^{-5}$. Recently,
\citet*{BahcallJ_03a} analysed 44 [O{\sc \,iii}] doublet QSO emission lines
from the Sloan Digital Sky Survey Early Data Release
\citep{SchneiderD_02a}. Their constraint, $\da = (-2 \pm 1.2) \times
10^{-4}$, is an order of magnitude weaker than equation
\ref{eq:Si}. However, if one allows or expects spatial variations in
$\alpha$, such a simplistic comparison is difficult since the two results
constrain $\da$ in very different environments (i.e. QSO emission regions
and intervening absorption clouds).

\subsubsection{The many-multiplet (MM) method}\label{sssec:mm_method}

The many-multiplet (MM) method was introduced in \citet{DzubaV_99a} and
\citet{WebbJ_99a} as a generalisation of the AD method. The MM method
constrains changes in $\alpha$ by utilizing many observed transitions from
different multiplets and different ions associated with each QSO absorption
system. The full details behind this technique are presented in
\cite{DzubaV_99b,DzubaV_01a,DzubaV_02a}. We presented a summary in
\citetalias{MurphyM_01a} and listed the many important advantages
the MM method holds over the AD method, including an effective
order-of-magnitude precision gain. Therefore, we only outline the salient
features of the method below.

For small shifts in $\alpha$ (i.e.~$\da \ll 1$), the rest wavenumber,
$\omega_z$, of a transition with a measured redshift, $z$, in a QSO
spectrum, can be written as
\begin{equation}\label{eq:omega}
\omega_z = \omega_0 + q_1x_z + q_2y_z\,,
\end{equation}
where $\omega_0$ is the wavenumber measured in the laboratory\footnote{This equation does not strictly hold
where level pseudocrossing occurs
\citep{DzubaV_01a,DzubaV_02a}. However, even in these cases, only the
derivative of $\omega$ with respect to $\alpha$ is required for $\da \la
10^{-3}$ (i.e.~for $\omega_z$ sufficiently close to $\omega_0$).}. If
$\alpha_z
\neq \alpha_0$ then
\begin{equation}\label{eq:xandy}
x_z\equiv\left(\frac{\alpha_z}{\alpha_0}\right)^2-1 \hspace{0.5cm}{\rm
and}\hspace{0.5cm} y_z\equiv\left(\frac{\alpha_z}{\alpha_0}\right)^4-1
\end{equation}
are non-zero and the magnitude and sign of $q_1$ and $q_2$ determine the
shift in the transition wavenumber. Since we only consider $\da \ll 1$, we
may write
\begin{equation}\label{eq:omega_z}
\omega_z = \omega_0 + qx_z\,,
\end{equation}
where we consolidate $q_1$ and $q_2$ into $q\equiv q_1+2q_2$. The $q$
coefficient represents all the relativistic corrections for the transition
of interest and varies both in magnitude and sign from transition to
transition. That is, if $\da\neq 0$, the QSO absorption lines will be
shifted in a distinct pattern with respect to their laboratory values.

\begin{figure}
\centerline{\psfig{file=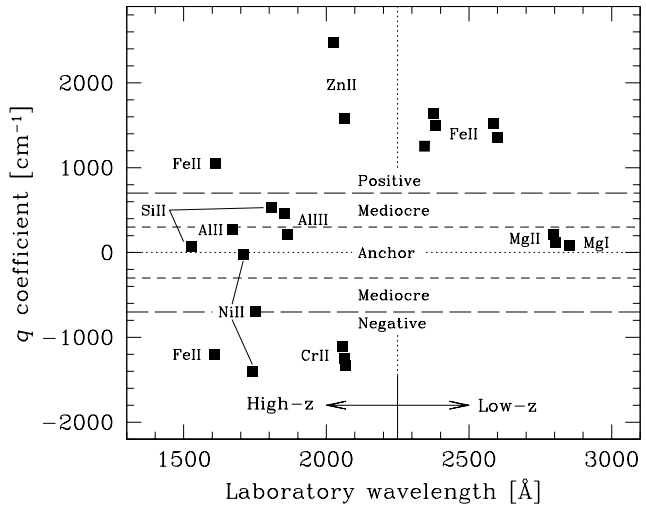,width=8.4cm}}
\caption{The distribution of $q$ coefficients in (rest) wavelength space
  for the low- and high-$z$ samples. Note the simple arrangement for the
  low-$z$ Mg/Fe{\sc \,ii} systems: the Mg transitions are used as anchors
  against which the large, positive shifts in the Fe{\sc \,ii} transitions
  can be measured. Compare this with the complex arrangement for the
  high-$z$ systems: low-order distortions of the wavelength scale will have
  a varied and complex effect on $\da$ depending on which transitions are
  fitted in a given absorption system. In general, the complexity at
  high-$z$ will yield more robust values of $\da$. For the discussion in
  Sections \ref{sssec:rem_comp} and \ref{sssec:rem_q} we define several
  different `$q$-types' as follows: positive-shifters ($q \ge 700{\rm
  \,cm}^{-1}$), negative-shifters ($q \le -700{\rm \,cm}^{-1}$), anchors
  ($\left|q\right| < 300{\rm \,cm}^{-1}$) and mediocre-shifters ($300 \le
  \left|q\right| < 700{\rm \,cm}^{-1}$).}
\label{fig:q_vs_wl}
\end{figure}

Fig.~\ref{fig:q_vs_wl} illustrates the distribution of $q$ coefficients for
the different transitions used in the MM method. The $q$ coefficients are
calculated in \citet{DzubaV_99b,DzubaV_01b,DzubaV_01a,DzubaV_02a} using
many-body techniques to include all dominant relativistic effects. The
values used in the present study are listed in Table
\ref{tab:atomdata}. The uncertainties in $q$ are typically $\la 30{\rm
\,cm}^{-1}$ for the Mg, Si, Al and Zn transitions but are $\la 300{\rm
\,cm}^{-1}$ for those of Cr, Fe and Ni due to the more complicated electronic
configurations involved. However, it is important to note that, in the
absence of systematic effects in the QSO spectra, the form of equation
\ref{eq:omega_z} ensures that errors in the $q$ coefficients can not lead
to a non-zero $\da$.

\citet{WebbJ_99a} first applied the MM method to 30 Keck/HIRES QSO
absorption systems (towards 14 QSOs) in the absorption redshift range $0.5
< z_{\rm abs} < 1.6$, fitting only the transitions of Mg{\sc \,i}, Mg{\sc
\,ii} and Fe{\sc \,ii}. The large positive $q$ coefficients for the Fe{\sc
\,ii} transitions are in sharp contrast to the small values for the Mg
transitions (Fig.~\ref{fig:q_vs_wl}): the Mg lines act as anchors against
which the large shifts in the Fe{\sc \,ii} lines can be used to constrain
$\da$. The difference in $q$ values ($\Delta q\approx 1200{\rm \,cm}^{-1}$)
is much larger than for the Si{\sc \,iv} transitions in the AD method
\citep[$\Delta q\approx 500{\rm \,cm}^{-1}$,][]{MurphyM_01c} and so the MM
method allows significantly increased precision. Additionally, analysing
the 5 Fe{\sc \,ii} and 3 Mg transitions allows a statistical gain over the
AD method. \citet{WebbJ_99a} explicitly demonstrated this increased
precision, obtaining tentative evidence for a smaller $\alpha$ in the
absorption clouds: $\da = (-1.09\pm 0.36)\times 10^{-5}$.

In \citetalias{MurphyM_01a} we analysed 18 independent Keck/HIRES
absorption systems in the range $1.8 < z_{\rm abs} < 3.5$. At these higher
redshifts, shorter (rest) wavelength transitions are shifted into the
optical bandpass. Fig.~\ref{fig:q_vs_wl} shows the range of $q$
coefficients available and their complex arrangement compared to the
low-$z$ Mg/Fe{\sc \,ii} systems. Such an arrangement should provide more
robust estimates of $\da$ since any systematic effects in the QSO spectra
will have a different effect on systems containing different
transitions. Despite this, the high-$z$ sample also yielded evidence for a
smaller $\alpha$ in the past, consistent with the low-$z$
systems. Combining this with a re-analysis of the
\citet{WebbJ_99a} data and including 3 low-$z$ Mg/Fe{\sc \,ii} systems
from the new data, we found 4.1\,$\sigma$ evidence for a variable $\alpha$
with 49 absorption systems over the range $0.5 < z_{\rm abs} < 3.5$:
\begin{equation}\label{eq:M01a_res}
\da = (-0.72 \pm 0.18)\times 10^{-5}\,.
\end{equation}

Such a potentially fundamental result requires extreme scrutiny and
warrants a thorough examination of possible systematic errors. We carried
out a detailed search for both instrumental and astrophysical systematic
effects in \citetalias{MurphyM_01b} but were unable to identify any which
explained the result in equation \ref{eq:M01a_res}. Upper limits were
placed on potential effects from atmospheric dispersion and possible
isotopic abundance evolution but we indicated that further exploration of
these possibilities was required.

In the present work we analyse a new, large sample of Keck/HIRES absorption
systems using the MM method. We also re-analyse the previous datasets and
significantly improve the analysis of potential systematic errors.

\section{New and previous data}\label{sec:data}

\subsection{Keck/HIRES QSO spectra}\label{ssec:keck_dat}

The QSO observations were carried out by three independent groups, all
using the high resolution spectrograph \citep[HIRES,][]{VogtS_94a} on the
Keck I 10-m telescope on Mauna Kea, Hawaii. We previously studied the first
two samples in \citetalias{MurphyM_01a} and we outline changes and
additions to these samples in Section \ref{sssec:prev_dat} below. The new,
large sample of 78 absorption systems is described in Section
\ref{sssec:new_dat}. Combining all three samples gives a total of 128
absorption systems observed towards 68 different QSOs. 8 absorption systems
(towards 8 QSOs) were observed by two different groups, bringing the total
number of independent Keck/HIRES spectra to 76. The absorption redshifts
cover the range $0.2 < z_{\rm abs} < 3.7$.

\subsubsection{Previous low- and high-$z$ samples}\label{sssec:prev_dat}

In \citetalias{MurphyM_01a} we analysed two independent Keck/HIRES
datasets, termed the low- and high-redshift samples. The previous low-$z$
sample was provided by C.~W. Churchill and comprises 27 Mg/Fe{\sc \,ii}
absorption systems in the spectra of 16 QSOs covering a redshift range $0.5
< z_{\rm abs} < 1.8$ \citep[e.g.][]{ChurchillC_00a}. An additional $z_{\rm
abs}=1.4342$ system was contributed by \citet{LuL_96a} but these data have
been re-reduced and are now included in the new sample. The previous
high-$z$ sample was provided by J.~X. Prochaska \& A.~M. Wolfe and
comprises 11 QSO spectra which contain 17 damped Lyman-$\alpha$ systems
(DLAs) in the range $1.8 < z_{\rm abs} < 3.5$ and 3 lower redshift
Mg/Fe{\sc \,ii} systems
\citep{ProchaskaJ_99b}. An additional DLA at $z_{\rm abs}=2.625$ was
contributed by \citet*{OutramP_99a}. In the present work we add 2 further
DLAs, again provided by J.~X. Prochaska \& A.~M. Wolfe, bringing the total
number of systems in the previous high-$z$ sample to 23. Details of the
data reduction for these previous samples are provided in
\citetalias{MurphyM_01a}. We have reanalysed these samples (Section
\ref{sec:analysis}) in a way consistent with the new, larger dataset
described in the next section.

\subsubsection{The new sample}\label{sssec:new_dat}

The main improvement on our previous results is due to a new dataset kindly
provided by W.~L.~W. Sargent. The observations were carried out during
numerous observing runs from 1993 November to 1999 November. The spectral
resolution for most spectra was ${\rm FWHM} \approx 6.6{\rm \,km\,s}^{-1}$
($R = 45000$). All spectra were reduced with the {\sc makee}\footnote{See
http://www2.keck.hawaii.edu:3636/inst/hires/makeewww.} package written by
T. Barlow. QSO exposure times ranged from 1000--6000\,s depending on the
QSO magnitude and between 3--16 separate exposures were combined to form
the final spectrum.

Thorium--argon (ThAr) emission lamp spectra, taken both before and after the
QSO exposures, were used to calibrate the wavelength scale. {\sc makee}
uses an internal list of ThAr air wavelengths derived using the
\citet{EdlenB_66a} formula from the measured vacuum thorium atlas of
\citet{PalmerB_83a} and the argon atlas of \citet{NorlenG_73a}. {\sc makee}
then converts the final wavelength scale to vacuum wavelengths using the
Cauchy formula from \citet{WeastR_79a} and not the (inverse of the) Edlen
formula. As we noted in \citetalias{MurphyM_01b}, this distortion of the
wavelength scale could lead to systematically non-zero values of $\da$. For
example, the effect on a $z_{\rm abs}=1.0$ Mg/Fe{\sc \,ii} system would be
a spurious $\da \approx -0.25\times 10^{-5}$. We have corrected the
wavelength scales of the reduced spectra to avoid such a distortion (see
\citetalias{MurphyM_01b} for details). We test the accuracy of this
wavelength scale in Section \ref{ssec:thar}. The effect of slight
miscalibrations (if any are present) on the overall $\da$ values is found
to be negligible.

\begin{figure}
\centerline{\psfig{file=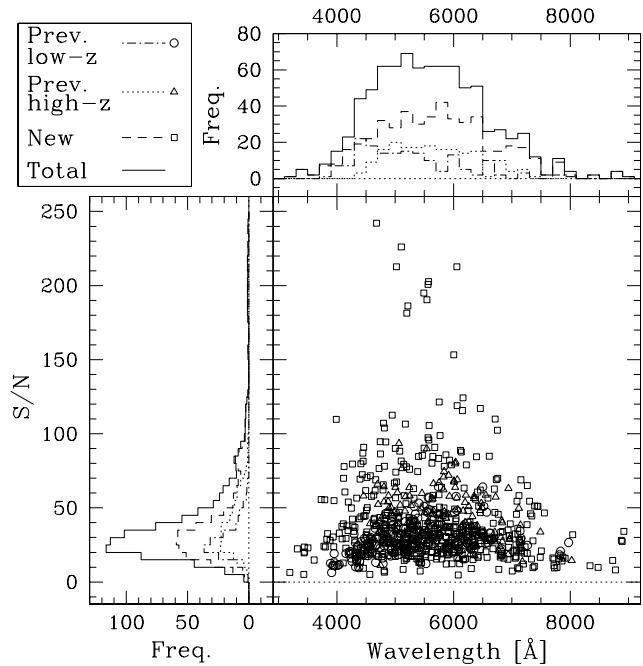,width=8.4cm}}
\caption{The distribution of S/N and central wavelength for the spectral
  regions of interest in the previous low-$z$, previous high-$z$ and new
  samples.}
\label{fig:cps_snr}
\end{figure}

For each QSO spectrum we identified all absorption systems which contained
enough MM transitions to yield meaningful constraints on $\da$. We selected
$\sim$500-km\,s$^{-1}$ sections around each transition and defined a
continuum by fitting a Legendre polynomial. 1\,$\sigma$ error arrays were
generated assuming Poisson counting statistics. The average S/N per pixel
for these spectral regions of interest ranged from 4--240, with most
sections having ${\rm S/N} \sim 30$ (see Fig.~\ref{fig:cps_snr}).

\begin{figure*}
\centerline{\psfig{file=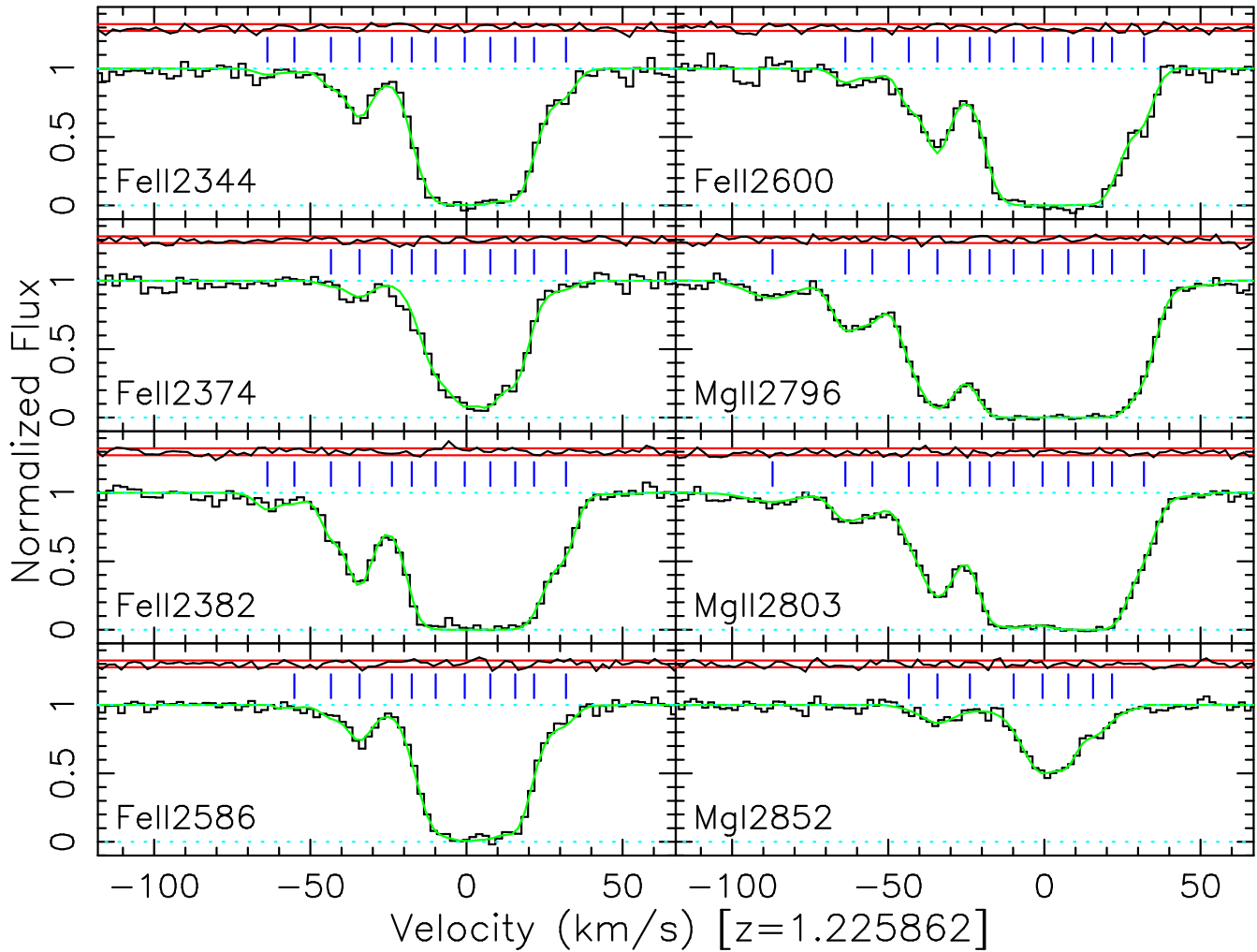,width=17.7cm,angle=270}}
\caption{Mg/Fe{\sc \,ii} absorption system towards Q1437$+$3007 at $z_{\rm
  abs}=1.2259$. The data have been normalized by a fit to the continuum and
  plotted as a histogram. Our Voigt profile fit (solid curve) and the
  residuals (i.e.~$[{\rm data}] - [{\rm fit}]$), normalized to the
  $1\sigma$ errors (horizontal solid lines), are also shown. The
  tick--marks above the continuum indicate individual velocity components.
  Note the range of linestrengths in Fe{\sc \,ii}, facilitating
  determination of the velocity structure. The large number of Fe{\sc \,ii}
  transitions and the large number of velocity components allows for tight
  constraints to be placed on $\da$. The Mg{\sc \,i} $\lambda$2852 and
  Fe{\sc \,ii} $\lambda$2374 transitions allow $\da$ to be constrained by
  the ${\rm velocity} \approx 0{\rm \,km\,s}^{-1}$ components whereas the
  Mg{\sc \,ii} and remaining Fe{\sc \,ii} transitions constrain $\da$ with
  the ${\rm velocity} \approx -35{\rm \,km\,s}^{-1}$ component.}
\label{fig:q1437}
\end{figure*}

\begin{figure*}
\centerline{\psfig{file=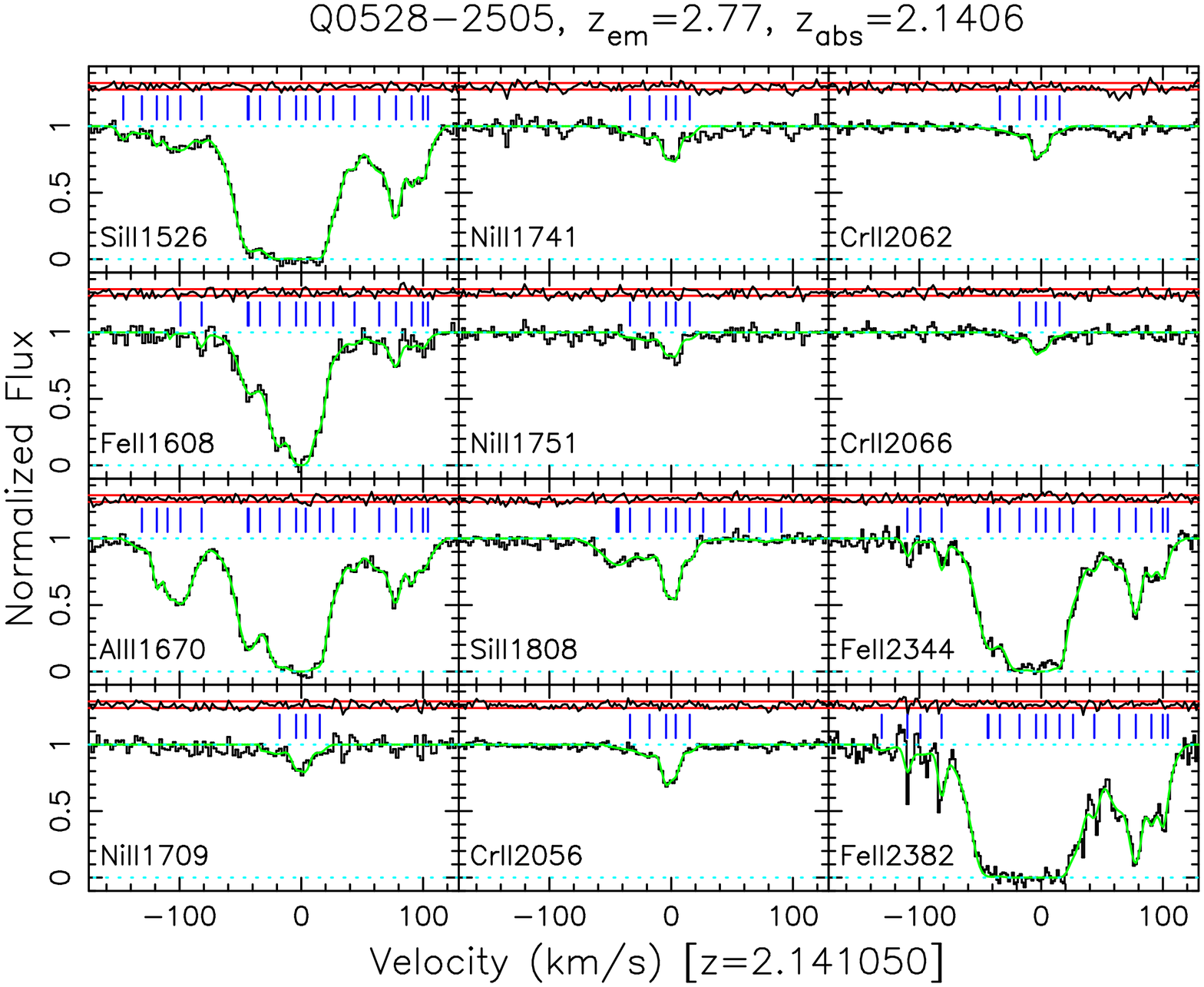,width=17.7cm,angle=270}}
\caption{Heavy element absorption lines in the damped Lyman-$\alpha$ system
  towards Q0528$-$2505 at $z_{\rm abs}=2.1406$. The data have been
  normalized by a fit to the continuum and plotted as a histogram. Our
  Voigt profile fit (solid curve) and the residuals (i.e.~$[{\rm data}] -
  [{\rm fit}]$), normalized to the $1\sigma$ errors (horizontal solid
  lines), are also shown. The tick--marks above the continuum indicate
  individual velocity components. Note the large range in linestrengths for
  the different transitions. The velocity structure is well-constrained by
  Si{\sc \,ii} $\lambda$1526, Al{\sc \,ii} $\lambda$1670 and Fe{\sc \,ii}
  $\lambda$1608 while the Ni{\sc \,ii} and Cr{\sc \,ii} transitions are
  strong enough to constrain $\da$ with the ${\rm velocity} \approx 0{\rm
  \,km\,s}^{-1}$ components. This system is at low enough redshift for the
  Fe{\sc \,ii} $\lambda\lambda$2344 and 2382 transitions to be
  detected. These combine with Si{\sc \,ii} $\lambda$1526, Al{\sc \,ii}
  $\lambda$1670 and Fe{\sc \,ii} $\lambda$1608 to constrain $\da$ with the
  components at ${\rm velocity} \approx 80{\rm \,km\,s}^{-1}$.}
\label{fig:q0528}
\end{figure*}

The new sample comprises 78 absorption systems, observed towards 46 QSOs,
covering a wide redshift range, $0.2 < z_{\rm abs} < 3.7$. Like the
previous samples, the new sample conveniently divides into low-$z$ ($z_{\rm
abs}<1.8$) and high-$z$ ($z_{\rm abs}>1.8$) subsamples. Table
\ref{tab:samples} shows the number of times each transition is used in the
different subsamples. Mg/Fe{\sc \,ii} systems dominate the low-$z$ region:
35 out of the 44 systems below $z_{\rm abs}=1.8$ contain only Fe{\sc \,ii}
and Mg{\sc \,i} or {\sc ii}, and, out of the remaining 9, only 1 system does
not contain transitions of these ions. We provide an example Mg/Fe{\sc
\,ii} absorption system, together with our Voigt profile fit, in
Fig.~\ref{fig:q1437}. Note that fitting a large number of transitions
allows the velocity structure of the absorption system to be reliably
determined. This is further facilitated by the range in Fe{\sc \,ii}
oscillator strengths and the presence of the weak Mg{\sc \,i} line. The
high-$z$ systems are characterized by a diverse range of transitions from
different ionic species. Table \ref{tab:samples} shows that the strong
Si{\sc \,ii} $\lambda$1526, Al{\sc \,ii} $\lambda$1670 and Fe{\sc \,ii}
$\lambda$1608 transitions are the most common. Fig.~\ref{fig:q0528}
illustrates a particularly high column density absorber which contains the
lower abundance species, Cr{\sc \,ii}, Ni{\sc \,ii} and Zn{\sc
\,ii}. Theoretically, the large magnitude of the $q$ coefficients for the
transitions of these species makes them very important for constraining
$\da$ (see Fig.~\ref{fig:q_vs_wl}). However, since their optical depths are
typically $\la 0.3$ and since they are not detected in all high-$z$
systems, their influence over $\da$ is reduced.

\begin{table}
\begin{center}
\caption{The frequency of occurrence for each transition in the different
samples of QSO absorbers. For the new sample, we define low-$z$ as $z_{\rm
abs}<1.8$ and high-$z$ as $z_{\rm abs}>1.8$. The previous low-$z$ sample
adheres to this definition but the previous high-$z$ sample does contain 3
low-$z$ systems. The low-$z$ samples are dominated by the Mg/Fe{\sc \,ii}
transitions. The high-$z$ systems contain a diverse range of transitions
and species but the strong Si{\sc \,ii} $\lambda$1526, Al{\sc \,ii}
$\lambda$1670 and Fe{\sc \,ii} $\lambda$1608 transitions are the most
common.}
\label{tab:samples}
\begin{tabular}{lccccccc}\hline
\multicolumn{1}{c}{Transition}&\multicolumn{7}{c}{Frequency of occurrence}\\
                              &\multicolumn{3}{c}{low-$z$ samples}&\multicolumn{3}{c}{high-$z$ samples}&Total\\
                              &Prev. &New &Tot.                   &Prev. &New &Tot.                    &     \\\hline 
Mg{\sc \,i} $\lambda$2852     &6     &21  &27                     &1     &0   &1                       &28\vspace{0.2cm}\\
Mg{\sc \,ii} $\lambda$2796    &25    &36  &61                     &2     &0   &2                       &63\\
Mg{\sc \,ii} $\lambda$2803    &26    &37  &63                     &3     &1   &4                       &67\vspace{0.2cm}\\
Al{\sc \,ii} $\lambda$1670    &0     &5   &5                      &11    &30  &41                      &46\vspace{0.2cm}\\
Al{\sc \,iii} $\lambda$1854   &0     &6   &6                      &6     &11  &17                      &23\\
Al{\sc \,iii} $\lambda$1862   &0     &6   &6                      &4     &9   &13                      &19\vspace{0.2cm}\\
Si{\sc \,ii} $\lambda$1526    &0     &3   &3                      &19    &26  &45                      &48\\
Si{\sc \,ii} $\lambda$1808    &0     &3   &3                      &15    &8   &23                      &26\vspace{0.2cm}\\
Cr{\sc \,ii} $\lambda$2056    &0     &2   &2                      &9     &7   &16                      &18\\
Cr{\sc \,ii} $\lambda$2062    &0     &1   &1                      &10    &7   &17                      &18\\
Cr{\sc \,ii} $\lambda$2066    &0     &0   &0                      &8     &7   &15                      &15\vspace{0.2cm}\\
Fe{\sc \,ii} $\lambda$1608    &0     &4   &4                      &19    &28  &47                      &51\\
Fe{\sc \,ii} $\lambda$1611    &0     &1   &1                      &9     &6   &15                      &16\\
Fe{\sc \,ii} $\lambda$2344    &21    &26  &47                     &5     &7   &12                      &59\\
Fe{\sc \,ii} $\lambda$2374    &10    &20  &30                     &3     &2   &5                       &35\\
Fe{\sc \,ii} $\lambda$2382    &22    &34  &56                     &3     &5   &8                       &64\\
Fe{\sc \,ii} $\lambda$2587    &20    &34  &54                     &3     &3   &6                       &60\\
Fe{\sc \,ii} $\lambda$2600    &25    &36  &61                     &3     &3   &6                       &67\vspace{0.2cm}\\
Ni{\sc \,ii} $\lambda$1709    &0     &0   &0                      &7     &7   &14                      &14\\
Ni{\sc \,ii} $\lambda$1741    &0     &1   &1                      &12    &6   &18                      &19\\
Ni{\sc \,ii} $\lambda$1751    &0     &1   &1                      &12    &8   &20                      &21\vspace{0.2cm}\\
Zn{\sc \,ii} $\lambda$2026    &0     &1   &1                      &7     &6   &13                      &14\\
Zn{\sc \,ii} $\lambda$2062    &0     &1   &1                      &7     &6   &13                      &14\\\hline
\end{tabular}
\end{center}
\end{table}

\subsection{Atomic data}\label{ssec:atom_dat}

We summarize all relevant atomic data for the MM transitions of interest in
Table \ref{tab:atomdata}. This is an updated version of table 1 in
\citetalias{MurphyM_01a}, the main changes being the following:
\begin{enumerate}
\item $q$ coefficients. As noted in Section \ref{ssec:qso_da}, the $q_1$
  and $q_2$ coefficients used in previous works are now replaced by a
  single value, $q \equiv q_1 + 2q_2$. The values of $q$ for Mg{\sc \,i}
  and Mg{\sc \,ii} are taken from \citet{DzubaV_99b,DzubaV_99a} and those
  for Al{\sc \,ii} and Al{\sc \,iii} are from \citet{DzubaV_01a}. All other
  $q$ coefficients are from updated calculations of the type detailed in
  \citet{DzubaV_02a}. Conservative uncertainties are estimated by comparing
  other calculated quantities (e.g.~energy intervals, $g$-factors) with
  their experimental values and by comparing the results of several
  different calculation techniques. Thus, the errors quoted in Table
  \ref{tab:atomdata} should be reliable but should not be treated as
  statistical.

  Of particular note are large changes to the $q$ coefficients for the
  Fe{\sc \,ii} $\lambda$1608 and Ni{\sc \,ii} lines. The $q$ coefficient
  used in \citetalias{MurphyM_01a} for Fe{\sc \,ii} $\lambda$1608 was
  $q=1002{\rm \,cm}^{-1}$ but this was based on an incorrect excited state
  electronic configuration reported in \cite{MooreC_71a}, $3d^64p~y^6{\rm
  P}_{7/2}^{\rm o}$. The configuration shown in Table \ref{tab:atomdata} is
  taken from \citet{DzubaV_02a} who also calculate the new $q$ coefficient,
  $q=-1200{\rm \,cm}^{-1}$. Since Fe{\sc \,ii} $\lambda$1608 is prominent
  in the high-$z$ systems (see Table \ref{tab:samples}), this change has an
  important effect on the values of $\da$ in the previous high-$z$
  sample. We discuss this in detail in Section \ref{sec:results}. The
  smaller changes to the Ni{\sc \,ii} $q$ coefficients are due to the
  complicated electronic configuration of Ni{\sc \,ii}
  \citep{DzubaV_01a,DzubaV_02a}. In this case, however, the effect on the
  values of $\da$ in the previous high-$z$ sample is very small since the
  Ni{\sc \,ii} transitions are weak and only appear in $\sim$25\,per cent
  of our high-$z$ systems (see Table \ref{tab:samples}).

\item Oscillator strengths. In order that the parameters derived from our
  Voigt profile fits to the QSO data can be compared with those of other
  studies, we have used the set of oscillator strengths recommended in the
  DLA database$^{\ref{footnote:tab1}}$ of \citet{ProchaskaJ_01a} (and
  references therein). These changes have a negligible effect on our values
  of $\da$. \addtocounter{footnote}{1}\footnotetext{Available at
  http://kingpin.ucsd.edu/$\sim$hiresdla.\label{footnote:tab1}}

\item Al{\sc \,iii} hyperfine structure. In \citetalias{MurphyM_01a} we
  used composite wavelengths to fit the Al{\sc \,iii} profiles. However,
  \citet{GriesmannU_00a} clearly resolved the hyperfine components in their
  laboratory Fourier transform spectra (FTS) and found that the hyperfine
  splitting is $\approx\!2.5{\rm \,km\,s}^{-1}$, comparable with the
  resolution of Keck/HIRES. In \citetalias{MurphyM_01b} we found that the
  composite values were adequate for our previous high-$z$ sample,
  primarily because Al{\sc \,iii} only appeared in 6 absorption
  systems. However, we now include the Al{\sc \,iii} hyperfine components
  listed in Table \ref{tab:atomdata} in our Voigt profile fits to the
  previous and new samples.
\end{enumerate}

Aside from Al, all the elements in Table \ref{tab:atomdata} have several
naturally occurring isotopes and so each QSO absorption line will be a
blend of absorption lines from each isotopic component. However, the
isotopic structures are only known for the transitions of Mg{\sc \,i}
\citep{HallstadiusL_79a} and Mg{\sc \,ii} \citep*{DrullingerR_80a}. The
latter authors also determined the hyperfine splitting constant for the
Mg{\sc \,ii} transitions. \cite{PickeringJ_98a} used laboratory FTS to
derive the wavelengths of the individual isotopes and main hyperfine
components for these transitions and these are the values given in Table
\ref{tab:atomdata}.

In \citetalias{MurphyM_01a} we estimated the isotopic structures for the
two Si{\sc \,ii} transitions by scaling the isotopic shifts for Mg{\sc
\,ii} $\lambda$2796 by the {\it mass} shift,
\begin{equation}\label{eq:mshift}
\Delta\omega_i \propto \omega_0/m^2\,,
\end{equation}
for $\Delta\omega_i$ the shift in wavenumber for isotope $i$ where $m$ is
the atomic mass. We give these estimated isotopic structures in Table
\ref{tab:atomdata}. Our approximation may break down if the {\it specific}
isotopic shifts for the Si{\sc \,ii} transitions are very different to
those of Mg{\sc \,ii} $\lambda$2796. Very recent theoretical calculations
\citep*{BerengutJ_03a} suggest that these are of opposite sign to the normal
mass shift and so the shifts in Table \ref{tab:atomdata} are probably
overestimates. We discuss the effect this may have on our values of $\da$
in Section \ref{ssec:iso}.

\begin{table*}
\centering
\vspace{-0.2cm}
\begin{minipage}{173mm}
\caption{Atomic data for the MM transitions in our analysis. Information
for isotopic and hyperfine components is given in italics. Column 2 shows
the mass number, $A$, for each species. The origin of the laboratory
wavenumbers ($\omega_0$) and wavelengths ($\lambda_0$) is summarized in
\citetalias{MurphyM_01a}. Columns 5 and 6 show the updated ground and
excited state electronic configurations. The ID letters in column 7 are
used in Table \ref{tab:da} to indicate the transitions used in our fits to
each absorption system. Column 8 shows the ionization potential for the
relevant ion, IP$^+$, and for the ion with a unit lower charge,
IP$^-$. Column 9 shows the oscillator strengths, $f$, from the DLA
database$^{\ref{footnote:tab1}}$ of \citet{ProchaskaJ_01a} or the relative
strengths of the isotopic \citep{RosmanK_98a} or hyperfine components
(italics). The $q$ coefficients are from
\citet{DzubaV_99b,DzubaV_99a,DzubaV_01a,DzubaV_02a} and the uncertainties
are discussed in the text. The Si{\sc \,ii}, Al{\sc \,ii} and Al{\sc \,iii}
wavenumbers have been scaled from their literature values due to the
\citeauthor{NorlenG_73a}/\citeauthor{WhalingW_95a} calibration difference.}
\label{tab:atomdata}
\begin{tabular}{lcllllcclr}\hline
\multicolumn{1}{c}{Ion}&\multicolumn{1}{c}{$A$}&\multicolumn{1}{c}{$\lambda_0$
(\AA)}&\multicolumn{1}{c}{$\omega_0$ (cm$^{-1}$)}&\multicolumn{1}{c}
{Ground}&\multicolumn{1}{c}{Upper}&\multicolumn{1}{c}{ID}&
\multicolumn{1}{c}{IP$^-$, IP$^+$ (eV)}&\multicolumn{1}{c}{$f$
or {\it \%}}&\multicolumn{1}{c} {$q$ (cm$^{-1}$)}\\\hline
Mg{\sc \,i}  &24.32   &2852.96310(8)       &35051.277(1)$^a$         &$3s^2~^1{\rm S}_0              $&$3s3p~^1{\rm P}_1                 $&a & --\,, 7.7&1.81        &     86(10)\\ 
             &{\it 26}&{\it ~2852.95977}   &{\it ~35051.318}$^a$     &$                              $&$                                 $&  &          &{\it ~11.0} &           \\
             &{\it 25}&{\it ~2852.96316}   &{\it ~35051.295}$^a$     &$                              $&$                                 $&  &          &{\it ~10.0} &           \\
             &{\it 24}&{\it ~2852.96359}   &{\it ~35051.271}$^a$     &$                              $&$                                 $&  &          &{\it ~79.0} &           \\ \\
Mg{\sc \,ii} &24.32   &2796.3543(2)        &35760.848(2)$^a$         &$3s~^2{\rm S}_{1/2}            $&$3p~^2{\rm P}_{3/2}               $&b & 7.7, 15.0&0.6123      &    211(10)\\
             &{\it 26}&{\it ~2796.3473}    &{\it ~35760.937}$^a$     &$                              $&$                                 $&  &          &{\it ~11.0} &           \\
             &{\it 25}&{\it ~2796.3492}    &{\it ~35760.913}$^a$     &$                              $&$                                 $&  &          &{\it  ~5.8} &           \\
             &{\it 25}&{\it ~2796.3539}    &{\it ~35760.853}$^a$     &$                              $&$                                 $&  &          &{\it  ~4.2} &           \\
             &{\it 24}&{\it ~2796.3553}    &{\it ~35760.835}$^a$     &$                              $&$                                 $&  &          &{\it ~79.0} &           \\
             &        &2803.5315(2)        &35669.298(2)$^a$         &$                              $&$3p~^2{\rm P}_{1/2}               $&c &          &0.3054      &    120(10)\\
             &{\it 26}&{\it ~2803.5244}    &{\it ~35669.388}$^a$     &$                              $&$                                 $&  &          &{\it ~11.0} &           \\
             &{\it 25}&{\it ~2803.5258}    &{\it ~35669.370}$^a$     &$                              $&$                                 $&  &          &{\it  ~3.0} &           \\
             &{\it 25}&{\it ~2803.5266}    &{\it ~35669.360}$^a$     &$                              $&$                                 $&  &          &{\it  ~2.8} &           \\
             &{\it 25}&{\it ~2803.5305}    &{\it ~35669.310}$^a$     &$                              $&$                                 $&  &          &{\it  ~1.5} &           \\
             &{\it 25}&{\it ~2803.5313}    &{\it ~35669.300}$^a$     &$                              $&$                                 $&  &          &{\it  ~2.7} &           \\
             &{\it 24}&{\it ~2803.5324}    &{\it ~35669.286}$^a$     &$                              $&$                                 $&  &          &{\it ~79.0} &           \\ \\
Al{\sc \,ii} &27.00   &1670.7887(1)        &59851.972(4)$^b$         &$3s^2~^1{\rm S}_0              $&$3s3p~^1{\rm P}_1                 $&d & 6.0, 18.9&1.88        &    270(30)\\ \\
Al{\sc \,iii}&27.00   &1854.71841(3)       &53916.540(1)$^b$         &$3s~^2{\rm S}_{1/2}            $&$3p~^2{\rm P}_{3/2}               $&e &18.9, 28.4&0.539       &    464(30)\\
             &        &{\it ~1854.70910(3)}&{\it ~53916.8111(8)}$^b$ &$                              $&$                                 $&  &          &{\it ~41.7} &           \\
             &        &{\it ~1854.72483(2)}&{\it ~53916.3536(6)}$^b$ &$                              $&$                                 $&  &          &{\it ~58.3} &           \\
             &        &1862.79126(7)       &53682.880(2)$^b$         &$                              $&$3p~^2{\rm P}_{1/2}               $&f &          &0.268       &    216(30)\\
             &        &{\it ~1862.78046(5)}&{\it ~53683.1915(15)}$^b$&$                              $&$                                 $&  &          &{\it ~41.7} &           \\
             &        &{\it ~1862.79871(4)}&{\it ~53682.6654(12)}$^b$&$                              $&$                                 $&  &          &{\it ~58.3} &           \\ \\
Si{\sc \,ii} &28.11   &1526.70709(2)       &65500.4492(7)$^b$        &$3s^23p~^2{\rm P}_{1/2}^{\rm o}$&$3s^24s~^2{\rm S}_{1/2}           $&g & 8.2, 16.3&0.127       &     68(30)\\
             &{\it 30}&{\it ~1526.7040}    &{\it ~65500.583}         &$                              $&$                                 $&  &          &{\it  ~3.1} &           \\
             &{\it 29}&{\it ~1526.7055}    &{\it ~65500.517}         &$                              $&$                                 $&  &          &{\it  ~4.7} &           \\
             &{\it 28}&{\it ~1526.7073}    &{\it ~65500.442}         &$                              $&$                                 $&  &          &{\it ~92.2} &           \\
             &        &1808.01301(1)       &55309.3365(4)$^b$        &$                              $&$3s3p^2~^2{\rm D}_{3/2}           $&h &          &0.00218     &    531(30)\\
             &{\it 30}&{\it ~1808.0094}    &{\it ~55309.446}         &$                              $&$                                 $&  &          &{\it  ~3.1} &           \\
             &{\it 29}&{\it ~1808.0113}    &{\it ~55309.390}         &$                              $&$                                 $&  &          &{\it  ~4.7} &           \\
             &{\it 28}&{\it ~1808.0132}    &{\it ~55309.330}         &$                              $&$                                 $&  &          &{\it ~92.2} &           \\ \\
Cr{\sc \,ii} &52.06   &2056.25693(8)       &48632.055(2)$^c$         &$3d^5~^6{\rm S}_{5/2}          $&$3d^44p~^6{\rm P}_{7/2}^{\rm o}   $&i & 6.8, 16.5&0.105       &$-$1107(150)\\
             &        &2062.23610(8)       &48491.053(2)$^c$         &$                              $&$3d^44p~^6{\rm P}_{5/2}^{\rm o}   $&j &          &0.078       &$-$1251(150)\\
             &        &2066.16403(8)       &48398.868(2)$^c$         &$                              $&$3d^44p~^6{\rm P}_{3/2}^{\rm o}   $&k &          &0.0515      &$-$1334(150)\\ \\
Fe{\sc \,ii} &55.91   &1608.45085(8)       &62171.625(3)$^d$         &$3d^64s~a^6{\rm D}_{9/2}       $&$3d^54s4p~y^6{\rm P}_{7/2}^{\rm o}$&l & 7.9, 16.2&0.0580      &$-$1200(300)\\
             &        &1611.20034(8)       &62065.528(3)$^d$         &$                              $&$3d^64p~y^4{\rm F}_{7/2}^{\rm o}  $&m &          &0.00136     &   1050(300)\\
             &        &2344.2130(1)        &42658.2404(2)$^e$        &$                              $&$3d^64p~z^6{\rm P}_{7/2}^{\rm o}  $&n &          &0.114       &   1254(150)\\
             &        &2374.4603(1)        &42114.8329(2)$^e$        &$                              $&$3d^64p~z^6{\rm F}_{9/2}^{\rm o}  $&o &          &0.0313      &   1640(150)\\
             &        &2382.7642(1)        &41968.0642(2)$^e$        &$                              $&$3d^64p~z^6{\rm F}_{11/2}^{\rm o} $&p &          &0.320       &   1498(150)\\
             &        &2586.6496(1)        &38660.0494(2)$^e$        &$                              $&$3d^64p~z^6{\rm D}_{7/2}^{\rm o}  $&q &          &0.06918     &   1520(150)\\
             &        &2600.1725(1)        &38458.9871(2)$^e$        &$                              $&$3d^64p~z^6{\rm D}_{9/2}^{\rm o}  $&r &          &0.23878     &   1356(150)\\ \\
Ni{\sc \,ii} &58.76   &1709.6042(1)        &58493.071(4)$^c$         &$3d^9~^2{\rm D}_{5/2}          $&$3d^84p~z^2{\rm F}_{5/2}^{\rm o}  $&s & 7.6, 18.2&0.0324      &  $-$20(250)\\
             &        &1741.5531(1)        &57420.013(4)$^c$         &$                              $&$3d^84p~z^2{\rm D}_{5/2}^{\rm o}  $&t &          &0.0427      &$-$1400(250)\\
             &        &1751.9157(1)        &57080.373(4)$^c$         &$                              $&$3d^84p~z^2{\rm F}_{7/2}^{\rm o}  $&u &          &0.0277      & $-$700(250)\\ \\
Zn{\sc \,ii} &65.47   &2026.13709(8)       &49355.002(2)$^c$         &$3d^{10}4s~^2{\rm S}_{1/2}     $&$3d^{10}4p~^2{\rm P}_{3/2}^{\rm o}$&v & 9.4, 17.8&0.489       &    2479(25)\\
             &        &2062.66045(9)       &48481.077(2)$^c$         &$                              $&$3d^{10}4p~^2{\rm P}_{1/2}^{\rm o}$&w &          &0.256       &    1577(25)\\\hline
\end{tabular}
{\footnotesize $^a$\citet*{PickeringJ_98a}; $^b$\citet{GriesmannU_00a};
$^c$\citet{PickeringJ_00a}; $^d$\citet{PickeringJ_02a};
$^e$\citet{NaveG_91a}.}
\end{minipage}
\end{table*}

At the time of our calculations, no measurements or theoretical estimates
of the isotopic structures existed for the other transitions (to our
knowledge) and so we used the composite values listed in Table
\ref{tab:atomdata}. Since that time, measurements \citep{MatsubaraK_02a}
and theoretical estimates \citep{BerengutJ_03a} of the Zn{\sc \,ii}
isotopic structure have become available. In Section \ref{ssec:iso} we
discuss the potential effects of differential isotopic saturation and
strong cosmological evolution of the isotopic abundances on our values of
$\da$.

Finally, we note that the \citet{WhalingW_95a} Ar{\sc \,ii} wavenumbers
used to calibrate the Si{\sc \,ii}, Al{\sc \,ii} and Al{\sc \,iii}
wavelengths are systematically larger than those of \citet{NorlenG_73a}
used to calibrate the wavelengths of the remaining lines. The difference
between the calibration scales is proportional to the wavenumber:
$\delta\omega = 7\times 10^{-8}\omega$. In \citetalias{MurphyM_01b} we note
that calibration errors of this type will be absorbed by the redshift
parameters of the QSO absorption lines and so, as long as all laboratory
wavelengths are normalized to the one calibration scale, values of $\da$
will be unaffected. We have therefore normalized all laboratory wavelengths
in Table \ref{tab:atomdata} to the
\citet{NorlenG_73a} calibration scale.

\section{Analysis}\label{sec:analysis}

We have detailed our analysis procedure in \citetalias{MurphyM_01a} and so
only provide a short summary here. For each absorption system we construct
multiple velocity component Voigt profile fits to the data using {\sc
vpfit} (v5)\footnote{Available at
http://www.ast.cam.ac.uk/$\sim$rfc/vpfit.html}. Each velocity component is
described by three parameters: the column density, the Doppler width or $b$
parameter and the redshift, $z_{\rm abs}$, of the absorbing gas. We have
modified {\sc vpfit} to include $\da$ as a free parameter: for each
transition we alter the rest wavelength using equation \ref{eq:omega_z} so
that all velocity components shift in concert. To constrain $\da$ one must
tie together the $z_{\rm abs}$ parameters of the corresponding velocity
components in different transitions. This assumes negligible proper motion
between the absorbing gas of different ionic species. We discuss this
assumption in detail in Section \ref{ssec:kin}.

\begin{table}
\caption{The raw results from the $\chi^2$ minimization procedure. For each
  absorption system we list the QSO emission redshift, $z_{\rm em}$, the
  nominal absorption system redshift, $z_{\rm abs}$, the transitions fitted
  and the value of $\da$ with associated 1\,$\sigma$ statistical error. A
  `*' after $\da$ indicates systems included in the `high-contrast'
  sample. An additional random error of $2.09 \times 10^{-5}$ should be
  added in quadrature to these systems to form the fiducial sample (see
  Section \ref{ssec:scat}).}
\label{tab:da}
\vspace{-0.5cm}
\begin{center}
\begin{tabular}{cllcr}\hline
Object     &\multicolumn{1}{c}{$z_{\rm em}$}&\multicolumn{1}{c}{$z_{\rm abs}$}&Transitions$^a$ &\multicolumn{1}{c}{$\da~(10^{-5})$}\\\hline
\multicolumn{3}{l}{Previous low-$z$ sample}\\ \\
0002$+$0507&1.90                            &0.85118                          &bcnopqr         &$-0.346 \pm 1.279\,\,\,           $\\
0117$+$2118&1.49                            &0.72913                          &abcqr           &$ 0.084 \pm 1.297\,\,\,           $\\
           &                                &1.0479                           &bcnpr           &$-0.223 \pm 2.200\,\,\,           $\\
           &                                &1.3246                           &bcpqr           &$ 0.695 \pm 0.803\,\,\,           $\\
           &                                &1.3428                           &cnpq            &$-1.290 \pm 0.948\,\,\,           $\\
0420$-$0127&0.915                           &0.63308                          &abcr            &$ 4.211 \pm 4.076\,\,\,           $\\
0450$-$1312&2.25                            &1.1743                           &bnopr           &$-3.070 \pm 1.098\,\,\,           $\\
           &                                &1.2294                           &bcnpqr          &$-1.472 \pm 0.836\,\,\,           $\\
           &                                &1.2324                           &bcp             &$ 1.017 \pm 2.752\,\,\,           $\\
0454$+$0356&1.34                            &0.85929                          &acnoprq         &$ 0.405 \pm 1.325\,\,\,           $\\
           &                                &1.1534                           &bcnqr           &$-0.749 \pm 1.782\,\,\,           $\\
0823$-$2220&0.91                            &0.91059                          &bcnpqr          &$-0.394 \pm 0.609\,\,\,           $\\
1148$+$3842&1.30                            &0.55339                          &bcqr            &$-1.861 \pm 1.716\,\,\,           $\\
1206$+$4557&1.16                            &0.92741                          &bcnopqr         &$-0.218 \pm 1.389\,\,\,           $\\
1213$-$0017&2.69                            &1.3196                           &abcnopqr        &$-0.738 \pm 0.760\,\,\,           $\\
           &                                &1.5541                           &bcnopqr         &$-1.268 \pm 0.892\,\,\,           $\\
1222$+$2251&2.05                            &0.66802                          &bcnpr           &$ 0.067 \pm 1.474\,\,\,           $\\
1225$+$3145&2.22                            &1.7954                           &abcnopr         &$-1.296 \pm 1.049\,\,\,           $\\
1248$+$4007&1.03                            &0.77292                          &bcnopqr         &$ 2.165 \pm 1.191\,\,\,           $\\
           &                                &0.85452                          &bcnpqr          &$-0.021 \pm 1.268\,\,\,           $\\
1254$+$0443&1.02                            &0.51934                          &abcqr           &$-3.371 \pm 3.247\,\,\,           $\\
           &                                &0.93426                          &bcnpqr          &$ 1.485 \pm 1.908\,\,\,           $\\
1317$+$2743&1.01                            &0.66004                          &bcnpqr          &$ 0.590 \pm 1.515\,\,\,           $\\
1421$+$3305&1.91                            &0.84324                          &bcnopqr         &$ 0.099 \pm 0.847\,\,\,           $\\
           &                                &0.90301                          &bcnopqr         &$-0.998 \pm 1.783\,\,\,           $\\
           &                                &1.1726                           &bcnpr           &$-2.844 \pm 1.448\,\,\,           $\\
1634$+$7037&1.34                            &0.99010                          &bcnpqr          &$ 1.094 \pm 2.459\,\,\,           $\\ \\ \\
\multicolumn{3}{l}{Previous high-$z$ sample}\\ \\
0019$-$1522&4.53                            &3.4388                           &ghl             &$ 0.925 \pm 3.958\,\,\,           $\\
0100$+$1300&2.68                            &2.3095                           &efgjklmvw       &$-3.941 \pm 1.368^*               $\\
0149$+$3335&2.43                            &2.1408                           &defghijklmstu   &$-5.112 \pm 2.118^*               $\\
0201$+$3634&2.49                            &1.4761                           &cnoqr           &$-0.647 \pm 1.219\,\,\,           $\\
           &                                &1.9550                           &ehil            &$ 1.989 \pm 1.048^*               $\\
           &                                &2.3240                           &deghl           &$ 0.758 \pm 1.592\,\,\,           $\\
           &                                &2.4563                           &dgl             &$-3.731 \pm 2.285\,\,\,           $\\
           &                                &2.4628                           &ghiltu          &$ 0.572 \pm 1.719^*               $\\
0347$-$3819&3.23                            &3.0247                           &gl              &$-2.795 \pm 3.429\,\,\,           $\\
0841$+$1256&2.55                            &2.3742                           &dghijtuvw       &$ 2.277 \pm 3.816^*               $\\
           &                                &2.4761                           &dghijklmtu      &$-4.304 \pm 1.944^*               $\\
1215$+$3322&2.61                            &1.9990                           &defghijlmtuvw   &$ 5.648 \pm 3.131^*               $\\
1759$+$7539&3.05                            &2.6253                           &eghklmstu       &$-0.750 \pm 1.387^*               $\\
           &                                &2.6253$^b$                       &dglmstu         &$-0.492 \pm 1.645^*               $\\
2206$-$1958&2.56                            &0.94841                          &bcnpqr          &$-3.659 \pm 1.855\,\,\,           $\\
           &                                &1.0172                           &abcnopqr        &$-0.322 \pm 0.732\,\,\,           $\\
           &                                &1.9204                           &dghijklmstuvw   &$ 1.878 \pm 0.702^*               $\\
2230$+$0232&2.15                            &1.8585                           &dghjlnpstu      &$-5.407 \pm 1.179^*               $\\
           &                                &1.8640                           &ghijklmnostuvw  &$-0.998 \pm 0.492^*               $\\
2231$-$0015&3.02                            &2.0653                           &ghjklmtuvw      &$-2.604 \pm 1.015^*               $\\
2348$-$1444&2.94                            &2.2794                           &ghl             &$ 1.346 \pm 4.180\,\,\,           $\\
2359$-$0216&2.31                            &2.0951                           &dghijklstuvw    &$-0.068 \pm 0.722^*               $\\
           &                                &2.1539                           &dfgl            &$ 4.346 \pm 3.338\,\,\,           $\\
\end{tabular}
\end{center}
\end{table}

\begin{table}
\contcaption{$\!\!.$ The new sample.}
\begin{center}
\begin{tabular}{cllcr}\hline
Object     &\multicolumn{1}{c}{$z_{\rm em}$}&\multicolumn{1}{c}{$z_{\rm abs}$}&Transitions$^a$ &\multicolumn{1}{c}{$\da~(10^{-5})$}\\\hline
\multicolumn{3}{l}{New sample}\\ \\
0000$-$2620&4.11                            &1.4342                           &bcpr            &$-1.256 \pm 1.167\,\,\,           $\\
           &                                &3.3897                           &dglm            &$-7.666 \pm 3.231\,\,\,           $\\
0002$+$0507&1.90                            &0.59137                          &bcnpqr          &$-3.100 \pm 2.428\,\,\,           $\\
           &                                &0.85118                          &efnopqr         &$ 0.494 \pm 1.021\,\,\,           $\\
0055$-$2659&3.66                            &1.2679                           &abcpqr          &$ 1.669 \pm 2.745\,\,\,           $\\
           &                                &1.3192                           &bcpqr           &$-2.642 \pm 2.457\,\,\,           $\\
           &                                &1.5337                           &abcnopqr        &$-1.319 \pm 1.072\,\,\,           $\\
0058$+$0155&1.96                            &0.61256                          &bcnopqr         &$ 0.374 \pm 1.189\,\,\,           $\\
           &                                &0.72508                          &bcnpqr          &$-2.637 \pm 3.522\,\,\,           $\\
0119$-$0437&1.95                            &0.65741                          &bcnopqr         &$ 7.123 \pm 4.599\,\,\,           $\\
0153$+$7427&2.33                            &0.74550                          &bcnpqr          &$-2.168 \pm 0.778\,\,\,           $\\
0207$+$0503&4.19                            &3.6663                           &dgl             &$-0.748 \pm 3.468\,\,\,           $\\
0216$+$0803&2.99                            &1.7680                           &eflnopqr        &$ 0.044 \pm 1.235\,\,\,           $\\
0237$-$2321&2.23                            &1.3650                           &deghlnop        &$-0.197 \pm 0.565\,\,\,           $\\
0241$-$0146&4.04                            &2.0994                           &dnop            &$-0.739 \pm 2.675\,\,\,           $\\
0302$-$2223&1.41                            &1.0092                           &abcnopqr        &$-0.189 \pm 1.008\,\,\,           $\\
0449$-$1325&3.09                            &1.2667                           &abcnopq         &$-1.212 \pm 1.430\,\,\,           $\\
0528$-$2505&2.77                            &0.94398                          &bcr             &$ 0.759 \pm 2.335\,\,\,           $\\
           &                                &2.1406                           &dghijklnpstu    &$-0.853 \pm 0.880^*               $\\
           &                                &2.8114                           &defgijklmstuvw  &$ 0.850 \pm 0.846^*               $\\
0636$+$6801&3.17                            &1.2938                           &bcpqr           &$-1.392 \pm 0.623\,\,\,           $\\
0741$+$4741&3.21                            &1.6112                           &abcnpq          &$-1.299 \pm 1.726\,\,\,           $\\
           &                                &3.0173                           &dghl            &$ 0.794 \pm 1.796\,\,\,           $\\
0757$+$5218&3.24                            &2.6021                           &defl            &$-1.396 \pm 1.955\,\,\,           $\\
           &                                &2.8677                           &dgl             &$ 3.837 \pm 3.288\,\,\,           $\\
0841$+$1256&2.55                            &1.0981                           &bnopr           &$-3.589 \pm 1.203\,\,\,           $\\
           &                                &1.1314                           &abcnpqr         &$ 0.562 \pm 0.787\,\,\,           $\\
           &                                &1.2189                           &abcnopqrvw      &$-0.522 \pm 0.542\,\,\,           $\\
           &                                &2.3742                           &deghijknqrtuvw  &$ 1.435 \pm 1.227^*               $\\
0930$+$2858&3.42                            &3.2351                           &dgl             &$ 0.867 \pm 1.777\,\,\,           $\\
0940$-$1050&3.05                            &1.0598                           &abcnpqr         &$-0.453 \pm 1.572\,\,\,           $\\
0956$+$1217&3.31                            &2.3103                           &defl            &$-2.161 \pm 5.977\,\,\,           $\\
1009$+$2956&2.62                            &1.1117                           &bcnpqr          &$-5.461 \pm 2.518\,\,\,           $\\
1011$+$4315&3.10                            &1.4162                           &bcnopqr         &$-0.892 \pm 0.552\,\,\,           $\\
           &                                &2.9587                           &deghlmsu        &$ 2.475 \pm 1.706^*               $\\
1055$+$4611&4.12                            &3.3172                           &dgl             &$ 2.706 \pm 5.677\,\,\,           $\\
1107$+$4847&2.97                            &0.80757                          &abcpq           &$ 1.199 \pm 1.222\,\,\,           $\\
           &                                &0.86182                          &abcqr           &$-2.030 \pm 1.632\,\,\,           $\\
           &                                &1.0158                           &abcp            &$-2.086 \pm 0.934\,\,\,           $\\
1132$+$2243&2.88                            &2.1053                           &defgl           &$ 6.323 \pm 3.622\,\,\,           $\\
1202$-$0725&4.70                            &1.7549                           &bcqr            &$-1.465 \pm 2.182\,\,\,           $\\
1206$+$4557&1.16                            &0.92741                          &abcnopqr        &$-0.275 \pm 0.776\,\,\,           $\\
1223$+$1753&2.94                            &2.4653                           &hijkmsuvw       &$ 1.635 \pm 1.919\,\,\,           $\\
           &                                &2.5577                           &dglnp           &$ 0.546 \pm 1.199\,\,\,           $\\
1225$+$3145&2.22                            &1.7954                           &defghl          &$ 1.352 \pm 1.388\,\,\,           $\\
1244$+$3142&2.95                            &0.85048                          &bcpr            &$-6.897 \pm 7.012\,\,\,           $\\
           &                                &2.7504                           &dgl             &$ 2.414 \pm 4.110\,\,\,           $\\
1307$+$4617&2.13                            &0.22909                          &abcr            &$ 2.551 \pm 5.392\,\,\,           $\\
1337$+$1121&2.97                            &2.7955                           &dhl             &$ 4.103 \pm 8.538\,\,\,           $\\
1425$+$6039&3.20                            &2.7698                           &dgl             &$-0.688 \pm 1.843\,\,\,           $\\
           &                                &2.8268                           &dgl             &$ 0.433 \pm 0.827\,\,\,           $\\
1437$+$3007&3.00                            &1.2259                           &abcnopqr        &$ 0.308 \pm 1.460\,\,\,           $\\
1442$+$2931&2.76                            &2.4389                           &dgl             &$-0.882 \pm 1.473\,\,\,           $\\
1549$+$1919&2.83                            &1.1425                           &cfnpqr          &$-0.076 \pm 0.671\,\,\,           $\\
           &                                &1.3422                           &nopq            &$-0.740 \pm 1.232\,\,\,           $\\
           &                                &1.8024                           &def             &$-3.050 \pm 2.473\,\,\,           $\\
1626$+$6433&2.32                            &0.58596                          &bcr             &$-1.977 \pm 4.529\,\,\,           $\\
           &                                &2.1102                           &defgl           &$-0.705 \pm 1.068\,\,\,           $\\
1634$+$7037&1.34                            &0.99010                          &abcdnopqr       &$-2.194 \pm 1.343\,\,\,           $\\
1850$+$4015&2.12                            &1.9900                           &ghijklmnopqrvw  &$-1.663 \pm 0.859^*               $\\
1946$+$7658&3.02                            &1.7385                           &efhijmntu       &$-0.212 \pm 1.857\,\,\,           $\\
           &                                &2.8433                           &dgl             &$-4.743 \pm 1.289\,\,\,           $\\
\end{tabular}			  
\end{center}			  
\end{table}			  

\begin{table}
\contcaption{$\!\!.$ The new sample (continued).}
\begin{center}
\begin{tabular}{cllcr}\hline
Object     &\multicolumn{1}{c}{$z_{\rm em}$}&\multicolumn{1}{c}{$z_{\rm abs}$}&Transitions$^a$ &\multicolumn{1}{c}{$\da~(10^{-5})$}\\\hline
2145$+$0643&1.00                            &0.79026                          &bcnopqr         &$ 0.087 \pm 0.589\,\,\,           $\\
2206$-$1958&2.56                            &1.0172                           &abcqr           &$ 1.354 \pm 0.883\,\,\,           $\\
           &                                &2.0762                           &dl              &$ 1.429 \pm 3.022\,\,\,           $\\
2231$-$0015&3.02                            &1.2128                           &abcopqr         &$ 1.223 \pm 1.465\,\,\,           $\\
           &                                &2.0653                           &efghijklmstuvw  &$ 1.707 \pm 1.249^*               $\\
           &                                &2.6532                           &dglstu          &$-3.348 \pm 1.904^*               $\\
2233$+$1310&3.30                            &2.5480                           &dgl             &$ 2.942 \pm 5.207\,\,\,           $\\
           &                                &2.5548                           &defgl           &$ 1.732 \pm 6.349\,\,\,           $\\
           &                                &3.1513                           &dgl             &$-4.005 \pm 3.301\,\,\,           $\\
2343$+$1232&2.52                            &0.73117                          &abcoqr          &$-1.211 \pm 0.975\,\,\,           $\\
           &                                &1.5899                           &acdefgqr        &$ 0.453 \pm 1.187\,\,\,           $\\
           &                                &2.1711                           &cdefnp          &$-0.961 \pm 1.295\,\,\,           $\\
           &                                &2.4300                           &dgijklnqrstuvw  &$-1.224 \pm 0.389^*               $\\
2344$+$1228&2.77                            &1.0465                           &abcinopqr       &$-0.747 \pm 1.530\,\,\,           $\\
           &                                &1.1161                           &abcopqr         &$ 0.009 \pm 1.963\,\,\,           $\\
           &                                &2.5378                           &dgl             &$-3.205 \pm 2.094\,\,\,           $\\\hline
\end{tabular}
\end{center}{\footnotesize
$^a$Transitions identified as in Table \ref{tab:atomdata}; $^b$This
absorber contributed by \citet{OutramP_99a}.}
\end{table}

We minimize $\chi^2$ simultaneously for all fitted transitions to find the
best fitting value of $\da$. Parameter errors are calculated from the
diagonal terms of the final parameter covariance matrix
\citep{FisherR_58a}. We have performed Monte Carlo simulations of
absorption systems of varying complexity (i.e.~different numbers of
velocity components and transitions fitted) and with a range of input
values of $\da$ to verify that {\sc vpfit} returns the correct values and
1\,$\sigma$ errors (see also Section \ref{ssec:remove}).

We impose two self-consistency checks on each absorption system before
accepting a value of $\da$. Firstly, we require the value of $\chi^2$ per
degree of freedom, $\chi^2_\nu$, to be $\sim$1. Secondly, we construct two
fits to the data with different constraints on the $b$ parameters: (i)
entirely thermal broadening and (ii) entirely turbulent
broadening\footnote{This is a minor departure from our previous
analysis. In \citetalias{MurphyM_01a} we constructed a third fit for each
absorption system where the thermal and turbulent component of the $b$
parameters were determined by goodness of fit. Such a fit is more
computationally intensive and, in general, the thermal and turbulent $b$
parameters are poorly determined. In \citetalias{MurphyM_01a} we found that
this fit did not lead to any inconsistent values of $\da$ and so fits (i)
and (ii) provide a sufficient consistency check.}. We require that the
values of $\da$ derived from both fits be consistent with each other since
the choice above should not greatly affect $\da$ (i.e.~the measured
position of velocity components should not change systematically). We
rejected only 4 systems in this way.

\begin{figure}
\centerline{\psfig{file=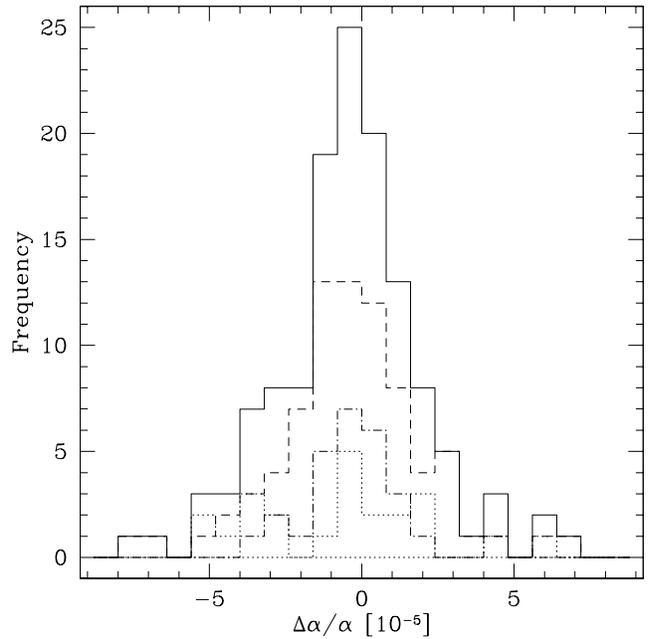,width=8.4cm}}
\caption{Histograms of raw results from Table \ref{tab:da}. The distribution of
  $\da$ for the previous low-$z$ (dot-dashed line), previous high-$z$
  (dotted line) and new (dashed line) samples are shown together with the
  overall distribution (solid line, binned differently in
  Fig.~\ref{fig:cps_thar}a).}
\label{fig:cps_hist}
\end{figure}

\section{Results}\label{sec:results}

\subsection{Raw QSO results}\label{ssec:qsores}

We present the $\chi^2$ minimization results in Table \ref{tab:da} for the
previous and new samples separately. We list the QSO (B1950) name, the
emission redshift, $z_{\rm em}$, the nominal absorption redshift, $z_{\rm
abs}$, and $\da$ for each absorption system with the associated 1\,$\sigma$
statistical error. The transitions fitted in each system are indicated with
the ID letters defined in Table \ref{tab:atomdata}. We present basic
statistics for the different samples in Table \ref{tab:stats}:
$\left<\da\right>_{\rm w}$ is the weighted mean with 1\,$\sigma$ error,
$\left<\da\right>$ is the unweighted mean, $S_{0}$ is the significance of
the departure of the weighted mean from zero, rms is the root-mean-square
deviation from the mean of $\da$, $\left<\delta(\da)\right>$ is the mean
1\,$\sigma$ error and $\chi^2_\nu$ is the value of $\chi^2$ per degree of
freedom with $\left<\da\right>_{\rm w}$ as the model.

The statistics in Table \ref{tab:stats} indicate a smaller $\alpha$ over
the redshift range $0.2 < z_{\rm abs} < 3.7$ at the 5.6\,$\sigma$
significance level:
\begin{equation}\label{eq:da}
\da = (-0.574 \pm 0.102)\times 10^{-5}\,.
\end{equation}
We note the consistency between the previous low-$z$, previous high-$z$ and
new samples. Breaking the new sample down into low-$z$ ($z_{\rm abs} <
1.8$) and high-$z$ ($z_{\rm abs} > 1.8$) subsamples also yields consistent
results. We also see an overall agreement between the weighted and
unweighted means, indicating that our results are not dominated by a small
number of highly significant points. This is supported by the histogram in
Fig.~\ref{fig:cps_hist} which shows a symmetric, roughly Gaussian
distribution for the values of $\da$.

\begin{table*}
\centering
\begin{minipage}{173mm}
\caption{Statistics for the raw QSO results and tests for
  systematic errors. The previous samples are described in Section
  \ref{sssec:prev_dat}. For the new and total samples, `low-$z$' means
  $z_{\rm abs}<1.8$ and `high-$z$' means $z_{\rm abs}>1.8$. For each
  sample, $\left<z_{\rm abs}\right>$ is the mean absorption redshift,
  $N_{\rm abs}$ is the number of absorption systems, $\left<\da\right>_{\rm
  w}$ is the weighted mean, $\left<\da\right>$ is the unweighted mean,
  $S_0$ is the significance of the weighted mean (with respect to zero),
  rms is the root-mean-square deviation from the mean,
  $\left<\delta(\da)\right>$ is the mean 1\,$\sigma$ error and $\chi^2_\nu$
  is $\chi^2$ per degree of freedom using $\left<\da\right>_{\rm w}$ as the
  model.}
\label{tab:stats}
\begin{tabular}{lcccccccc}\hline
Sample                        &$\left<z_{\rm abs}\right>$&$N_{\rm abs}$&$\left<\da\right>_{\rm w}$&$\left<\da\right>        $&$S_{0}                   $&rms        &$\left<\delta(\da)\right>$&$\chi^2_\nu$\\\hline
\multicolumn{9}{l}{\bf QSO results}\vspace{0.2cm}\\
\multicolumn{9}{l}{~~Raw results (Fig.~\ref{fig:cps})}\\
Prev. low-$z$                 &1.00                      &27           &$-0.513 \pm 0.224        $&$-0.305 \pm 0.304        $&2.28\,$\sigma            $&1.579      &$1.554 \pm 0.153         $&0.981       \\
Prev. high-$z$                &2.17                      &23           &$-0.672 \pm 0.244        $&$-0.656 \pm 0.601        $&2.76\,$\sigma            $&2.880      &$1.951 \pm 0.232         $&3.082       \vspace{0.1cm}\\
New low-$z$                   &1.11                      &44           &$-0.537 \pm 0.159        $&$-0.652 \pm 0.321        $&3.37\,$\sigma            $&2.131      &$1.757 \pm 0.204         $&1.123       \\
New high-$z$                  &2.60                      &34           &$-0.623 \pm 0.224        $&$-0.053 \pm 0.473        $&2.78\,$\sigma            $&2.755      &$2.604 \pm 0.318         $&1.533       \\
New total                     &1.76                      &78           &$-0.566 \pm 0.130        $&$-0.391 \pm 0.276        $&4.36\,$\sigma            $&2.441      &$2.126 \pm 0.186         $&1.284       \vspace{0.1cm}\\
low-$z$ total                 &1.07                      &74           &$-0.539 \pm 0.130        $&$-0.562 \pm 0.226        $&4.24\,$\sigma            $&1.941      &$1.663 \pm 0.134         $&1.051       \\
high-$z$ total                &2.50                      &54           &$-0.637 \pm 0.172        $&$-0.227 \pm 0.389        $&3.72\,$\sigma            $&2.862      &$2.400 \pm 0.224         $&2.179       \\
{\bf Raw sample}              &{\bf 1.67}                &{\bf 128}    &$\bmath{-0.574 \pm 0.102}$&$\bmath{-0.421 \pm 0.210}$&{\bf 5.62}\,$\bmath{\sigma}$&{\bf 2.379}&$\bmath{1.974 \pm 0.127} $&{\bf 1.515} \vspace{0.25cm}\\
\multicolumn{9}{l}{~~Removing extra scatter at high-$z$ (Figs.~\ref{fig:cps_hiz_alt} and \ref{fig:cps_fiducial})}\\
High-contrast sample removed  &2.65                      &32           &$-0.518 \pm 0.347        $&$-0.011 \pm 0.517        $&1.49\,$\sigma            $&2.926      &$3.070 \pm 0.308         $&1.207       \\
High-contrast errors increased&2.50                      &54           &$-0.563 \pm 0.291        $&$-0.227 \pm 0.389        $&1.94\,$\sigma            $&2.862      &$2.881 \pm 0.191         $&1.103       \\
All high-$z$ errors increased &2.50                      &54           &$-0.601 \pm 0.328        $&$-0.227 \pm 0.389        $&1.83\,$\sigma            $&2.862      &$2.986 \pm 0.194         $&1.000       \\
High-$z$ $\sigma$-clip        &2.54                      &44           &$-0.489 \pm 0.191        $&$ 0.273 \pm 0.341        $&2.57\,$\sigma            $&2.261      &$2.559 \pm 0.263         $&0.997       \\
{\bf Fiducial sample}         &{\bf 1.67}                &{\bf 128}    &$\bmath{-0.543 \pm 0.116}$&$\bmath{-0.421 \pm 0.210}$&{\bf 4.66}\,$\bmath{\sigma}$&{\bf 2.379}&$\bmath{2.177 \pm 0.124} $&{\bf 1.065} \vspace{0.25cm}\\
\multicolumn{9}{l}{\bf Systematic error tests}\vspace{0.2cm}\\
\multicolumn{9}{l}{~~ThAr calibration test: $(\da)_{\rm ThAr}$ (Fig.~\ref{fig:cps_thar})}\\
Prev. low-$z$                 &1.00                      &27           &$0.016 \pm 0.021         $&$0.034 \pm 0.045         $&0.74\,$\sigma            $&0.234      &$0.129 \pm 0.008         $&3.697       \\
Prev. high-$z$                &2.10                      &18           &$-0.013 \pm 0.020        $&$-0.015 \pm 0.059        $&0.66\,$\sigma            $&0.250      &$0.165 \pm 0.043         $&3.212       \vspace{0.1cm}\\
New low-$z$                   &1.14                      &41           &$-0.008 \pm 0.014        $&$-0.003 \pm 0.036        $&0.59\,$\sigma            $&0.229      &$0.123 \pm 0.016         $&3.095       \\
New high-$z$                  &2.63                      &32           &$0.019 \pm 0.014         $&$0.016 \pm 0.052         $&1.38\,$\sigma            $&0.293      &$0.148 \pm 0.022         $&2.714       \\
New total                     &1.79                      &73           &$0.006 \pm 0.010         $&$0.005 \pm 0.030         $&0.57\,$\sigma            $&0.259      &$0.134 \pm 0.013         $&2.915       \vspace{0.1cm}\\
low-$z$ total                 &1.08                      &71           &$-0.006 \pm 0.011        $&$0.006 \pm 0.027         $&0.54\,$\sigma            $&0.230      &$0.124 \pm 0.010         $&3.286       \\
high-$z$ total                &2.52                      &47           &$0.014 \pm 0.012         $&$0.013 \pm 0.041         $&1.24\,$\sigma            $&0.284      &$0.159 \pm 0.022         $&2.827       \\
{\bf Total}                   &{\bf 1.66}                &{\bf 118}    &$\bmath{0.004 \pm 0.008} $&$\bmath{0.009 \pm 0.023}$&{\bf 0.48}\,$\bmath{\sigma}$&{\bf 0.253}&$\bmath{0.138 \pm 0.011} $&{\bf 3.091}  \vspace{0.25cm}\\
\multicolumn{9}{l}{~~Composite wavelengths only: $(\da)_{\rm comp}$ (Fig.~\ref{fig:cps_noiso_comp}, top panel)}\\
Prev. low-$z$                 &1.00                      &27           &$-0.802 \pm 0.227        $&$-0.557 \pm 0.334        $&3.54\,$\sigma            $&1.736      &$1.567 \pm 0.156         $&1.094       \\
Prev. high-$z$                &2.17                      &23           &$-0.736 \pm 0.243        $&$-0.731 \pm 0.608        $&3.03\,$\sigma            $&2.918      &$1.929 \pm 0.230         $&3.178       \vspace{0.1cm}\\
New low-$z$                   &1.11                      &44           &$-0.550 \pm 0.161        $&$-0.609 \pm 0.329        $&3.42\,$\sigma            $&2.184      &$1.779 \pm 0.209         $&1.060       \\
New high-$z$                  &2.60                      &34           &$-0.650 \pm 0.224        $&$-0.085 \pm 0.456        $&2.90\,$\sigma            $&2.660      &$2.636 \pm 0.326         $&1.457       \\
New total                     &1.76                      &78           &$-0.584 \pm 0.131        $&$-0.380 \pm 0.274        $&4.47\,$\sigma            $&2.417      &$2.152 \pm 0.191         $&1.218       \vspace{0.1cm}\\
low-$z$ total                 &1.07                      &74           &$-0.650 \pm 0.128        $&$-0.648 \pm 0.238        $&5.08\,$\sigma            $&2.045      &$1.681 \pm 0.138         $&1.098       \\
high-$z$ total                &2.50                      &54           &$-0.667 \pm 0.172        $&$-0.251 \pm 0.380        $&3.89\,$\sigma            $&2.792      &$2.411 \pm 0.229         $&2.129       \\
{\bf Total}                   &{\bf 1.67}                &{\bf 128}    &$\bmath{-0.656 \pm 0.103}$&$\bmath{-0.481 \pm 0.212}$&{\bf 6.39}\,$\bmath{\sigma}$&{\bf 2.397}&$\bmath{1.989 \pm 0.129} $&{\bf 1.520} \vspace{0.25cm}\\
\multicolumn{9}{l}{~~Strong isotopes only: $(\da)_{\rm iso}$ (Fig.~\ref{fig:cps_noiso_comp}, middle panel)}\\
Prev. low-$z$                 &1.00                      &27           &$-1.204 \pm 0.228        $&$-0.885 \pm 0.353        $&5.29\,$\sigma            $&1.834      &$1.567 \pm 0.157         $&1.257       \\
Prev. high-$z$                &2.17                      &23           &$-0.738 \pm 0.244        $&$-0.734 \pm 0.619        $&3.03\,$\sigma            $&2.967      &$1.928 \pm 0.226         $&3.210       \vspace{0.1cm}\\
New low-$z$                   &1.11                      &44           &$-0.860 \pm 0.159        $&$-0.834 \pm 0.323        $&5.40\,$\sigma            $&2.145      &$1.768 \pm 0.206         $&1.279       \\
New high-$z$                  &2.60                      &34           &$-0.634 \pm 0.224        $&$-0.077 \pm 0.463        $&2.84\,$\sigma            $&2.700      &$2.626 \pm 0.325         $&1.494       \\
New total                     &1.76                      &78           &$-0.784 \pm 0.130        $&$-0.504 \pm 0.275        $&6.04\,$\sigma            $&2.432      &$2.142 \pm 0.190         $&1.363       \vspace{0.1cm}\\
low-$z$ total                 &1.07                      &74           &$-0.984 \pm 0.128        $&$-0.913 \pm 0.239        $&7.71\,$\sigma            $&2.055      &$1.674 \pm 0.137         $&1.300       \\
high-$z$ total                &2.50                      &54           &$-0.638 \pm 0.171        $&$-0.231 \pm 0.384        $&3.72\,$\sigma            $&2.824      &$2.404 \pm 0.228         $&2.143       \\
{\bf Total}                   &{\bf 1.67}                &{\bf 128}    &$\bmath{-0.861 \pm 0.102}$&$\bmath{-0.626 \pm 0.215}$&{\bf 8.41}\,$\bmath{\sigma}$&{\bf 2.433}&$\bmath{1.982 \pm 0.128} $&{\bf 1.663} \vspace{0.25cm}\\
\multicolumn{9}{l}{~~Atmospheric dispersion corrected: $(\da)_{\rm adc}$ (Fig.~\ref{fig:cps_ad_corr})}\\
Prev. low-$z$                 &1.00                      &27           &$-0.098 \pm 0.225        $&$ 0.142 \pm 0.339        $&0.43\,$\sigma            $&1.762      &$1.554 \pm 0.153         $&1.208       \\
Prev. high-$z$                &2.17                      &23           &$-0.637 \pm 0.244        $&$-0.668 \pm 0.583        $&2.61\,$\sigma            $&2.794      &$1.951 \pm 0.232         $&2.979       \vspace{0.1cm}\\
New low-$z$                   &1.11                      &44           &$-0.416 \pm 0.159        $&$-0.557 \pm 0.319        $&2.61\,$\sigma            $&2.118      &$1.757 \pm 0.204         $&1.050       \\
New high-$z$                  &2.60                      &34           &$-0.740 \pm 0.224        $&$-0.189 \pm 0.472        $&3.31\,$\sigma            $&2.755      &$2.604 \pm 0.318         $&1.545       \\
New total                     &1.76                      &78           &$-0.525 \pm 0.130        $&$-0.397 \pm 0.274        $&4.48\,$\sigma            $&2.423      &$2.126 \pm 0.186         $&1.266       \vspace{0.1cm}\\
low-$z$ total                 &1.07                      &74           &$-0.288 \pm 0.127        $&$-0.309 \pm 0.232        $&2.27\,$\sigma            $&1.999      &$1.663 \pm 0.134         $&1.118       \\
high-$z$ total                &2.50                      &54           &$-0.764 \pm 0.172        $&$-0.363 \pm 0.386        $&4.45\,$\sigma            $&2.837      &$2.400 \pm 0.224         $&2.099       \\
{\bf Total}                   &{\bf 1.67}                &{\bf 128}    &$\bmath{-0.457 \pm 0.102}$&$\bmath{-0.332 \pm 0.211}$&{\bf 4.48}\,$\bmath{\sigma}$&{\bf 2.389}&$\bmath{1.974 \pm 0.127} $&{\bf 1.558} \\\hline
\end{tabular}
\end{minipage}
\end{table*}

\begin{figure*}
\centerline{\psfig{file=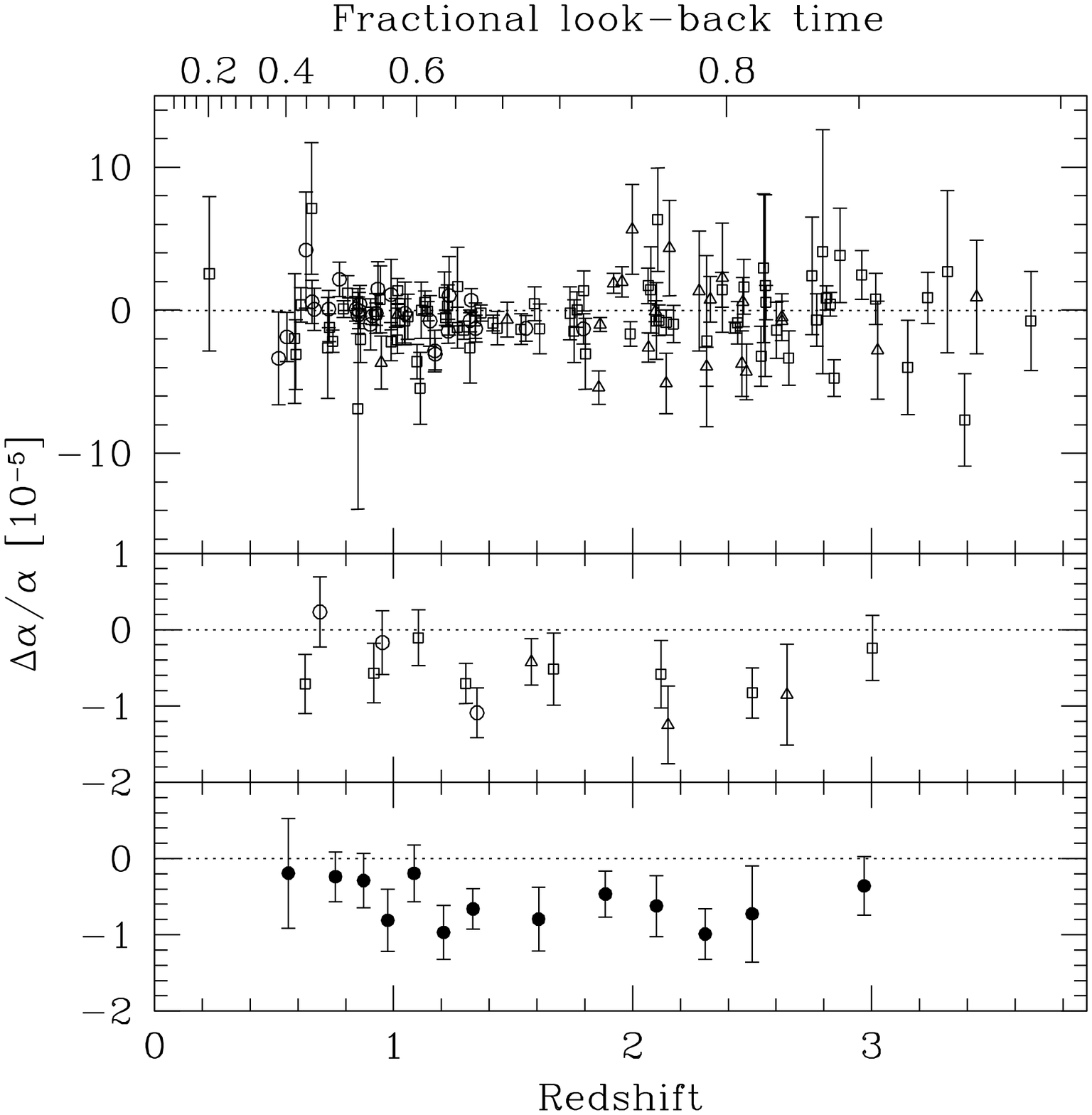,width=14.0cm}}
\caption{$\da$ versus absorption redshift for the previous low-$z$ (open
  circles), previous high-$z$ (open triangles) and new (open squares)
  samples. The upper panel shows the raw results and $1\sigma$ error
  bars. The middle panel shows an arbitrary binning of each sample and the
  lower panel combines all three samples. The redshifts of the binned
  points are taken as the mean absorption redshift within each bin and the
  value of $\da$ is the weighted mean with its associated $1\sigma$ error
  bar. The upper scale is the look-back time to the absorber as a fraction
  of the age of the universe ($H_0=70{\rm \,km\,s}^{-1}{\rm Mpc}^{-1}$,
  $\Omega_{\rm m}=0.3$, $\Omega_\Lambda=0.7 \rightarrow t_0=13.47{\rm
  \,Gyr}$).}
\label{fig:cps}
\end{figure*}

We illustrate the distribution of $\da$ over redshift and cosmological time
in Fig.~\ref{fig:cps}. The upper panel shows the raw values of $\da$ (with
their 1\,$\sigma$ error bars) for each sample separately. The middle panel
shows an arbitrary binning of each sample such that each bin has an equal
number of points (aside from the highest $z$ bin which also contains the
remaining points after dividing the total number by the number of bins). We
plot the weighted mean for each bin with the associated 1\,$\sigma$ error
bar. The lower panel combines all three samples. Note the expanded vertical
scales on the lower two panels.

\subsection{Extra scatter at high-$\bmath{z}$}\label{ssec:scat}

In Table \ref{tab:stats} we note that the scatter in the total low-$z$
sample is consistent with that expected from the size of the error bars
[i.e.~${\rm rms} \approx \left<\delta(\da)\right>$ and $\chi^2_\nu \approx
1$]. However, at higher $z$, Fig.~\ref{fig:cps} shows several values of
$\da$ which are significantly different from one another, particularly
around $z_{\rm abs} \approx 1.9$. Indeed, Table \ref{tab:stats} shows that
${\rm rms} > \left<\delta(\da)\right>$ and $\chi^2_\nu > 1$ for the total
high-$z$ sample and that the previous high-$z$ sample dominates these
statistics. Therefore, the weighted mean will not indicate the true
significance of $\da$ at high-$z$. Note that the binned points in the lower
panels of Fig.~\ref{fig:cps} will also appear too significant.

There are several reasons why {\it we expect extra scatter at high
$z$}. Consider fitting two transitions, arising from different species,
with significantly different linestrengths (e.g.~Al{\sc \,ii} $\lambda$1670
and Ni{\sc \,ii} $\lambda$1709 in Fig.~\ref{fig:q0528}). Weak components in
the high optical depth portions of the strong transition's profile are not
necessary to obtain a good fit to the data. Even though the {\sc vpfit}
$\chi^2$ minimization ensures that constraints on $\da$ derive primarily
from the optically thin velocity components, these weak components missing
from the fit will cause small line shifts. The resulting shift in $\da$ is
random from component to component and from system to system: the effect of
missing components will be to increase the random scatter in the individual
$\da$ values. We expect this effect to be greater in the high-$z$ sample
for several reasons:
\begin{enumerate}
\item At high $z$, a larger number of different species are generally
available for fitting compared to lower $z$. The range in optical depths
for corresponding velocity components is therefore significantly larger
than for the low-$z$ Mg/Fe{\sc \,ii} systems (compare Figs.~\ref{fig:q1437}
and \ref{fig:q0528}), complicating determination of the velocity structure
(see below for details). Abundance and ionization variations will also be
more important due to the diversity of species at high-$z$.

\item The high-$z$ sample is dominated by DLAs. If these have a more
complex velocity structure, i.e.~the number of absorbing components per
km\,s$^{-1}$ is higher than for the Me/Fe{\sc \,ii} systems, Voigt profile
decomposition would be more difficult, increasing the scatter in $\da$.

\item In the high-$z$ systems we fit more heavy, i.e.~low $b$, species.
Absorption features are therefore closer to the resolution of the
instrument, increasing the systematic bias against finding the weaker
components.
\end{enumerate}

\begin{figure}
\centerline{\psfig{file=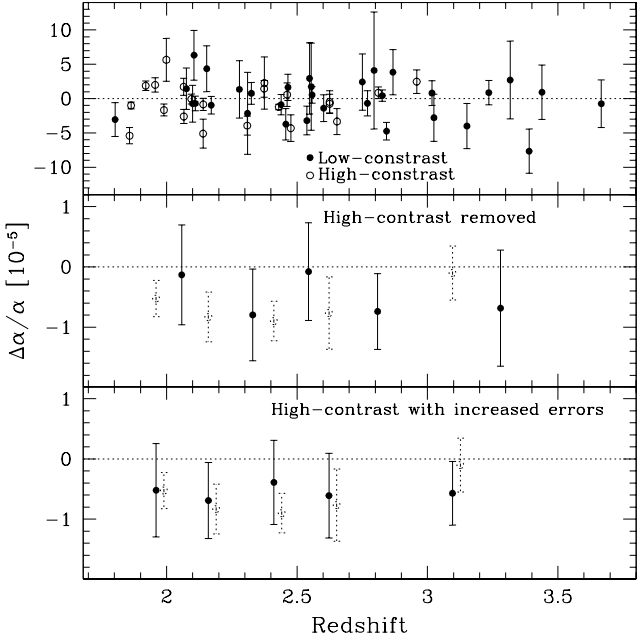,width=8.4cm}}
\caption{Reducing the extra scatter at high-$z$. The upper panel identifies
the `high-contrast' systems, i.e.~where any of the strong Al{\sc
\,ii}, Si{\sc \,ii} or Fe{\sc \,ii} transitions {\it and} any of the weak
Cr{\sc \,ii}, Ni{\sc \,ii} or Zn{\sc \,ii} transitions are fitted. These
systems are expected to exhibit additional scatter. In the middle panel the
high-contrast systems are removed and the binned low-contrast values (solid
circles) are compared with the raw values (dotted open circles). In the
lower panel the high-contrast error bars have been increased by adding
$2.09 \times 10^{-5}$ in quadrature to the raw errors to match the observed
additional scatter. The solid circles are binned values which include the
(unaltered) low-contrast systems.}
\label{fig:cps_hiz_alt}
\end{figure}

If these effects cause the extra scatter we observe, the most affected
systems will be those with a large range of transition linestrengths. We
form such a `high-contrast' sample from the high-$z$ (i.e.~$z_{\rm abs} >
1.8$) sample by selecting those systems in which any of the strong Al{\sc
\,ii}, Si{\sc \,ii} or Fe{\sc \,ii} transitions {\it and} any of the weak
Cr{\sc \,ii}, Ni{\sc \,ii} or Zn{\sc \,ii} transitions are fitted. These
high-contrast systems are marked with a `*' in Table \ref{tab:da}. In the
top panel of Fig.~\ref{fig:cps_hiz_alt} these systems (open circles) are
delineated from the rest of the high-$z$ sample, i.e.~the `low-contrast'
systems (solid circles). One immediately notes that the high-contrast
sample dominates the extra scatter. The middle panel in
Fig.~\ref{fig:cps_hiz_alt} shows the effect of removing the high-contrast
systems from the analysis. We give relevant statistics in Table
\ref{tab:stats}: the high-contrast sample comprises 22 systems and, upon
removal, $\chi^2_\nu$ at high-$z$ drops from 2.179 to 1.207, indicating
that the extra scatter is indeed due to the effects mentioned above.

\begin{figure}
\centerline{\psfig{file=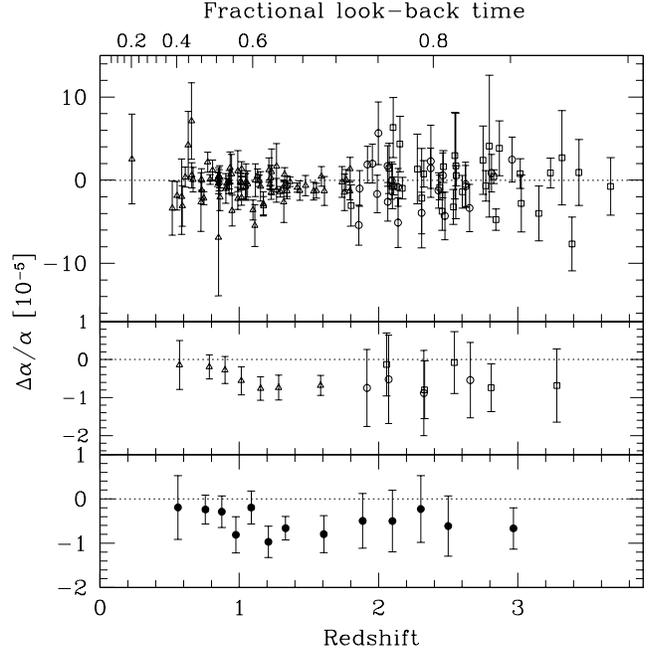,width=8.4cm}}
\caption{The fiducial sample. The upper panel shows the raw low-$z$ results
(solid triangles), the raw values for the low-contrast sample (open
squares) and the high-contrast sample with increased error bars (open
circles). The middle panel shows a binning of each sample taken separately
while the lower panel bins all samples together. The weighted mean for this
fiducial sample is $\da = (-0.543 \pm 0.116) \times 10^{-5}$ and is our
most robust estimate of $\da$ from the QSO data.}
\label{fig:cps_fiducial}
\end{figure}

To obtain a more robust estimate of the weighted mean $\da$ at high-$z$, we
may increase the individual 1\,$\sigma$ errors on $\da$ until $\chi^2_\nu =
1$ about the weighted mean for the high-contrast sample. We achieve this by
adding $2.09 \times 10^{-5}$ in quadrature to the error bars of the 22
relevant systems. The resulting binned values of $\da$ for the entire
high-$z$ sample are plotted in the lower panel of
Fig.~\ref{fig:cps_hiz_alt} together with the raw results (shifted for
clarity). The statistics in Table \ref{tab:stats} indicate that the best
estimate of $\da$ at high-$z$ is $\da = (-0.563 \pm 0.291) \times
10^{-5}$. Table \ref{tab:stats} also gives the statistics for two other
tests: (i) increasing the error bars for all high-$z$ systems until
$\chi^2_\nu = 1$ and (ii) removing points which deviate significantly from
the weighted mean (i.e.~$\sigma$-clipping) in the high-$z$ sample until
$\chi^2_\nu = 1$. All tests yield similar results.

Throughout the rest of this paper, `fiducial sample' will refer to the raw
sample (i.e.~the values and 1\,$\sigma$ errors in Table \ref{tab:da}) but
with increased errors on the high-contrast systems at high-$z$, as
described above. We plot the fiducial sample in Fig.~\ref{fig:cps_fiducial}
for comparison with the raw sample in Fig.~\ref{fig:cps}. Table
\ref{tab:stats} shows the statistics for the fiducial sample:
\begin{equation}\label{eq:fiducial}
\da = (-0.543 \pm 0.116) \times 10^{-5}
\end{equation}
is our most reliable estimate of the weighted mean $\da$ over the redshift
range $0.2 < z_{\rm abs} < 3.7$. This represents 4.7\,$\sigma$ statistical
evidence for a varying $\alpha$.

\subsection{Temporal variations in $\bmath{\alpha}$?}\label{ssec:tempvar}

\begin{figure}
\centerline{\psfig{file=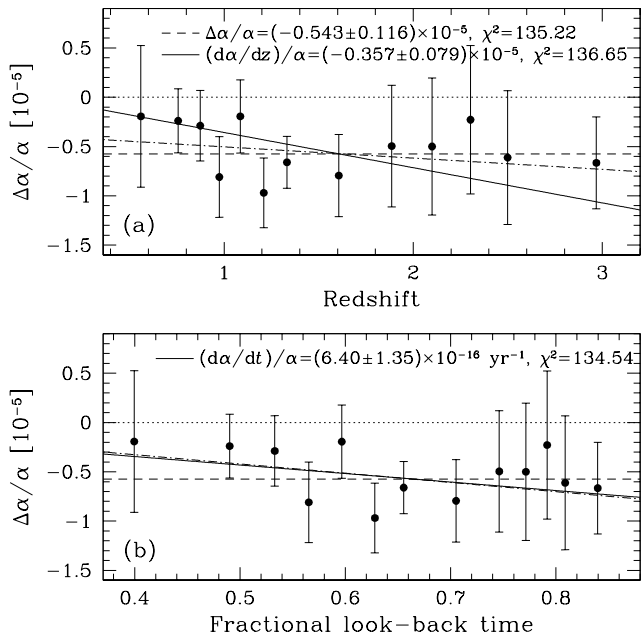,width=8.4cm}}
\caption{Temporal variation in $\alpha$. {\bf (a)} The points are the binned
values of $\da$ for the fiducial sample plotted versus redshift. The dashed
line is the weighted mean and the solid line is a fit to the unbinned data
fixed to $\da = 0$ at $z_{\rm abs}=0$. The dot-dashed line is a fit with
the $z_{\rm abs}=0$ value allowed to vary. The values of $\chi^2$ indicate
that the constant $\da$ model is preferred. {\bf (b)} Same as (a) but
versus fractional look-back time. The values of $\chi^2$ indicate that a
linear increase in $\alpha$ with time is preferred.}
\label{fig:cps_fits}
\end{figure}

In Fig.~\ref{fig:cps_fits} we overlay the binned fiducial data from
Fig.~\ref{fig:cps_fiducial} with several fits to the unbinned values of
$\da$ versus $z_{\rm abs}$ and cosmological time. Note that for the
constant $\dota \equiv ({\rm d}\alpha/{\rm d}t)/\alpha$ and constant $({\rm
d}\alpha/{\rm d}z)/\alpha$ models we fix $\da$ to zero at $z_{\rm
abs}=0$. The values of $\chi^2$ for the constant $\da$ and constant $\dota$
fits are $\chi^2=135.22$ and 134.54 respectively. This indicates that a
linear increase in $\alpha$ with time,
\begin{equation}\label{eq:dota}
\dota = (6.40 \pm 1.35)\times 10^{-16}{\rm \,yr}^{-1}\,,
\end{equation}
is preferred over a constant $\da$. We estimate the significance of this
preference using a bootstrap technique. Each value of $\da$ is randomly
assigned to one of the measured absorption redshifts and the best fitting
constant $\dota$ is determined. Constructing many such bootstrap samples
yields a distribution of $\chi^2$ for the constant $\dota$ fits. In this
procedure we construct low- and high-$z$ bootstrap samples separately since
the distribution of $\da$ in the real low- and high-$z$ samples may differ
(Section \ref{ssec:scat}). The $\chi^2$ distribution indicates that 31\,per
cent of $\chi^2$ values were $<\! 135.22$ and 15\,per cent were $<\!
134.54$. Thus, the constant $\dota$ model is preferred over the constant
$\da$ model only at the 50\,per cent confidence level. The latter is
preferred over a linear evolution of $\alpha$ with $z_{\rm abs}$ at the
63\,per cent level. These confidence intervals are only modestly affected
if we analyse the raw sample rather than the fiducial sample. Not
distinguishing the low- and high-$z$ samples gives lower confidence levels,
i.e.~$\approx$38\,per cent for both cases.

Furthermore, the constant $\dota$ and $({\rm d}\alpha/{\rm d}z)/\alpha$
models should be treated with caution. Fixing $\da$ to zero (i.e.~fixing
$\alpha$ to the laboratory value) at $z_{\rm abs}=0$ may not be valid
because of the potential for spatial variations in $\alpha$. For example,
if $\alpha$ varies over $\sim$100\,Mpc spatial scales and our Galaxy
resides in a region with a slightly larger $\alpha$ than the `universal
mean', then we should expect to measure a constant negative $\da$ rather
than an evolution of $\alpha$ with time. \citet{BekensteinJ_79a} and
\citet{BarrowJ_01a} also note that it is difficult to compare values of
$\da$ in areas of different gravitational potential without a detailed
theory giving both time {\it and} space variations of $\alpha$. If we allow
the $z_{\rm abs}=0$ value of $\da$ to vary, the linear fit against
cosmological time is barely altered (see Fig.~\ref{fig:cps_fits}), though
somewhat poorly constrained.

\subsection{Angular variations in $\bmath{\alpha}$?}\label{ssec:angvar}

\begin{figure*}
\centerline{\psfig{file=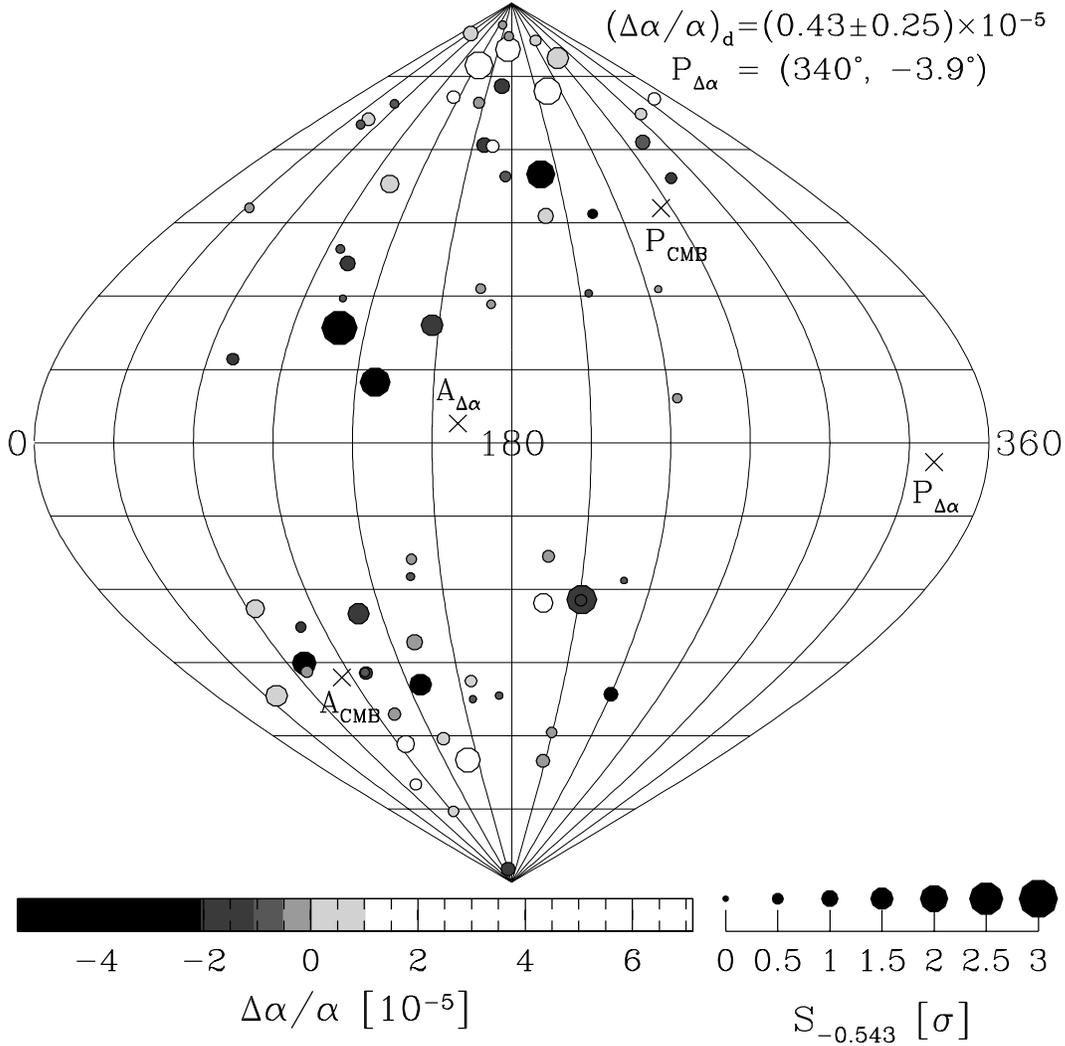,width=14.0cm}}
\caption{The distribution of $\da$ in Galactic coordinates $(l,b)$. All
  fiducial values along a single QSO sight-line (i.e.~for many absorption
  clouds and/or repeated, independent QSO observations) are collapsed to a
  single weighted mean. The value of $\da$ is given by the grey-scale and
  the size of each point scales with the significance, $S_{\rm -0.543}$, of
  the departure from the overall weighted mean of the fiducial sample
  (equation \ref{eq:fiducial}). The best-fit dipole has an amplitude
  $(\da)_{\rm d} = (0.43 \pm 0.25)\times 10^{-5}$ with its pole and
  anti-pole in the directions $P_{\Delta\alpha}$ and
  $A_{\Delta\alpha}$. However, a bootstrap analysis indicates that the
  dipole model is preferred over the constant $\da$ model at only the
  60\,per cent confidence level. We plot the CMB pole and anti-pole for
  comparison. (The on-line version of this figure is colour-coded and is
  also available from http://www.ast.cam.ac.uk/$\sim$mim/pub.html)}
\label{fig:cps_aitoff}
\end{figure*}

The fiducial sample can also be used to search for spatial variations in
$\alpha$. If very large scale ($\sim$10\,Gpc) spatial variations exist, one
might expect $\da$ to be different in different directions on the sky. For
the fiducial sample, we plot the distribution of $\da$ over Galactic
coordinates ($l,b$) in Fig.~\ref{fig:cps_aitoff}. All redshift information
is removed by taking all absorption systems along a single QSO sight-line
and collapsing them to a single weighted mean value of $\da$. We also
combine independent values from the same QSO in different observational
samples (see Section \ref{ssec:repeats}). The grey-scale indicates the
weighted mean $\da$ and the size of each point scales with the significance
with respect to the overall weighted mean of the fiducial sample, $S_{\rm
-0.543}$ (equation \ref{eq:fiducial}).

\begin{figure}
\centerline{\psfig{file=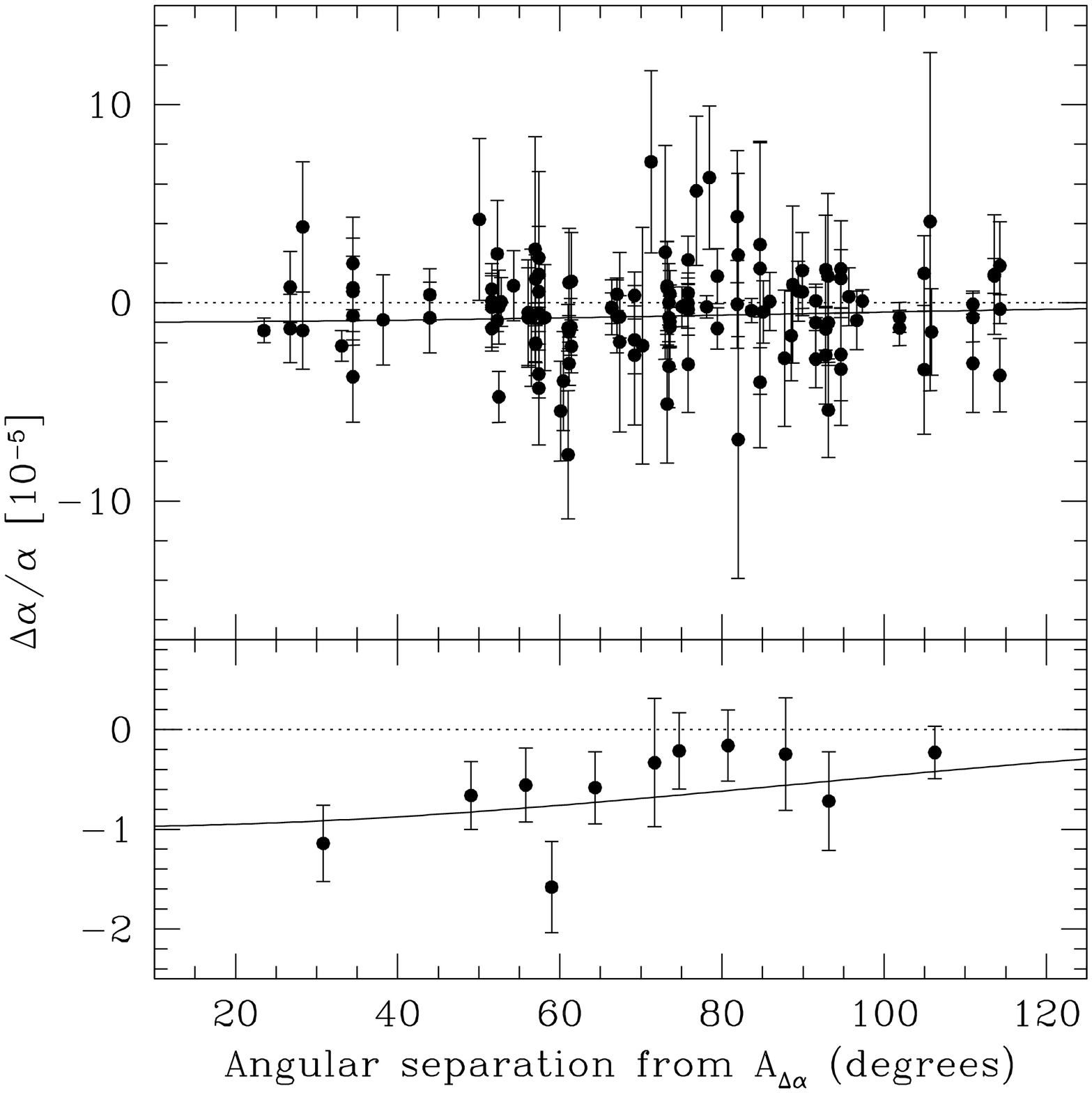,width=8.4cm}}
\caption{Best-fit dipole to the angular distribution of $\da$ in
Fig.~\ref{fig:cps_aitoff}. The pole lies in the direction
$P_{\Delta\alpha}(l,b)=(340\degr,-3.9\degr)$. The upper panel shows the
distribution of $\da$ with angular separation from the anti-pole,
$A_{\Delta\alpha}$. The lower panel shows a binning of the results. The
solid line is the best-fit cosine with an amplitude $(\da)_{\rm d}=(0.43
\pm 0.25)\times 10^{-5}$. However, the limited angular coverage of QSO
sight-lines clearly undermines confidence in this dipole interpretation.}
\label{fig:cps_dipole}
\end{figure}

We used a $\chi^2$ minimization algorithm to find the best-fit dipole in
the angular distribution of $\da$. We first constructed a grid of
directions in equatorial coordinates, (RA, DEC). For each direction, a
cosine fit to $\da$ as a function of the angular distance, $\phi$, to each
absorption system, yields an amplitude for the dipole, $(\da)_{\rm d}$,
defined by
\begin{equation}\label{eq:cosine}
\frac{\Delta\alpha}{\alpha}(\phi) = \left<\frac{\Delta\alpha}{\alpha}\right>_{\rm w} + \left(\frac{\Delta\alpha}{\alpha}\right)_{\rm d}\cos{\phi}\,.
\end{equation}
Fig.~\ref{fig:cps_dipole} shows the best-fit dipole which has an amplitude
$(\da)_{\rm d}=(0.43 \pm 0.25)\times 10^{-5}$. The pole is in the direction
$P_{\Delta\alpha}({\rm RA},\rm{\rm DEC})=(5.0{\rm \,hr}, -48\degr)$ or
$P_{\Delta\alpha}(l,b)=(340\degr, -3.9\degr)$ as marked on
Fig.~\ref{fig:cps_aitoff}. We also mark the anti-pole, $A_{\Delta\alpha}$,
and the cosmic microwave background (CMB) pole and anti-pole directions
\citep[$P_{\rm CMB}$ and $A_{\rm CMB}$,][]{LineweaverC_96a} for comparison.

The limited range of angular separations in Fig.~\ref{fig:cps_dipole}
severely limits the dipole interpretation. Although the dipole amplitude is
significant at the 1.7\,$\sigma$ level, a bootstrap analysis similar to
that in the previous section indicates that the dipole model is not
significantly preferred over the constant $\da$ (i.e.~monopole)
model. Bootstrap samples were formed by randomizing the values of $\da$
over the QSO sight-line directions. Again, we treated the low- and high-$z$
samples separately, combining them only to find the best-fit dipole
directions. The resulting probability distribution for $(\da)_{\rm d}$
indicates that values at $>$1.7\,$\sigma$ significance occur 40\,per cent
of the time by chance alone. Analysing the raw sample gives a similar
confidence level. Therefore, the data do not support significant angular
variations in $\alpha$. There is also no evidence for angular variations in
the low-$z$ and high-$z$ samples taken separately.

\subsection{Spatial correlations in $\bmath{\alpha}$?}\label{ssec:spatvar}

If spatial variations in $\alpha$ do exist at the $\da \sim 10^{-5}$ level
then one expects to see additional scatter in the raw values of $\da$ in
Fig.~\ref{fig:cps}. However, as noted previously, the statistics in Table
\ref{tab:stats} indicate that the scatter at low $z$ is consistent with the
size of the error bars and so we have no evidence for spatial variations in
$\alpha$ for $z_{\rm abs}<1.8$. The additional scatter at higher $z$ may
provide evidence for spatial variations but this seems unlikely given the
discussion in Section \ref{ssec:scat}.

We performed a more thorough search for spatial variations by calculating
the two-point correlation function for $\alpha$. Consider two absorption
systems with fiducial values of $\alpha$, $\alpha_1$ and $\alpha_2$. The
two-point correlation function, $C_{12}$, is then
\begin{equation}\label{eq:C}
C_{12}=\frac{\left(\alpha_1-\alpha_{\rm m}\right)\left(\alpha_2-\alpha_{\rm
m}\right)}{\alpha^2_{\rm m}}\,,
\end{equation}
where $\alpha_{\rm m}$ is the mean value of $\alpha$ over the entire
sample.

\begin{figure}
\centerline{\psfig{file=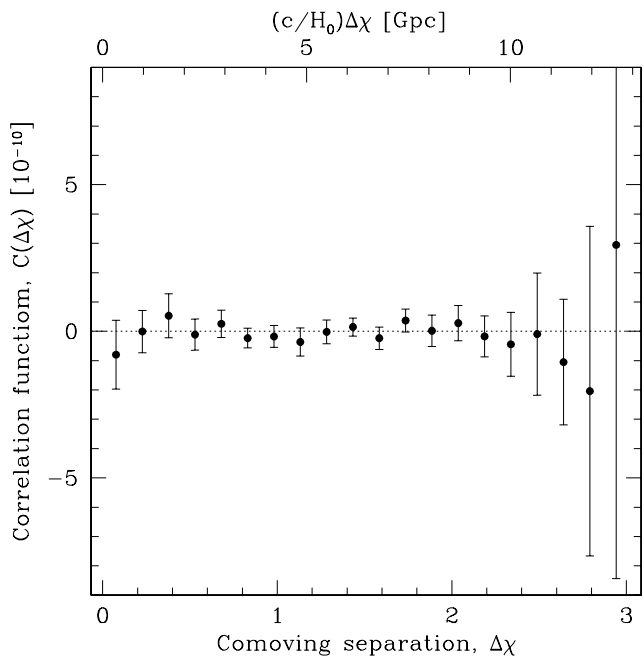,width=8.4cm}}
\caption{The two-point correlation function for $\alpha$,
  $C(\Delta\chi)$, as a function of comoving separation between
  absorption clouds, $\Delta\chi$ ($H_0=70{\rm \,km\,s}^{-1}{\rm Mpc}^{-1}$,
  $\Omega_{\rm m}=0.3$, $\Omega_\Lambda=0.7$). The values and errors were
  obtained with a Monte Carlo technique. However, since each value of $\da$
  may contribute to many bins, the bins are not independent.}
\label{fig:cps_2ptcorr}
\end{figure}

In Fig.~\ref{fig:cps_2ptcorr} we plot the correlation function versus the
comoving separation between the absorption systems
\citep[e.g.][]{LiskeJ_03a}, $\Delta\chi$. We used a Monte Carlo technique to
estimate the mean two-point correlation function, $C(\Delta\chi)$, and
68\,per cent confidence interval for each bin. Synthetic values of $\alpha$
were drawn randomly from Gaussian distributions centred on the measured
values with 1\,$\sigma$ widths equal to the fiducial 1\,$\sigma$
errors. Where independent observations of an absorption system exist
(Section \ref{ssec:repeats}), we centred the Gaussian on the weighted mean
value and used the 1\,$\sigma$ error in the weighted mean. For each bin,
the mean $C(\Delta\chi)$ was found from all pairs of absorption systems
with comoving separations within the bin. In Fig.~\ref{fig:cps_2ptcorr} we
plot the mean and rms of this quantity, represented by a point and error
bar. Fig.~\ref{fig:cps_2ptcorr} shows no evidence for significant spatial
correlations in $\alpha$ over 0.2--13\,Gpc (comoving) scales.

\subsection{Comparison with previous analysis}\label{ssec:prevanal}

\begin{figure}
\centerline{\psfig{file=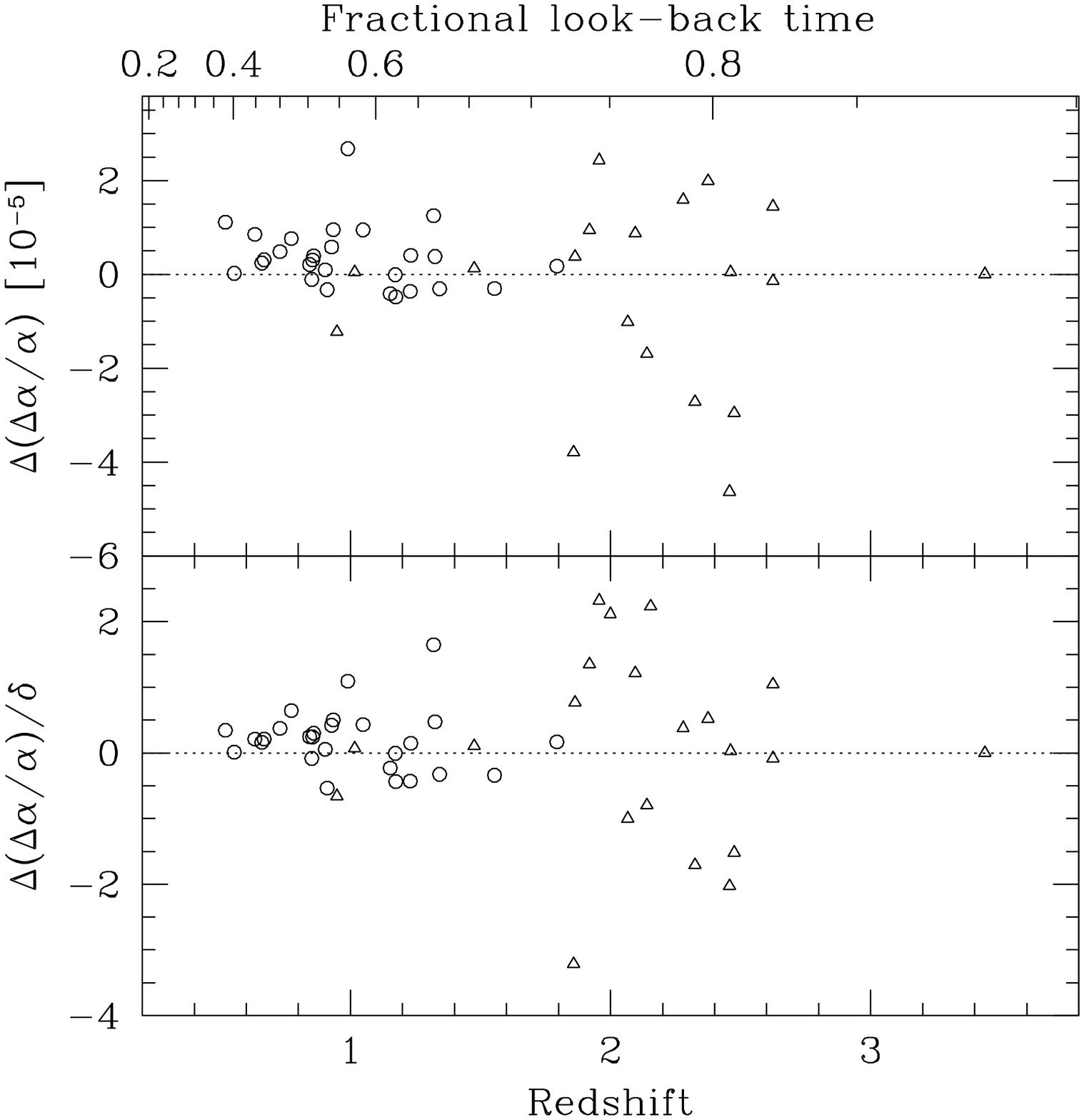,width=8.4cm}}
\caption{Comparison with previous analysis in \citetalias{MurphyM_01a}. The
  upper panel compares the raw values of $\da$ found here with those in
  \citetalias{MurphyM_01a} $[\Delta(\da) \equiv \da - (\da)_{\rm M01a}]$
  for the previous low-$z$ (open circles) and previous high-$z$ (open
  triangles) samples. The lower panel shows a measure of the significance
  of this difference: $\Delta(\da)$ normalized by the 1\,$\sigma$ error in
  Table \ref{tab:da}. For the low-$z$ data we find a slightly more positive
  $\da$ on average because the $q$-coefficients for the $\lambda_0 >
  2000{\rm \AA}$ Fe{\sc \,ii} transitions have been revised to slightly
  higher values. The larger scatter seen in $\Delta(\da)$ for the high-$z$
  systems is due mainly to the large change in the Fe{\sc \,ii}
  $\lambda$1608 $q$ coefficient (see Section \ref{ssec:atom_dat}).}
\label{fig:cp_new_old}
\end{figure}

The majority of velocity structures fitted to the previous samples are
unchanged from \citetalias{MurphyM_01a}. For the low-$z$ sample, the raw
values of $\da$ are similar to those found in \citetalias{MurphyM_01a}, as
shown in Fig.~\ref{fig:cp_new_old}. In \citetalias{MurphyM_01a}, the
weighted mean was $\da = (-0.70 \pm 0.23)\times 10^{-5}$ for this
sample. The primary reason for the slightly less negative value of $\da$ we
find here [from Table \ref{tab:stats}, $\da = (-0.51 \pm 0.22)\times
10^{-5}$] is the increased $q$ coefficients for the low-$z$ Fe{\sc \,ii}
transitions (i.e.~those with $\lambda_0 > 2000{\rm \AA}$ in Table
\ref{tab:atomdata}). For the previous high-$z$ sample,
Fig.~\ref{fig:cp_new_old} shows that the raw values of $\da$ found here
are, in individual cases, somewhat different to those found in
\citetalias{MurphyM_01a}. This is primarily due to the large change in the
Fe{\sc \,ii} $\lambda$1608 $q$ coefficient (see Section
\ref{ssec:atom_dat}). However, since the high-$z$ sample is characterized
by many different transitions with a diverse range of $q$ coefficients, the
overall weighted mean $\da$ found in \citetalias{MurphyM_01a} compares well
with the value found here: $\da = (-0.67 \pm 0.33)\times 10^{-5}$ and
$(-0.67 \pm 0.24)\times 10^{-5}$ respectively.

\subsection{Repeated observations}\label{ssec:repeats}
\begin{figure}
\centerline{\psfig{file=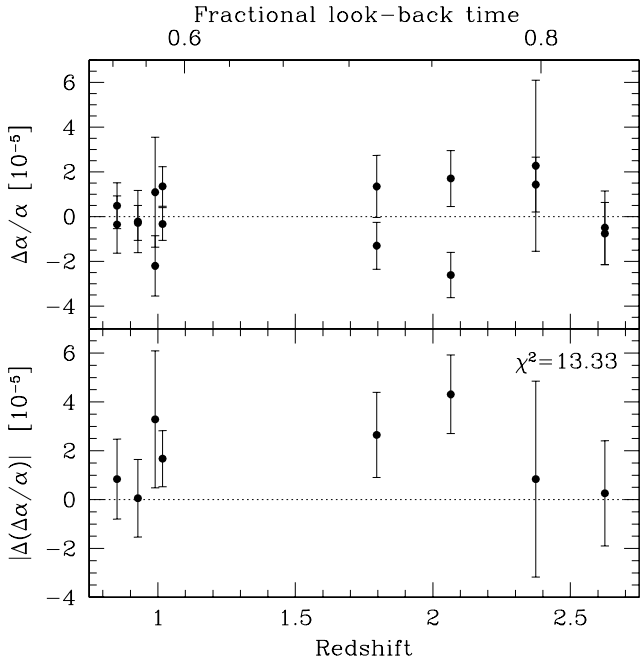,width=8.4cm}}
\caption{Repeated absorption systems. The upper panel shows the raw values of
  $\da$ for those systems which have been observed independently by
  different groups. The absolute difference is shown in the lower
  panel. For 7 degrees of freedom, the indicated value of $\chi^2$ will be
  exceeded 6.4\,per cent of the time by chance alone, providing only
  marginal evidence for significant additional random errors.}
\label{fig:cps_repeats}
\end{figure}

Several QSOs in our sample have been observed independently by two
different groups: the new sample contains 7 absorption systems which are
also (independently) contained in the previous samples and two independent
spectra of the $z_{\rm abs}=2.6253$ system towards GB 1759+7539 are
contained in the previous high-$z$ sample. We compare the values of $\da$
from the raw sample for these repeated systems in
Fig.~\ref{fig:cps_repeats}. The lower panel may indicate some evidence for
systematic errors in individual points (i.e.~some additional scatter may
exist). However, the value of $\chi^2$ with respect to zero difference is
$\chi^2 = 13.33$ which, for 7 degrees of freedom, has a 6.4\,per cent
probability of being exceeded by chance. Evidence for additional scatter is
therefore rather marginal.

\section{Systematic effects}\label{sec:syserr}

The above results represent strong statistical evidence for a smaller
$\alpha$ in the QSO absorption clouds. The fact that three different
samples show the same effect leaves little doubt that we do detect small
line shifts in the data. The central question is now whether these line
shifts are due to systematic errors or really due to varying $\alpha$. In
\citetalias{MurphyM_01b} we considered potential systematic effects due to
laboratory wavelength errors, wavelength miscalibration, atmospheric
dispersion effects, unidentified blending transitions, isotopic and/or
hyperfine structure effects, intrinsic instrumental profile variations,
spectrograph temperature variations, heliocentric velocity corrections,
kinematic effects and large scale magnetic fields. Below we extend this
initial search to include the entire Keck/HIRES sample.

\subsection{Kinematic effects}\label{ssec:kin}

In order to measure $\da$ with the MM method one must assume that
corresponding velocity components of different ionic species have the same
redshift. Velocity segregation could arise through chemical abundance
gradients combined with differential velocity fields. These effects may
also generate departures from Voigt profiles. The low-$z$ Mg/Fe{\sc \,ii}
systems would be more sensitive to any velocity segregation between species
due to the simple arrangement of $q$ coefficients (see
Fig.~\ref{fig:q_vs_wl}). For the high-$z$ systems, the effect on $\da$ for
an individual system would be minimized due to the diversity and
complicated arrangement of $q$ coefficients.

Inhomogeneous photoionization throughout the absorption complex is also a
possibility. With the exception of Mg{\sc \,i} and Al{\sc \,iii}, the
species in our analysis are singly ionized, have very similar ionization
potentials, IP$^+$, all above the Lyman limit and the corresponding neutral
ions have similar ionization potentials, IP$^-$, all below the Lyman limit
(see Table \ref{tab:atomdata}). Thus, the absorption from different ionic
species, associated with a given velocity component, should arise
co-spatially. However, variations in the incident radiation field could
cause weak components in the singly ionized species to appear very strong
in Mg{\sc \,i} or Al{\sc \,iii}. Spurious shifts in $\da$ could be
introduced if a strong Mg{\sc \,i} or Al{\sc
\,iii} component was mistakenly tied to a nearby ($\left|\Delta v\right|
\sim 0.5{\rm \,km\,s}^{-1}$) singly ionized component\footnote{We find no
additional scatter in $\da$ for systems containing Mg{\sc \,i} or Al{\sc
\,iii} and so have no evidence for such effects. We discussed the specific
case of Al{\sc \,iii} in \citetalias{MurphyM_01b}, finding that its
velocity structure followed very closely that of the singly ionized
species. See also discussion in
\citet{WolfeA_00a}}.

However, as we pointed out in \citetalias{MurphyM_01b}, these kinematic
shifts in $\da$ are themselves random over a large ensemble of random
sight-lines. If kinematic effects are significant, they act only to
increase the scatter in the values of $\da$ beyond what is expected on the
basis of the statistical errors alone. The value of $\chi^2_\nu$ for the
low-$z$ sample in Table \ref{tab:stats} is $\approx$1 and so the low-$z$
data present no evidence for significant kinematic effects. One may
interpret the extra scatter in $\da$ observed at high-$z$ as evidence for
such effects, though we discuss other, more likely, mechanisms for
producing this scatter in Section \ref{ssec:scat}.

Despite the random nature of kinematic effects over a large number of
velocity components and absorption systems, we discuss below the
contributory factors to potential effects on small velocity scales for
individual velocity components.

\subsubsection{Small-scale velocity structure and equilibrium in QSO absorbers}

Detailed kinematic studies of QSO absorbers indicate that a disk+halo model
provides a reasonable description of the absorption systems
\citep{BriggsF_85a,LanzettaK_92a,WolfeA_00b,ChurchillC_01a}. Alternative
models exist, including multiple merging clumps bound to dark matter halos
(\citealt*{HaehneltM_98a}; \citealt{McDonaldP_99a,MallerA_99a}) and
outflows from supernovae winds \citep*{NulsenP_98a,SchayeJ_01a}. For
disk+halo models of Mg/Fe{\sc \,ii} systems, the disk component is strongly
saturated and spread over small velocity scales whereas the halo component
is broadly spread in velocity space and causes the lower column density
absorption. The constraints on $\da$ in the Mg/Fe{\sc \,ii} systems are
therefore dominated by velocity components arising in the outer parts of
galaxy halos. The constraints on $\da$ from DLAs arise either from low
abundance, hence unsaturated, species (e.g.~Ni{\sc \,ii}, Cr{\sc \,ii},
Zn{\sc \,ii}), or from optically thin components flanking the saturated
ones (e.g.~Fe{\sc \,ii} $\lambda$1608 or Al{\sc \,ii} $\lambda$1670).

However, the {\it large-scale properties of the absorbing gas have no
influence on estimates of $\da$}. Galactic rotation or large-scale galactic
winds are unimportant. The empirical fact that we find excellent agreement
between the redshifts of individual velocity components down to $\la
0.3{\rm \,km\,s}^{-1}$ illustrates this. Any contribution to the scatter in
$\da$ comes only from the detailed properties of gas on velocity scales
typical of the observed $b$ parameters, i.e.~$\la 5{\rm \,km\,s}^{-1}$.

Given the importance of small-scale properties in the determination of
$\da$, it is relevant to ask whether the gas is in dynamic equilibrium on
these scales. For the low-$z$ Mg/Fe{\sc \,ii} systems, a recent detailed
study (Churchill, in preparation) suggests that the column density ratio
$N$(Mg{\sc \,ii})/$N$(Fe{\sc \,ii}) appears not to change systematically
across an absorption complex. A similar chemical uniformity is found for
the majority of high-$z$ systems \citep{ProchaskaJ_03a}. This implies that,
if not in equilibrium, the gas is well mixed, i.e.~absorption lines of
different species arise co-spatially. Photoionization models also suggest
gas in photoionization equilibrium with an ambient extra-galactic UV
background. If this is correct, local equilibrium may be valid. Any
redshift evolution of the number of absorption lines per unit redshift
interval, over and above that expected due to cosmology alone, could be
explained by cosmological evolution of the integrated background UV flux
\citep[e.g.][]{RaoS_00a}.

Thus, there is evidence to suggest that no gross changes in physical
conditions occur over large velocity scales across an absorption
complex. This may imply that we should not expect to find abundance
variations and non-equilibrium on small scales. If this is incorrect, one
expects departures from a Maxwellian velocity distribution in the
absorbers, the assumption of which is inherent in fitting Voigt
profiles. We note in passing that no evidence currently exists for
non-Voigt profiles
\citep[e.g.][]{OutramP_99a}, despite expectations from hydrodynamic
simulations of large-scale structure formation \citep{OutramP_00a}.

\subsubsection{Comparison with the ISM}

In contrast to the previous section, analogy with the interstellar medium
(ISM) in our own Galaxy suggests that non-equilibrium could apply on very
small scales. \citet*{AndrewsS_01a} used stars in a background globular
cluster, M92, to probe the kinematics on scales defined by the separation
between the lines of sight at the absorber. They find significant
variations in Na{\sc \,i} column densities in the ISM on scales as small as
1600\,AU (or $\sim 0.01{\rm \,pc}$). Even smaller-scale details come from
measurements of temporal variation of Na{\sc \,i} and K{\sc \,i} absorption
lines, implying non-equilibrium scales $\sim$10--100\,AU
(\citealt{CrawfordI_00a}; \citealt*{LauroeschJ_00a,PriceR_00a}). Although
these ISM cloud sizes are small compared to estimates for individual QSO
absorption cloud components, $\sim$10--100\,pc \citep{ChurchillC_01a}, the
characteristic size for a QSO continuum emission region may be
$\sim\!10^{-3}{\rm \,pc}$. The lines of sight to opposite edges of the QSO
therefore probe similar size scales as the Galactic ISM studies. However,
it should be noted that the gas densities are very different and so the
comparison should be treated with caution.

The ISM analogy may also allow us to estimate an upper limit on the
velocity segregation in QSO absorption clouds. Some theories argue that
gravity may be important for cloud confinement on small size scales in the
Galactic ISM \citep[e.g.][]{WalkerM_98a}. If gravity plays a significant
role in QSO absorption systems on similar scales, we could apply a simple
stability condition on the velocity dispersion, $v^2 = GM/R$, where $M$ and
$R$ are the cloud mass and radius. Estimates for individual cloud sizes
vary but adopting $R=10{\rm \,pc}$ and $M=30\,M_{\odot}$
\citep{ChurchillC_01a} we find $\left|v\right| \la 0.1{\rm
\,km\,s}^{-1}$. This provides an upper limit on the velocity shift between
different species. For one single Mg/Fe{\sc \,ii} velocity component, this
translates into an error on an individual $\da$ measurement of roughly
$\left|\da\right| \sim 0.3\times 10^{-5}$. However, this would be
randomized over $\sim$100 observations (and over many velocity components),
producing a maximum overall effect which is 20 times smaller than that
observed.

\subsection{Line blending}\label{ssec:blend}

We distinguished between {\it random} and {\it systematic} line blending in
\citetalias{MurphyM_01b}. We discuss both cases and their effect on $\da$
below.

\subsubsection{Random blends}\label{sssec:random_blends}

Random blends can occur when many absorption clouds lie along the line of
sight to the QSO, including any interstellar material and the Earth's
atmosphere. In general, random blends can only have a random effect on
$\da$ and so can not have caused the systematically non-zero values
observed. Two distinct categories of random blend can be identified:
\begin{enumerate}
\item Strong blends. Lyman-$\alpha$ forest absorption lines would cause
  numerous strong blends. However, if MM transitions fell into the
  Lyman-$\alpha$ forest region, we generally did not fit them. If the
  velocity structure of one of our MM transitions is obviously affected by
  a random blend then we modify our fit for that transition according to
  the identity of the interloper. If the interloper is a MM transition
  (i.e.~one with a precise laboratory wavelength listed in Table
  \ref{tab:atomdata}), and its velocity structure can be constrained using
  other associated MM transitions at the interloper's redshift, then we
  include the interloper in the fit. In general, the laboratory wavelengths
  of non-MM transitions (e.g.~C{\sc \,iv} $\lambda\lambda$1548 and 1550)
  are not known to high precision. Therefore, if the interloper and/or the
  associated transitions are not MM transitions, possible errors in the
  laboratory wavelengths could have a significant effect on $\da$. In these
  cases, and if the velocity structure of the interloper was both simple
  and clear, we freed its redshift parameters, not tieing it to any
  associated transitions. In all other cases, we either masked the blended
  data out of the fit or simply rejected that MM transition altogether.

\item Weak blends. It is possible that weak, random interlopers exist in
  our data but were not identified in our analysis. For example, numerous
  broad (${\rm FWHM}=20$--$2000{\rm \,km\,s}^{-1}$), weak (${\rm
  equivalent~width} \la 0.1{\rm \,\AA}$), diffuse interstellar absorption
  bands have been discovered in stellar spectra
  (e.g.~\citealt{HerbigG_75a,JenniskensP_94a}). The narrower of these lines
  may blend with some velocity components of individual MM transitions,
  causing slight, apparent line shifts. However, since our absorption
  systems lie at a range of redshifts and since many different transitions
  are fitted to arrive at a final value of $\da$, it is unlikely that such
  weak interlopers have significantly affected our results.
\end{enumerate}

\subsubsection{Systematic blends}\label{sssec:sys_blends}

A systematic blend occurs when two ionic species are in the same absorption
cloud and have transitions with similar rest wavelengths. Such a blend
could mimic a systematic shift in $\alpha$. We have explored two approaches
to this problem:
\begin{enumerate}

\item In \citetalias{MurphyM_01b} we attempted to identify candidate blends
  by searching atomic line databases for transitions lying close to the
  laboratory wavelengths of the MM transitions. Simulations of various
  blends indicated that, in order to cause a significant and systematic
  shift in $\alpha$, the interloping transition must lie within
  $\left|\Delta v\right|\approx 10{\rm \,km\,s}^{-1}$ of the `host'
  line. No possible blending transitions satisfying this criterion could be
  found in \citet{MooreC_71a} or the Vienna Atomic Line
  Database\footnote{Available at http://www.astro.univie.ac.at/$\sim$vald}
  \citep[VALD,][]{PiskunovN_95a,KupkaF_99a}.

  We have now extended our search criterion to include potential
  interlopers lying up to $100{\rm \,km\,s}^{-1}$ away from the host MM
  transitions. We have identified two possible interlopers lying to the red
  of the Zn{\sc \,ii} $\lambda$2026 transition:
  \begin{enumerate}

  \item Cr{\sc \,ii} $\lambda$2026.269 ($\Delta v\approx 19{\rm
    \,km\,s}^{-1}$) has an oscillator strength $f=0.0047$ (VALD; correcting a
    typographical error in \citetalias{MurphyM_01b}). Using the largest
    value of $N$(Cr{\sc \,ii})/$N$(Zn{\sc \,ii}) from our data, simulations
    of this blend suggest a negligible effect on $\da$ (see
    \citetalias{MurphyM_01b} for details).

  \item Mg{\sc \,i} $\lambda$2026.477 ($\Delta v\approx 50{\rm km\,s}^{-1}$)
    has an oscillator strength $f=0.1154$ \citep{MortonD_91a} and was
    apparent in six of the high-$z$ systems. However, in four cases the
    velocity structure is narrow enough that the Mg{\sc \,i} interloper
    does not overlap and blend with the Zn{\sc \,ii} line. In the remaining
    two cases we fitted the Mg{\sc \,i} interloper and left the redshift
    parameters free (i.e.~we did not tie the Mg{\sc \,i} velocity structure
    to that of the other MM transitions). This has the effect of removing
    any constraints the blended Zn{\sc \,ii} components may have on
    $\da$. We checked the resulting values of $\da$ against those found
    after completely masking out the blended portion of the Zn{\sc \,ii}
    velocity structure, finding them to be consistent. Also, completely
    removing the Zn{\sc \,ii} $\lambda$2026 transition from the analysis of
    these systems produced very little change in $\da$.
  \end{enumerate}

Note that the atomic line databases searched are not complete and that we
have not considered molecular interlopers. We therefore compliment the
treatment above with the following test.

\item To asses the significance of any unknown blends, we may separately
  remove most transitions from the analysis and re-fit to find a new value
  of $\da$. If the new and old values of $\da$ are significantly different
  then one interpretation is that the transition removed is affected by an
  unknown systematic blend. In \citetalias{MurphyM_01b} we found no
  evidence for blends in this way. We update this line-removal analysis in
  Section \ref{sssec:rem_single} and reach the same conclusion.
\end{enumerate}

In summary, we find that neither random or systematic line blending has
significantly affected the values of $\da$.

\subsection{Wavelength miscalibration}\label{ssec:thar}

The QSO CCD images are wavelength calibrated by comparison with ThAr
emission lamp spectra which are generally taken both before and after each
QSO exposure. A polynomial wavelength solution is obtained by centroiding a
standard set of ThAr lines which are strong and (appear to be) unblended in
the particular ThAr lamp in use. If one or more of the ThAr lines in this
standard set are misidentified or, in a more subtle way, lead to long range
miscalibrations of the wavelength scale, then a systematically non-zero
$\da$ would be found. This effect would be particularly problematic in the
low-$z$ absorption systems where all values of $\da$ would be affected in
the same way by such a low-order distortion of the wavelength scale (see
Fig.~\ref{fig:q_vs_wl}).

\subsubsection{The ThAr calibration test}\label{sssec:thar_test}

The effect of any ThAr miscalibrations on the values of $\da$ can be
estimated directly using a technique proposed and demonstrated in
\citetalias{MurphyM_01b}: we analyse sets of ThAr emission lines in the
same way we analyse sets of QSO absorption lines. If no calibration error
exists then one should obtain $(\da)_{\rm ThAr}=0$ for each absorption
system.

We provide a detailed description of this ThAr test in
\citetalias{MurphyM_01b}, the main steps being as follows. First we select
several independent sets of ThAr emission lines from sections of spectra
which correspond to the observed wavelengths of the QSO absorption
lines. Each set of lines therefore corresponds to, and has been selected in
an analogous way to, one QSO absorption system. Selecting several
(typically 3--7) different sets of ThAr lines allows a more robust estimate
of $(\da)_{\rm ThAr}$ for each absorption system and provides a cross-check
on its 1\,$\sigma$ uncertainty.

We fit Gaussian profiles to the emission lines with the same modified
version of {\sc vpfit} used to fit Voigt profiles to the QSO absorption
lines. In order to relate any measured ThAr line shifts directly to a value
of $(\da)_{\rm ThAr}$, we assign to each ThAr line the $q$ coefficient
corresponding to the QSO absorption line falling in that part of the
spectrum. That is, we treat the ThAr lines as if they {\it were} QSO
absorption lines and derive a value of $(\da)_{\rm ThAr}$ using the MM
method.

\subsubsection{Results and discussion}\label{sssec:thar_res}

\begin{figure}
\centerline{\psfig{file=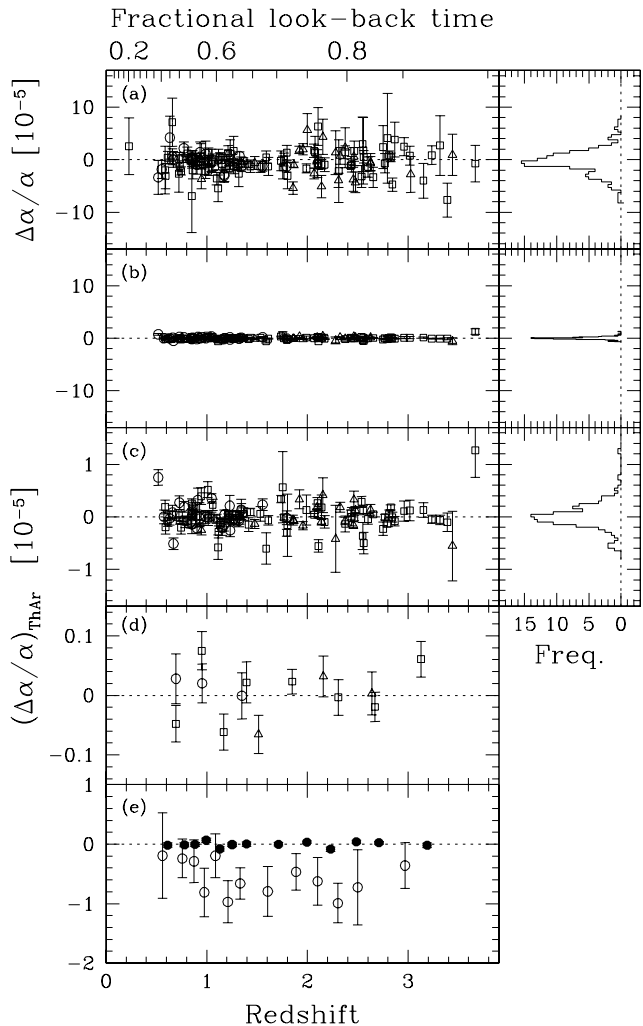,width=8.4cm}}
\caption{Comparison between the raw QSO results [panel (a)] and the ThAr
  calibration error [panel (b)] for the previous low-$z$ (open circles),
  previous high-$z$ (open triangles) and new (open squares) samples. The
  $(\da)_{\rm ThAr}$ scale is expanded by a factor of 10 in panel
  (c). Panel (d) shows binned values for the three sub-samples and panel
  (e) compares the binned values for the total ThAr (solid circles) and raw
  QSO (open circles) samples. The weighted mean of the ThAr results is
  $(\da)_{\rm ThAr}=(0.4\pm 0.8)\times 10^{-7}$.}
\label{fig:cps_thar}
\end{figure}

ThAr spectra were readily available for 118 of the 128 absorption systems
in our total sample. We applied the above ThAr test to these spectra and
obtain the results presented in Fig.~\ref{fig:cps_thar}. Comparing the QSO
results (raw sample) in panel (a) with the ThAr results in panel (b) we
immediately see that miscalibrations of the wavelength scale did not
systematically affect the overall value of $\da$. This is particularly
emphasized in panel (e) which compares binned raw values of $\da$ and
$(\da)_{\rm ThAr}$. We tabulate various statistics for the ThAr test in
Table \ref{tab:stats} for direct comparison with the QSO results. Notably,
the overall weighted mean value, $(\da)_{\rm ThAr}=(0.4 \pm 0.8)\times
10^{-7}$, is two orders of magnitude below the observed value of $\da$.

Panel (c) of Fig.~\ref{fig:cps_thar} shows an expanded view of the ThAr
results which clearly demonstrate a similar `extra scatter' to the high-$z$
QSO sample (see Section \ref{ssec:scat}). The extra scatter has a similar
origin as for the high-$z$ QSO results: close inspection of any portion of
ThAr spectrum reveals a plethora of weak emission lines and so the fit to
any given strong ThAr emission line will be affected by weak blends. The
ThAr lines are of such high S/N that significant extra scatter ensues. This
effect would produce uncorrelated deviations in $(\da)_{\rm ThAr}$ from
point to point, as observed. Another interpretation may be that the extra
scatter presents evidence for ThAr line misidentifications. However, if
these were important, one would expect a correlation between $(\da)_{\rm
ThAr}$ and $\da$. We observe no such correlation.

The ThAr results in Fig.~\ref{fig:cps_thar} also illuminate discussion of
potential instrumental profile (IP) variations and their effect on
$\da$. \citet*{ValentiJ_95a} have derived the HIRES IP for several
positions along a single echelle order, finding that the IP is slightly
asymmetric and that this asymmetry varies slightly along the
order. However, to a first approximation, the ThAr and QSO light paths
through the spectrograph are the same. The ThAr results in
Fig.~\ref{fig:cps_thar} therefore strongly suggest that IP asymmetries (and
variations thereof) do not contribute significantly or systematically to
$\da$. One possibility is that the ThAr and QSO light follow very different
and possibly wavelength dependent paths through the spectrograph to the
CCD. We explore this possibility in Section \ref{ssec:atdisp} in the
context of atmospheric dispersion effects.

We conclude this section by emphasizing the reliability of the wavelength
scale as derived from the ThAr spectra: we directly measure the effect of
any ThAr miscalibrations on $\da$ and find it to be negligible. One
objection to the reliability of the MM method has been that fitting
transitions which fall on different echelle orders may lead to systematic
errors \citep{VarshalovichD_00a,IvanchikA_02a}. The ThAr test in
Fig.~\ref{fig:cps_thar} \citepalias[and figure 2 in][]{MurphyM_01b} clearly
demonstrates that these concerns are unfounded.

\subsection{Spectrograph temperature variations}\label{ssec:temps}

The refractive index of air within the spectrograph depends on temperature
(and also on pressure, but this is a smaller effect). If the temperature
during the QSO exposure, $T_{\rm QSO}$, is $\sim\!15{\rm \,K}$ higher than
that during the ThAr calibration exposure, $T_{\rm ThAr}$, the wavelength
scale applied to the QSO frame would be distorted, resulting in $\da \sim
+1\times 10^{-5}$ for a typical Mg/Fe{\sc \,ii} system (see
\citetalias{MurphyM_01b} for details). Note that a systematically non-zero
$\da$ can only be introduced for the Mg/Fe{\sc \,ii} systems if $T_{\rm
QSO}$ is systematically higher or lower than $T_{\rm ThAr}$.

Such an effect can be minimized if the ThAr calibration exposures are taken
immediately before and/or after the QSO exposures or if the spectrograph is
temperature stabilized (e.g.~VLT/UVES). The former was generally the case
for our observations. We have used image header information to calculate
$\Delta T = \left<T_{\rm QSO}\right>-\left<T_{\rm ThAr}\right>$ for each
QSO in each sample, where the average is taken over all QSO and ThAr
exposures. We find mean values of $\Delta T = 0.04 \pm 0.02{\rm \,K}$ and
$\Delta T = 0.2 \pm 0.1{\rm \,K}$ for the previous low- and high-$z$
samples, and $\Delta T = 0.013 \pm 0.006{\rm \,K}$ for the new sample. It
is therefore clear that our values of $\da$ are not affected by
spectrograph temperature variations.

\subsection{Isotopic and hyperfine structure effects}\label{ssec:iso}

As stated in Section \ref{ssec:atom_dat}, the only species for which full
isotopic and hyperfine structures are known are Mg{\sc \,i}, Mg{\sc \,ii}
and Al{\sc \,iii}. We obtained an estimate of the Si{\sc \,ii} isotopic
structures by scaling the Mg{\sc \,ii} $\lambda$2796 structure by the mass
shift, equation \ref{eq:mshift}. In this section we investigate systematic
errors that could result from ignorance of the isotopic/hyperfine
structures for other species and from any evolution of the isotopic ratios
and/or hyperfine level populations.

\subsubsection{Differential isotopic saturation}\label{sssec:iso_diffsat}

For the transitions of Cr{\sc \,ii}, Fe{\sc \,ii}, Ni{\sc \,ii} and Zn{\sc
\,ii}, the laboratory wavelengths in Table \ref{tab:atomdata} are {\it
composite} values only. The composite wavelengths are only strictly
applicable in the optically thin regime (linear part of the curve of
growth). Consider an absorption line with several isotopes. As the dominant
(highest abundance) isotope saturates, the weaker isotopes will have
increased influence on the fitted line centroid. This could lead to
systematic errors in $\da$.

Table \ref{tab:iso} shows the terrestrial isotopic abundances for the atoms
used in the MM method. Note that the abundance is quite distributed and
asymmetric for Ni, Zn and, to a lesser extent, Cr. If the isotopic
components of the transitions of these species were widely spaced in
wavelength then differential isotopic saturation may be an important
effect. However, the Cr, Ni and Zn lines are always weak in our QSO
spectra, typically absorbing only $\sim$20\,per cent of the
continuum. Using only the composite wavelengths for transitions of these
species is therefore justified.

\begin{table}
\centering
\caption{The percentage terrestrial isotopic abundances of the atoms used
in our analysis \citep{RosmanK_98a}. The second column shows the mass
number, $A$, of the isotope with the highest abundance and $\Delta A$ is
defined relative to this (negative values representing lighter isotopes).}
\label{tab:iso}
\begin{tabular}{lcccccccc}\hline
Atom    &$A     $& \multicolumn{7}{c}{$\Delta A$}\\
        &        &$-2$ &$0$   &$+1$ &$+2$ &$+3$&$+4$&$+6$\\\hline
Mg      &24      &     &79.0  &10.0 &11.0 &    &    &    \\
Al      &27      &     &100.0 &     &     &    &    &    \\
Si      &28      &     &92.229&4.683&3.087&    &    &    \\
Cr      &52      &4.3  &83.8  &9.5  &2.4  &    &    &    \\
Fe      &56      &5.8  &91.8  &2.1  &0.3  &    &    &    \\
Ni      &58      &     &68.08 &     &26.22&1.14&3.63&0.93\\ 
Zn      &64      &     &49    &     &28   &4   &19  &1 \\\hline
\end{tabular}
\end{table}

Although the Fe abundance is strongly centred on the $^{56}$Fe isotope,
the Fe{\sc \,ii} transitions in our QSO spectra are often saturated. The
possible effect of differential isotopic saturation for Fe should therefore
be tested. The mass isotopic shift (equation \ref{eq:mshift}) for Fe
transitions should be $\ga 5$ times smaller than for those of
Mg. Therefore, using only composite wavelengths of Mg and Si transitions in
our analysis (instead of the full isotopic structures) should place a firm
upper limit on the possible effect from Fe{\sc \,ii}.

\begin{figure}
\centerline{\psfig{file=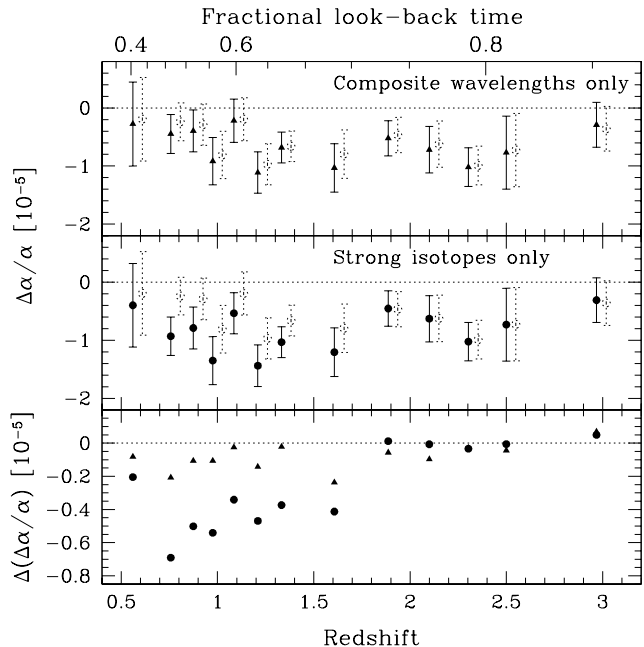,width=8.4cm}}
\caption{Upper limits on the effects of differential
  isotopic saturation and isotopic abundance evolution. The top panel
  compares the binned values of $\da$ derived using only the composite
  laboratory wavelengths (solid triangles) from Table \ref{tab:atomdata}
  and the values from the raw sample (dotted circles, slightly shifted for
  clarity). The solid circles in the middle panel are derived using only
  the isotopic components of the Mg{\sc \,i}, Mg{\sc \,ii} and Si{\sc \,ii}
  transitions with the highest terrestrial abundance. The lower panel shows
  the differences between these composite and single isotope values of
  $\da$ and the values from the raw sample.}
\label{fig:cps_noiso_comp}
\end{figure}

We have conducted such a test with our total sample: we use composite
wavelengths for the Mg{\sc \,i}, Mg{\sc \,ii} and Si{\sc \,ii} isotopic
structures and refit our spectra to find values of $(\da)_{\rm comp}$. We
also use composite wavelengths for the Al{\sc \,iii} hyperfine structures
(see Section \ref{sssec:hyp_diffsat}). We provide a detailed discussion of
the expected magnitude of $\Delta(\da) = (\da)_{\rm comp}-\da$ in
\citetalias{MurphyM_01b} and find values of $(\da)_{\rm comp}$ for the
previous low- and high-$z$ samples. The top panel of
Fig.~\ref{fig:cps_noiso_comp} compares values of $(\da)_{\rm comp}$ (solid
triangles) for the total sample with those found using the full
isotopic/hyperfine structures (dotted circles). We see that isotopic
saturation has only a small effect on $\da$, even for the Mg/Fe{\sc \,ii}
systems at low-$z$. This is confirmed when comparing the statistics for
this test in Table \ref{tab:stats} with the values from the raw sample. It
is therefore clear that differential isotopic saturation of Fe{\sc \,ii}
transitions is unlikely to have significantly affected our results.

\subsubsection{Differential hyperfine saturation}\label{sssec:hyp_diffsat}

Hyperfine splitting of energy levels occurs only in species with odd proton
or neutron numbers (note that most isotopes in Table \ref{tab:iso} are
even). Different hyperfine components will have different transition
probabilities but the composite wavelength of a line will be unchanged by
the splitting (i.e.~the centre of gravity of the hyperfine components is
constant with increased splitting). However, a similar differential
saturation effect will occur for the hyperfine components as for the
isotopic components discussed in Section
\ref{sssec:iso_diffsat}.

By far the most prominent hyperfine structure is that of Al{\sc \,iii}
which is clearly resolved in the laboratory experiments of
\citet{GriesmannU_00a} and which we take into account in our fits to the
QSO data (see Section \ref{ssec:atom_dat}). The high-$z$ composite
wavelength results in Fig.~\ref{fig:cps_noiso_comp} therefore provide an
upper limit on the effect of hyperfine saturation effects for all other
transitions. The only other known hyperfine structures are those of the
$^{25}$Mg transitions. These are so closely separated that they would have
a negligible effect on $\da$ (especially considering the small effect of
differential {\it isotopic} saturation in
Fig.~\ref{fig:cps_noiso_comp}). In \citetalias{MurphyM_01b} we ruled out
significant effects on $\da$ from saturation effects in other (odd) species
due either to their low relative abundances, low magnetic moments or the
nature of the ground- and excited-state wavefunctions. All the above
arguments imply a negligible effect due to differential hyperfine
saturation.

\subsubsection{Isotopic abundance variations}\label{sssec:iso_ratvar}

We assumed the terrestrial isotopic abundances in Table \ref{tab:iso} when
fitting Mg and Si absorption lines. However, if the isotopic abundances in
the QSO absorbers are different, small apparent shifts in the absorption
lines would be introduced, potentially mimicking a non-zero $\da$. This
effect will be greatest for changes in the isotopic abundances of Mg{\sc
\,i}, Mg{\sc \,ii} and Si{\sc \,ii} since the mass shift (equation
\ref{eq:mshift}) implies these ions will have the largest isotopic
separations.

Observations of Mg \citep{GayP_00a} and theoretical estimates for Si
\citep{TimmesF_96a} in stars clearly show a decrease in the isotopic
abundances with decreasing metallicity. For example, at relative metal
abundances ${\rm [Fe/H]} \sim -1$, $^{25}{\rm Mg}/^{24}{\rm Mg}
\approx\,^{26}{\rm Mg}/^{24}{\rm Mg} \approx 0.1$, which is about 30\,per
cent below terrestrial values (see Table \ref{tab:iso}). Theoretical
estimates suggest even larger decreases \citep*{TimmesF_95a}. The
metallicities of our absorption systems are all likely to be sub-solar: the
Mg/Fe{\sc \,ii} systems (previous low-$z$ sample) have $Z=-2.5$--0.0
\citep{ChurchillC_98a,ChurchillC_00a,ChurchillC_00b} and the DLAs in the
previous high-$z$ sample have $Z \approx -1.0$
\citep{ProchaskaJ_99b,ProchaskaJ_00a}. Therefore, we expect significantly
lower isotopic abundances of $^{25,26}$Mg and $^{29,30}$Si in our QSO
absorption systems.

Considering the above, refitting the QSO spectra using only the strong
isotopes of Mg and Si (i.e.~$^{24}$Mg and $^{28}$Si) should place an upper
limit on the effect of possible isotopic abundance variations on $\da$. We
first demonstrated this test on the previous samples in
\citetalias{MurphyM_01b} and have now applied it to the total sample. We
find similar results.

Our new results, $(\da)_{\rm iso}$ (solid circles), are compared with the
values from the raw sample (dotted circles) in the middle panel of
Fig.~\ref{fig:cps_noiso_comp}. The lower panel also shows the size of the
correction, $\Delta(\da) = (\da)_{\rm iso}-\da$.
Fig.~\ref{fig:cps_noiso_comp} clearly demonstrates that strong isotopic
abundance evolution in Mg and Si can not explain our results. Indeed, $\da$
becomes more negative when we remove this potential systematic effect. Note
that removing $^{25,26}$Mg affects only the low-$z$ Mg/Fe{\sc
\,ii} systems whereas removing $^{29,30}$Si affects only the
high-$z$ points. The largest effect is for the low-$z$ Mg/Fe{\sc \,ii}
systems. This is expected because (a) the isotopic separations for Mg
transitions are larger than those for Si transitions (equation
\ref{eq:mshift}), (b) the relative terrestrial abundances of $^{25}$Mg and
$^{26}$Mg are larger than $^{29}$Si and $^{30}$Si (Table \ref{tab:iso}) and
(c) a large number of different species are fitted in the high-$z$ DLAs so
a systematic effect in just one species (in this case Si{\sc \,ii}) is
unlikely to significantly affect $\da$.

The above test estimates the maximum effect of strong isotopic abundance
evolution for Mg and Si transitions. Although we emphasize that the effect
for heavier species should be reduced by the mass shift, we point out that
isotopic (and hyperfine) structures for transitions of other species are
not known. It is possible that large specific isotopic shifts exist for
some transitions. If this is combined with strong isotopic ratio evolution
then small apparent line shifts would be measured in the QSO
spectra. However, this possibility is unlikely given the results of
removing single transitions or entire species in Section
\ref{sssec:rem_single}.

One possibility is that the isotopic abundances do not follow the above
expected trend based on the Galactic observations and theoretical
models. For example, {\it higher} $^{25,26}$Mg abundances in the absorption
clouds would mimic $\da < 0$ in the low-$z$ Mg/Fe{\sc \,ii} systems. The
results above allow us to estimate the Mg isotopic ratios required to
explain the observed $\da$ at low-$z$. The approximate shift in $\da$ will
be $\Delta(\da) \approx \beta\domg$ where $\domg$ is the fractional change
in the abundance weighted mean transition wavenumber. Applying this
relation to the test results above, we find that the proportionality
constant $\beta = 12.2$ using an average $\Delta(\da) = -0.445 \times
10^{-5}$ (from Table
\ref{tab:stats}) and $\domg = -3.6 \times 10^{-7}$, the fractional
difference between the $^{24}$Mg{\sc \,ii} and composite Mg{\sc \,ii}
$\lambda$2796 wavenumbers. To shift $\da$ to an average of zero in the
low-$z$ sample, we require $\Delta(\da) = +0.539 \times 10^{-5}$. Using the
$\beta$ just derived, this implies $\domg = 4.4 \times 10^{-7}$ which could
be achieved if the ($^{25}$Mg+$^{26}$Mg)/$^{24}$Mg abundance ratio is
$\approx$0.58 in the absorption clouds (compared to the terrestrial value
of 0.27).

\subsubsection{Hyperfine level population variations}\label{sssec:hyp_popvar}

Once again, consider the prominent hyperfine structure of the Al{\sc \,iii}
transitions. If the populations of the ground-state hyperfine levels are
not equal in the absorption clouds, as they are in the laboratory
measurements of \citet{GriesmannU_00a}, we would measure small apparent
shifts for the Al{\sc \,iii} transitions. In \citetalias{MurphyM_01b} we
pointed out that a lower bound on the relative population is set by
interaction with CMB photons, $\exp[-h\Delta
\omega/k_{\rm B}T_{\rm CMB}(z)]$, where $\Delta \omega \approx 0.5{\rm
\,cm}^{-1}$ is the hyperfine splitting for the Al{\sc \,iii}
$\lambda\lambda$1854 and 1862 transitions. At $z_{\rm abs} \sim 2.0$, the
relative population is $\approx\! 0.9$, leading to a shift in the line
centroid of $\sim\! 0.01{\rm \,cm}^{-1}$ ($\sim\! 5\times 10^{-4}{\rm
\,\AA}$). This corresponds to $\da = +1\times 10^{-5}$ for the Al{\sc
\,iii} lines alone. Clearly, including other transitions will reduce the
effect on $\da$ for a particular system.

Two caveats significantly lessen concern in this case: (i) collisional
excitation processes should drive the relative populations to equality;
(ii) removing either or both of the Al{\sc \,iii} transitions from our
analysis has a very small effect on $\da$ (see Section
\ref{sssec:rem_single}). We therefore consider systematic errors from
variations in the hyperfine level populations to be negligible.

\subsection{Atmospheric dispersion effects}\label{ssec:atdisp}

Below we describe how atmospheric dispersion could have distorted the
wavelength scale and instrumental profile (IP) of the QSO spectra. This was
the largest potential systematic effect identified in
\citetalias{MurphyM_01b} for the previous samples. We model the {\it
potential} effects on our QSO spectra in Section
\ref{sssec:atdisp_mod}, significantly improving the model in
\citetalias{MurphyM_01b}. We search for such effects in the QSO data in 
Section \ref{sssec:atdisp_evi}, finding that {\it the data themselves
suggest negligible effects on $\da$ from atmospheric dispersion}.

\subsubsection{Modelling possible distortions from atmospheric
dispersion}\label{sssec:atdisp_mod}

The Keck/HIRES was only fitted with an image rotator in 1996
August. Therefore, prior to this time, the spectrograph slit could not be
held perpendicular to the horizon during observations and so the atmosphere
will have dispersed QSO light with some component across the slit
\citepalias[see figure 5 of][]{MurphyM_01b}. All of the 27 absorption
systems in the previous low-$z$ sample were observed before 1996
August. Similarly, 11 systems in the previous high-$z$ sample and half (39)
of the new sample were observed `pre-rotator'. Therefore, for the total
sample, 77 of the 128 absorption systems could be affected by atmospheric
dispersion effects.

Before the image rotator was installed, the angle of the Keck/HIRES slit to
the vertical, $\theta$, was directly related to the zenith angle by $\theta
= \xi$. That is, the slit lies with its axis along the horizon at $\xi =
90\degr$. The component of atmospheric dispersion along the spectral
direction of the slit (i.e.~across its axis) therefore increases with
increasing $\xi$ because atmospheric dispersion {\it and} $\theta$ increase
with $\xi$. This will lead to two effects on the QSO wavelength scale:
\begin{enumerate}
\item Compression. Due to the angular separation of different
  wavelengths as they enter the spectrograph, the QSO spectrum will be
  distorted relative to the ThAr calibration frames. Consider two
  wavelengths, $\lambda_2 > \lambda_1$, entering the spectrograph slit
  separated by an angle $\Delta\psi$ in the vertical direction. The
  spectral separation of the two wavelengths in the extracted, calibrated
  spectrum will be \begin{equation}\label{eq:atdisp} \Delta\lambda \approx
  \lambda_2 - \lambda_1 - \frac{a\Delta\psi\sin{\theta}}{\delta}\,,
  \end{equation} where $a$ is the CCD pixel size in angstroms and $\delta$
  is the projected slit width in arcseconds per pixel (for HIRES, $\delta =
  0\farcs287{\rm \,per~pixel}$). We checked the accuracy of this equation
  with {\sc zemax} models of the Keck/HIRES provided by S.~Vogt. If
  $\Delta\psi \neq 0$ then the spectral separation between any two
  wavelengths will decrease, i.e.~the spectrum is compressed\footnote{Note
  that in \citetalias{MurphyM_01b} we argued that this effect amounted to
  an expansion of the spectrum rather than a compression. The change in
  sign here is due to the optics of the CCD camera on Keck/HIRES which
  reverses the image on the CCD (T.~Bida \& S.~Vogt, private
  communication). This has now been checked with {\sc zemax} models of the
  HIRES.}. If $\Delta\psi$ is due to atmospheric dispersion then it is a
  function of the atmospheric conditions and can be estimated using the
  refractive index of air at the observer and the zenith angle of the QSO
  \citep[e.g.][]{FilippenkoA_82a}. Note that equation \ref{eq:atdisp}
  assumes that the seeing profiles at $\lambda_1$ and $\lambda_2$ are not
  truncated by the slit edges.

\item Wavelength dependent IP asymmetry. If tracking errors or seeing
  effects do cause profile truncation at the slit edges then the truncation
  will be asymmetric and this asymmetry will be wavelength dependent. For
  example, the optical design of Keck/HIRES is such that a blue spectral
  feature will have its red wing truncated and a red feature will be
  truncated towards the blue. The asymmetry and the severity of its
  wavelength dependence will increase with increasing $\Delta\psi$ and with
  $\theta$ (and therefore, $\xi$). Note that when we centroid absorption
  features in the QSO spectra, we will find larger wavelength separations due
  to this effect, i.e.~a positive term is effectively added to the
  right-hand-side of equation \ref{eq:atdisp}.
\end{enumerate}

For the low-$z$ Mg/Fe{\sc \,ii} systems, these two effects will have an
opposite effect on $\da$: the compression will produce $\da < 0$ and the
wavelength dependent asymmetry will result in $\da > 0$. It is important to
note that both effects rely mainly on the same parameters for each
observation (i.e.~$\Delta\psi$ and $\theta=\xi$) and so their relative
strength is fixed by our model. The one free parameter is the seeing, which
determines the strength of the wavelength asymmetry effect and not the
compression effect. We discuss this below.

For each QSO observed without the image rotator we have calculated an
effective seeing profile at the spectrograph slit. In general, each QSO
spectrum is the combination of several exposures and each exposure was
taken at a different $\xi$. For each exposure we assume typical observing
conditions at the telescope\footnote{We assumed the following observational
parameters in our atmospheric dispersion model: temperature = 280\,K,
atmospheric pressure = 600\,mbar and relative humidity = 10\,per cent.} and
use the mean value of $\xi$ calculated from the recorded image headers to
obtain the relative angular separations, $\Delta\psi$, between all observed
wavelengths. Then, to reconstruct the full dispersion pattern relative to
the slit edges, we assumed that light of wavelength $\lambda = 5500{\rm
\,\AA}$ was centred on the slit axis. This corresponds to the mean
wavelength of the response curve for the acquisition camera used to guide
Keck/HIRES on the QSO.

For each exposure we assumed a seeing of ${\rm FWHM} = 0\farcs75$ to
produce a Gaussian intensity profile at the slit for each observed
wavelength. Although this is a realistic assumption, we have no detailed
information about the seeing conditions for each observation\footnote{In
principle, one could find a reliable estimate of the seeing for each
exposure by integrating the object profile in the spatial direction along
the echelle orders. In general, we did not have access to the un-extracted,
2-dimensional object exposures and so could not perform such an
analysis.}. Tracking errors will broaden the intensity profile so the above
assumption is effectively a conservative one: if ${\rm FWHM} > 0\farcs75$
then the effect of wavelength dependent IP asymmetry on $\da$ will reduce
because each wavelength will illuminate the slit more uniformly.

For each QSO observation and at each observed wavelength we average the
slit intensity profiles for all exposures. For each observed wavelength
(i.e.~each observed MM transition) a high S/N, high spectral dispersion
synthetic spectrum is constructed based on the Voigt profile fits to the
QSO data. We convolve each spectrum with the corresponding intensity
profile, truncated by the slit edges, and then we convolve with a Gaussian
instrumental response of width ${\rm FHWM} \approx 2.2{\rm \,km\,s}^{-1}$
\citep{ValentiJ_95a}. We vary this parameter slightly depending on the
observed wavelength so as to ensure constant velocity resolution to match
the real QSO data. Finally, we re-sample the simulated spectra to match the
real QSO spectral dispersion.

\begin{figure}
\centerline{\psfig{file=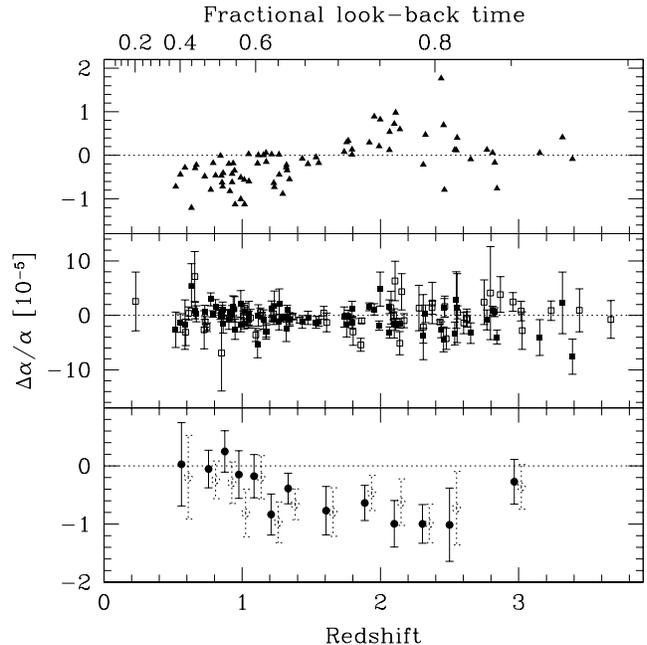,width=8.4cm}}
\caption{Correcting for potential atmospheric dispersion effects. The upper
  panel shows values of $(\da)_{\rm ad}$ derived from simulated spectra
  with our model for atmospheric dispersion applied. We have corrected the
  relevant raw values of $\da$ in the middle panel (solid squares) which
  also shows the unaffected raw values (open squares). The lower panel
  compares the binned values from the middle panel (solid circles) with
  those of the raw sample (open circles). The new overall weighted mean is
  $\da = (-0.46 \pm 0.10)\times 10^{-5}$.}
\label{fig:cps_ad_corr}
\end{figure}

The above procedure provides synthetic QSO spectra with $\da = 0$ but which
contain effects due to our model of atmospheric dispersion. We fit these
spectra and determine $(\da)_{\rm ad}$, an estimate of the effect of
atmospheric dispersion on the values of $\da$ derived from the real QSO
spectra. The results are plotted in the top panel of
Fig.~\ref{fig:cps_ad_corr} for the 77 systems observed without the image
rotator. In the low-$z$ Mg/Fe{\sc \,ii} systems, atmospheric dispersion
causes $\da < 0$ (i.e.~the compression effect dominates the wavelength
dependent IP asymmetry effect) and so, once the effect is modelled and
removed, $\da$ increases. At high $z$ not all values of $\da$ react the
same way to atmospheric dispersion since different transitions are fitted
in each absorption system. However, on average, the high-$z$ values of
$\da$ decrease after removing atmospheric dispersion effects.

The middle panel of Fig.~\ref{fig:cps_ad_corr} shows the corrected values
of $\da$, i.e.~$(\da)_{\rm adc} \equiv \da - (\da)_{\rm ad}$ (solid
squares), including those systems not affected by atmospheric dispersion
effects [$(\da)_{\rm ad} = 0$, open squares]. The lower panel compares the
binned values of $(\da)_{\rm adc}$ with the raw values from
Fig.~\ref{fig:cps}. Clearly, atmospheric dispersion effects (as modelled
above) can not explain our results. Indeed, on average, the overall
compression of the wavelength scale has an {\it opposite effect on the
low-$z$ and high-$z$ systems}. This is further borne out by the simulations
in Section \ref{sssec:rem_comp} (see Fig.~\ref{fig:cps_simplot_res}).

In summary, it is clear that atmospheric dispersion effects can not explain
the observed values of $\da$. Assuming an effective seeing, our model of
atmospheric dispersion provides a realistic upper limit to the potential
effects on $\da$. Additional tracking/guiding errors will significantly
reduce these effects by smearing the QSO light more evenly across the
spectrograph slit.

\subsubsection{Evidence for atmospheric dispersion
  effects?}\label{sssec:atdisp_evi}

\begin{figure}
\centerline{\psfig{file=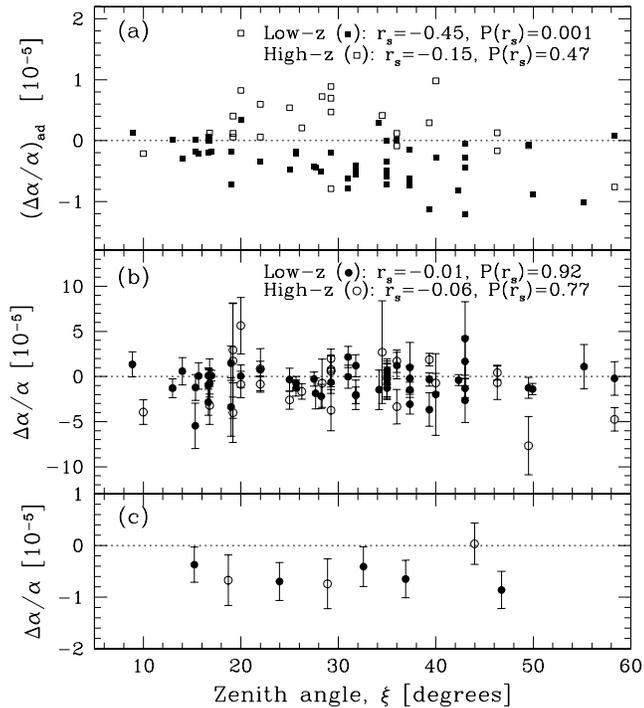,width=8.4cm}}
\caption{Is $\da$ correlated with zenith angle? {\bf (a)} The estimates of
  the effect of atmospheric dispersion on $\da$ as a function of the mean
  zenith angle, $\xi$, for each observation. The Spearman rank
  correlation coefficient, $r_{\rm s}$, and associated probability,
  $P(r_{\rm s})$, indicate a clear anti-correlation for the low-$z$ systems
  but no clear correlation for the high-$z$ systems. {\bf (b)} The real raw
  values of $\da$. No correlation is seen for either the low- or high-$z$
  systems. {\bf (c)} A binning of the real values to clarify the lack of
  correlation with $\xi$.}
\label{fig:cps_davxi}
\end{figure}

\begin{figure}
\centerline{\psfig{file=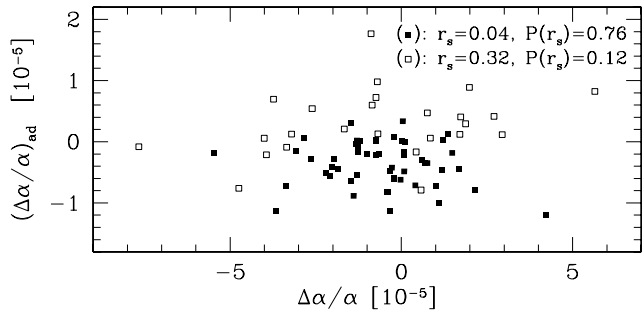,width=8.4cm}}
\caption{Is $(\da)_{\rm
  ad}$ correlated with $\da$? The Spearman rank correlation coefficient and
  associated probability indicate no clear correlation for the low-$z$
  systems (solid squares). A weak correlation may exist in the high-$z$
  systems (open squares) but only at low significance.}
\label{fig:cps_atdispcorr}
\end{figure}

\begin{figure}
\centerline{\psfig{file=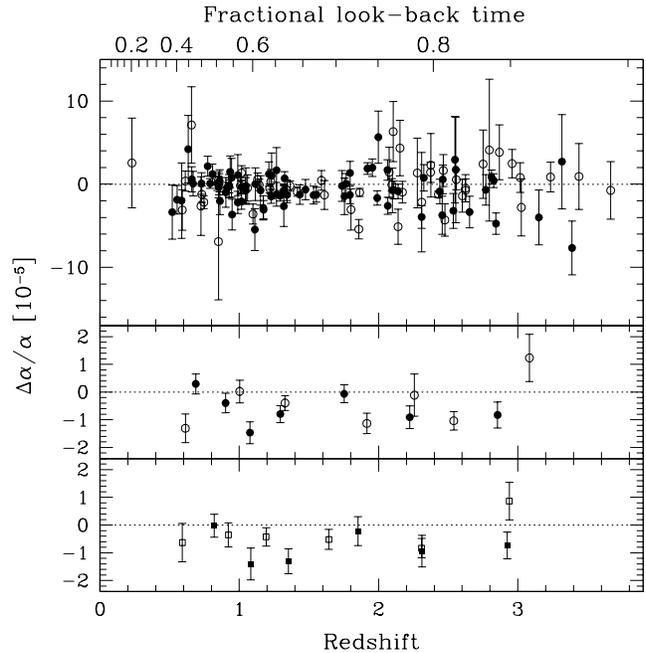,width=8.4cm}}
\caption{Comparison of the pre- and post-rotator systems. The upper panel
  compares raw values of $\da$ for the pre-rotator (solid circles, 77
  points) and post-rotator (open circles, 51 points) samples. The weighted
  means are consistent: $\da = (-0.54 \pm 0.14)\times 10^{-5}$ and $\da =
  (-0.62 \pm 0.15)\times 10^{-5}$ respectively. The middle panel bins the
  results in the same fashion as for the main results in
  Fig.~\ref{fig:cps}. We see no evidence for a significant systematic
  effect due to atmospheric dispersion. The lower panel compares the
  pre-rotator (solid squares, 39 points) and post-rotator (open squares, 39
  points) subsamples of the new sample alone. Again, we see no evidence for
  the atmospheric dispersion effect. Table \ref{tab:ad} presents the
  weighted mean values of $\da$.}
\label{fig:cps_ad}
\end{figure}

\begin{table}
\centering
\begin{minipage}{75mm}
\caption{Comparison of weighted mean values of $\da$ for the pre- and
  post-rotator systems.}
\label{tab:ad}
\begin{tabular}{lcccc}\hline
Sample  &\multicolumn{2}{c}{Pre-rotator}&\multicolumn{2}{c}{Post-rotator}\\
        &$N_{\rm abs}$&$\da$            &$N_{\rm abs}$&$\da$             \\\hline
\multicolumn{5}{l}{Total sample (Fig.~\ref{fig:cps_ad}, middle panel)}\vspace{0.1cm}\\
low-$z$ &52           &$-0.60 \pm 0.16$ &22           &$-0.45 \pm 0.20$  \\
high-$z$&25           &$-0.39 \pm 0.26$ &29           &$-0.83 \pm 0.23$  \\
Total   &77           &$-0.54 \pm 0.14$ &51           &$-0.62 \pm 0.15$  \\ \\
\multicolumn{5}{l}{New sample only (Fig.~\ref{fig:cps_ad}, lower panel)}\vspace{0.1cm}\\
low-$z$ &22           &$-0.67 \pm 0.25$ &22           &$-0.45 \pm 0.20$  \\
high-$z$&17           &$-0.77 \pm 0.33$ &17           &$-0.50 \pm 0.31$  \\
Total   &39           &$-0.71 \pm 0.20$ &39           &$-0.46 \pm 0.17$  \\\hline
\end{tabular}
\end{minipage}
\end{table}

If atmospheric dispersion significantly affects $\da$ then, from the
discussion above, we expect to find a correlation between $\da$ and the
zenith distance, $\xi$. We search for a such a correlation in
Fig.~\ref{fig:cps_davxi}. In panel (a) we plot the estimates of $(\da)_{\rm
ad}$ from the previous section versus the mean value of $\xi$ for each
absorption system affected by atmospheric dispersion. The Spearman rank
correlation coefficient, $r_{\rm s}$, and associated probability, $P(r_{\rm
s})$\footnote{$P(r_{\rm s})$ is the probability that $\left|r_{\rm
s}\right|$ could have been exceeded by chance.}, indicate a clear
anti-correlation for the low-$z$ systems (solid squares). However, we see
no evidence for such a correlation in the real low-$z$ values which are
shown in panel (b). Panel (c) shows a binning of the results for
clarity. This indicates that atmospheric dispersion has had no significant
effect on the raw values of $\da$. For the high-$z$ systems, we also see no
correlation between $\da$ and $\xi$ but, since the upper panel shows no
clear correlation for the estimates of $(\da)_{\rm ad}$, one cannot take
this as evidence against atmospheric dispersion. One also expects a
correlation between $(\da)_{\rm ad}$ and $\da$ if atmospheric dispersion
effects are significant. However, in Fig.~\ref{fig:cps_atdispcorr} we see
no clear evidence for such a correlation for either the low- or high-$z$
samples (solid and open squares respectively).

We may also test for atmospheric dispersion effects by comparing the value
of $\da$ for the (77) pre- and (51) post-rotator values of $\da$. In the
top panel of Fig.~\ref{fig:cps_ad} we identify these subsamples and plot
binned values in the middle panel. There is no discrepancy between the pre-
and post-rotator samples. This is confirmed in Table \ref{tab:ad} where we
compare the weighted mean values of $\da$. Even for the low-$z$ systems,
which should be most susceptible to atmospheric dispersion effects, we see
no evidence for a discrepancy. We also consider the new sample separately
in the lower panel of Fig.~\ref{fig:cps_ad} where exactly half of the
systems are pre-rotator observations. Again, in Table
\ref{tab:ad}, the weighted mean values of $\da$ are consistent.

To summarize this section, we emphasize that the QSO data themselves
suggest that atmospheric dispersion had an insignificant effect on
$\da$. No correlation between $\da$ and the zenith angle of the QSO
observations is observed (Fig.~\ref{fig:cps_davxi}) and the pre- and
post-rotator samples give completely consistent results
(Fig.~\ref{fig:cps_ad} and Table \ref{tab:ad}).

\subsection{Line--removal tests}\label{ssec:remove}

A non-zero $\da$ manifests itself as a distinct pattern of line shifts (see
Fig.~\ref{fig:q_vs_wl}). For a given absorption system, with a given set of
fitted transitions, we can remove one or more transitions and still obtain
a well-constrained value of $\da$. In the absence of systematic errors
associated with the transition removed or with the wavelength scale of the
QSO spectra, $\da$ should not change systematically over all absorption
systems. Removing a single transition or distinct groups of transitions
from our analysis therefore allows us to search for systematic errors
without specific knowledge of their origin. Note that removing one or more
transitions will result in a slightly modified estimate of the velocity
structure, revising the estimate of $\da$ for each system. However, this
will have a random effect on $\da$; the real question is, how robust are
the values of $\da$, averaged over the entire sample, to this line removal
process?

Below we construct three different line removal tests, each sensitive to
different types of systematic errors.

\subsubsection{Single transition and species removal}\label{sssec:rem_single}

Removing a single transition (e.g.~Mg{\sc \,ii} $\lambda$2796) from the
fits to the QSO data provides a direct means to search for systematic
errors due to systematic line blending (Section \ref{sssec:sys_blends}),
isotopic and hyperfine structure effects (Section \ref{ssec:iso}) and large
errors in the measured laboratory wavelengths in Table
\ref{tab:atomdata}. Removing entire species (e.g.~all transitions of Si{\sc
\,ii}) is also a test for isotopic and hyperfine structure effects.

Given an absorption system, with a certain set of fitted transitions, we
only remove transitions that {\it can} be removed. To clarify this,
consider removing the Mg{\sc \,ii} $\lambda$2796 transition. For the
low-$z$ Mg/Fe{\sc \,ii} systems, a well-constrained value of $\da$ can only
be obtained when at least one Mg line is fitted since there is only a small
difference between the $q$ coefficients for the different (low-$z$) Fe{\sc
\,ii} transitions (see Fig.~\ref{fig:q_vs_wl}). Therefore, if the only
anchor line present is Mg{\sc \,ii} $\lambda$2796 then we can not remove it
from the system. Similarly, we only remove the entire Mg{\sc \,ii} species
if the Mg{\sc \,i} $\lambda$2852 line is fitted.

\begin{figure}
\centerline{\psfig{file=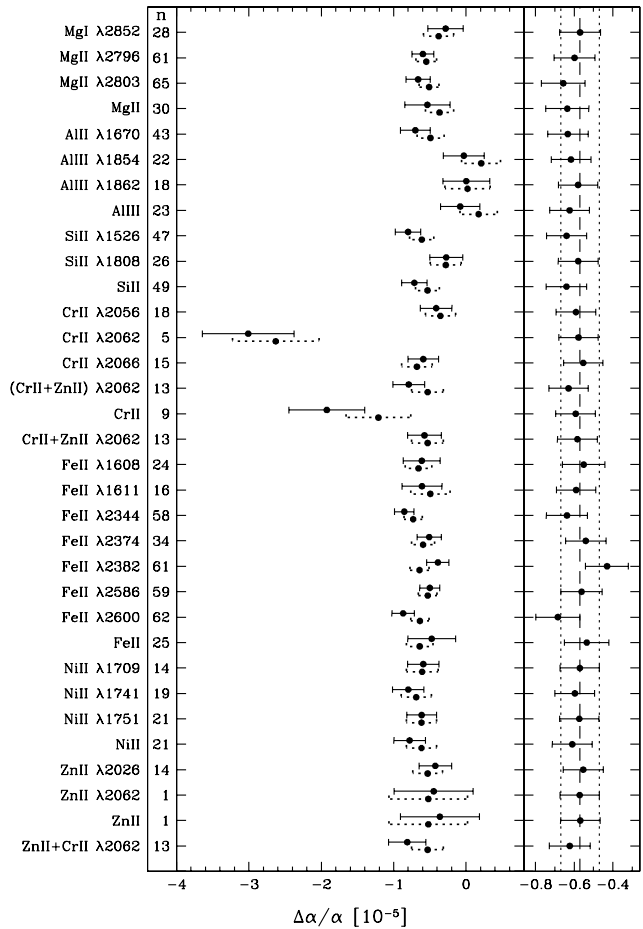,width=8.4cm}}
\caption{Single transition and species removal. The left panel compares the
  raw values of $\da$ before (dotted error bar) and after (solid error bar)
  line removal. The transitions or species removed are listed on the left
  together with the number of systems, $n$, for which line removal was
  possible. The right panel shows the impact of line removal on the value
  of $\da$ for the entire sample. The vertical lines indicate the raw
  weighted mean value of $\da$ (dashed line) and 1\,$\sigma$ error range
  (dotted lines). Note that some confusion may arise due to the occasional
  blending of the Cr{\sc \,ii} and Zn{\sc \,ii} $\lambda$2062 lines:
  `(Cr{\sc \,ii} + Zn{\sc \,ii}) $\lambda$2062' refers to cases where both
  transitions had to be removed simultaneously; `Cr{\sc \,ii} + Zn{\sc
  \,ii} $\lambda$2062' refers to cases when all Cr{\sc \,ii} transitions
  were removed along with the blended Zn{\sc ii} $\lambda$2062 line; a
  similar definition applies to `Zn{\sc \,ii} + Cr{\sc \,ii}
  $\lambda$2062'; `Cr{\sc \,ii}' and `Zn{\sc \,ii}' refer only to removal
  of the Cr{\sc \,ii} and Zn{\sc \,ii} species in cases where the Cr{\sc
  \,ii} and Zn{\sc \,ii} $\lambda$2062 lines were not blended.}
\label{fig:cps_rmline}
\end{figure}

We presented the results of removing single transitions and entire species
for the previous samples in \citetalias{MurphyM_01b}. In
Fig.~\ref{fig:cps_rmline} we present line removal results for the total
sample. The transition/species removed is given on the vertical axis
together with the number of systems, $n$, where removal was possible. The
left panel compares the weighted mean value of $\da$ before removing the
transition/species (dotted error bar) with that obtained afterwards (solid
error bar) for these $n$ systems. The right panel shows the effect of
removing the transition/species on the weighted mean $\da$ for the sample
as a whole (i.e.~including those systems where the transition in question
could not be removed). Raw values of $\da$ are used in all cases (i.e.~the
1\,$\sigma$ errors returned from {\sc vpfit} are not increased to take
account of extra scatter in high-$z$ sample).

Fig.~\ref{fig:cps_rmline} provides no evidence to suggest that systematic
errors associated with any one transition or species have significantly
affected $\da$. This strongly suggests that isotopic/hyperfine saturation
or evolution effects are not important for the transitions of
interest. This is particularly important information for the majority of
transitions for which the isotopic structures are not known (see Section
\ref{ssec:iso}). The line removal results also confirm that random and
systematic blending with unidentified transitions has a negligible effect
on $\da$.

\subsubsection{High-$z$ compression test}\label{sssec:rem_comp}

As previously noted, the arrangement of the $q$ coefficients for the
low-$z$ Mg/Fe{\sc \,ii} systems (see Fig.~\ref{fig:q_vs_wl}) implies that a
compression of the wavelength scale will systematically lead to $\da <
0$. The dependence of $\da$ on the $q$ coefficients is considerably more
complicated for the high-$z$ systems: a compression of the wavelength scale
will have a different effect on each value of $\da$ depending on which
transitions are fitted. That is, compression is not degenerate with $\da$
at high-$z$. If compression is responsible for the observed $\da < 0$ at
low $z$, can evidence for it be found in the high-$z$ absorption systems?

\begin{figure}
\centerline{\psfig{file=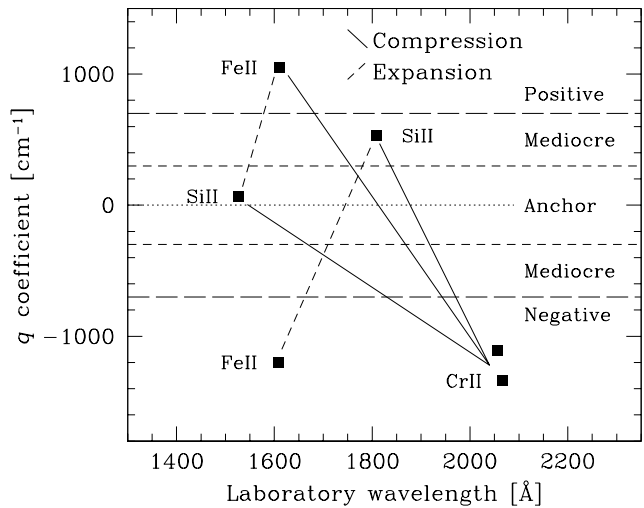,width=8.4cm}}
\caption{The high-$z$ compression test. A compression of the spectrum will
  systematically produce $\da < 0$ in a high-$z$ absorption system if only
  certain combinations of transitions (solid diagonal lines) are fitted
  simultaneously. This mimics the situation for the low-$z$ Mg/Fe{\sc \,ii}
  absorption systems and allows us to search for a compression-type
  systematic effect in the high-$z$ systems. Combinations of transitions
  which will give $\da < 0$ for an expansion of the spectrum can be
  selected in a similar way (dashed diagonal lines).}
\label{fig:q_vs_wl_eg}
\end{figure}

We search for compression in the high-$z$ sample by fitting combinations of
transitions for which the $q$ coefficients are arranged in a similar way to
those for the low-$z$ Mg/Fe{\sc \,ii} systems. For example, consider a
system where the following transitions are present in the QSO spectrum:
Si{\sc \,ii} $\lambda\lambda$1526 \& 1808, Fe{\sc \,ii}
$\lambda\lambda$1608 \& 1611 and Cr{\sc \,ii} $\lambda\lambda$2056 \&
2066. The arrangement of $q$ coefficients for such a system is shown in
Fig.~\ref{fig:q_vs_wl_eg}. Fitting the Si{\sc \,ii} $\lambda$1526 and
Cr{\sc \,ii} transitions will yield $\da < 0$ if a compression of the
spectrum exists. Note that several different combinations could be used to
the same effect, as indicated by solid diagonal lines. We treat such cases
separately in our analysis since the lower wavelength transitions
(i.e.~Si{\sc \,ii} $\lambda\lambda$1526 \& 1808 and Fe{\sc \,ii}
$\lambda$1608) are of different $q$-type. The values of $\da$ and the
1\,$\sigma$ errors in these separate cases will not be independent. Note
also that we can form combinations of transitions which mimic $\da < 0$ for
an {\it expansion} of the spectrum.

From the high-$z$ sample, 47 systems contribute 76 different, but not
independent, combinations to the compression sample. Similarly, 59 systems
contribute 126 combinations to the expansion sample. In Table
\ref{tab:comp} we calculate the raw values of the weighted and unweighted
means for each sample both before and after removing transitions to form
the compression and expansion combinations. Note that the number of
compression combinations formed in a given absorption system is the number
of times that system contributes to the pre-removal values. Table
\ref{tab:comp} shows no clear evidence for compression (or expansion) in
the high-$z$ sample. However, the weighted and unweighted means are not
adequate statistics because each absorption system can contain many
combinations. Furthermore, the pre- and post-removal values and their
1\,$\sigma$ errors are not independent of each other. Also of note are the
large errors in the unweighted mean post-removal values, indicating a large
scatter in the results and, therefore, that the weighted means are
unreliable.

\begin{figure}
\centerline{\psfig{file=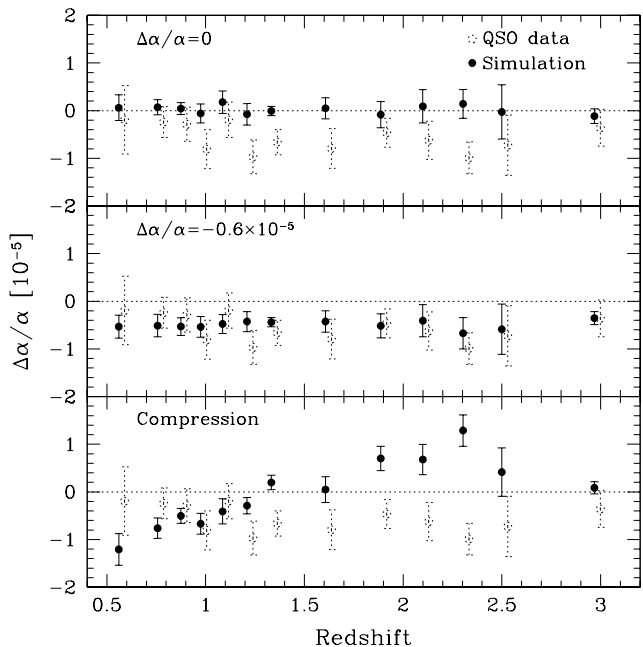,width=8.4cm}}
\caption{$\da$ from simulations of the QSO spectra. The
weighted mean from 20 simulations (solid circles) is compared with the real
raw values (dotted circles, shifted for clarity) in each panel. The rms
from the simulations is represented by the error bar. The upper panel has
no input value of $\da$ whereas $\da=-0.6\times 10^{-5}$ was introduced
into the spectra for the middle panel. Note that we recover the input
values and that the rms error bars match the 1\,$\sigma$ errors in the real
data. A compression was introduced into the spectra for the lower panel
(equation \ref{eq:comp}) to allow interpretation of the high-$z$
compression and $q$-type removal test results.}
\label{fig:cps_simplot_res}
\end{figure}

\begin{table*}
\centering
\begin{minipage}{133mm}
\caption{Coarse results from the high-$z$ compression test. Columns 4 and 5
  show the raw values of the weighted and unweighted mean $\da$ (in units
  of $10^{-5}$) for the relevant absorption systems before removing the
  transitions. Each absorption system may contribute many times to these
  values, depending on how many compression/expansion combinations can be
  formed from the fitted transitions. Columns 6 and 7 give the weighted and
  unweighted mean after transitions are removed to produce the
  compression/expansion combinations. Note the large errors in the
  post-removal unweighted means. Other caveats of interpreting this table
  are discussed in the text.}
\label{tab:comp}
\begin{tabular}{lcccccc}\hline
Sample      &$N_{\rm abs}$&Combinations&\multicolumn{2}{c}{Pre-removal}              &\multicolumn{2}{c}{Post-removal}             \\
            &             &            &$\left<\da\right>_{\rm w}$&$\left<\da\right>$&$\left<\da\right>_{\rm w}$&$\left<\da\right>$\\\hline
Compression &47           &76          &$-0.74 \pm 0.16$          &$-0.88 \pm 0.34$  &$-1.56 \pm 0.31$          &$-0.82 \pm 0.86$  \\
Expansion   &59           &126         &$-0.52 \pm 0.10$          &$-0.34 \pm 0.26$  &$-0.38 \pm 0.17$          &$-0.93 \pm 0.72$  \\\hline
\end{tabular}
\end{minipage}
\end{table*}

\begin{figure*}
  \hbox{
     \psfig{file=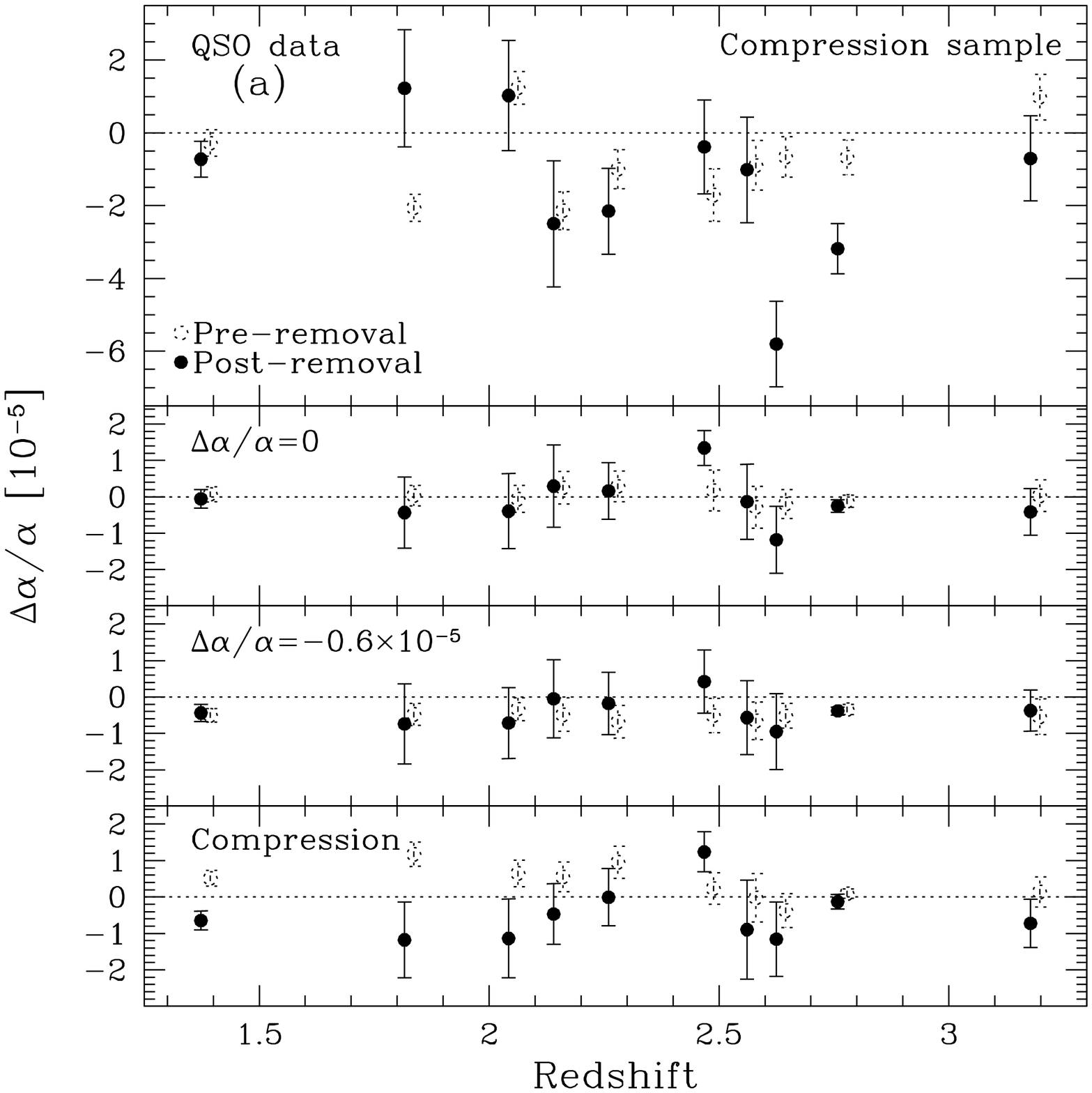,width=8.4cm}
     \hspace{0.5cm}
     \psfig{file=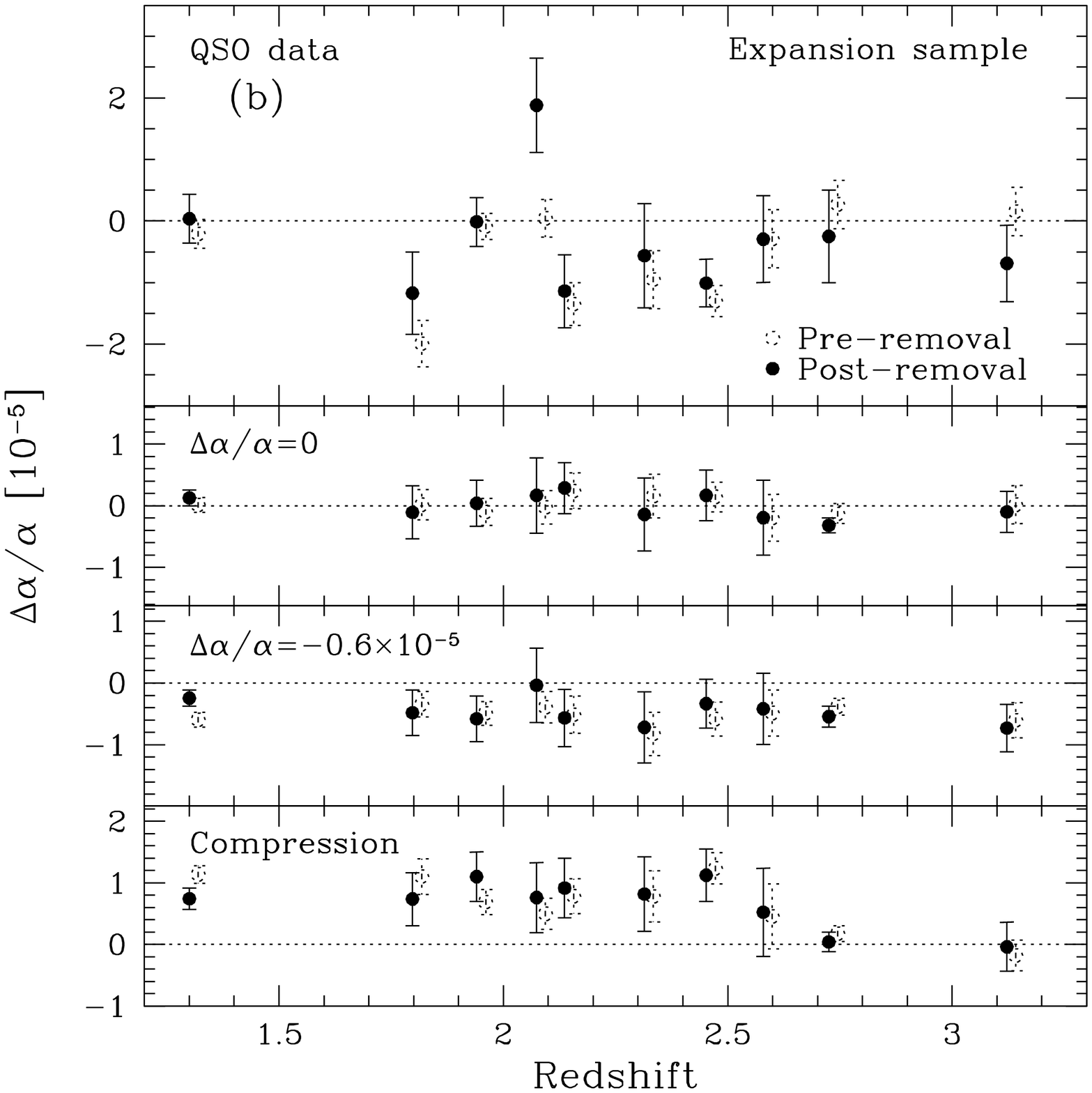,width=8.4cm}
  }
  \caption{Detailed results of the high-$z$ compression test for the QSO
  data (upper panels) and simulations (lower 3 panels). The error bars for
  the simulations represent the rms from 20 synthetic spectra. {\bf (a)}
  Compression sample. Each panel shows binned values of $\da$ before and
  after transitions are removed to form the 76 compression
  combinations. The bottom panel shows that the compression sample is
  sensitive to the synthetic compression introduced to the simulated
  spectra. No clear trend is seen in the real QSO data due to the large
  scatter in the post-removal values (see text for discussion). {\bf (b)}
  Expansion sample. 126 combinations contribute to the binned values of
  $\da$ shown. No significant difference is observed between the pre- and
  post-removal values in any of the simulations. The lower panel implies
  that the expansion sample is surprisingly insensitive to simple
  systematic errors. Significant scatter is also observed in the post-removal
  QSO results.}
\label{fig:cps_comp_exp}
\end{figure*}

To properly interpret results from the compression test, a comparison must
be made with detailed simulations of the QSO data. We synthesized each
absorption system from the Voigt profile fit to the real QSO data, adding
Gaussian noise appropriate to the measured S/N. From each synthetic
spectrum we constructed 3 different simulations:
\begin{enumerate}
\item $\da = 0$. No line shifts were introduced into the synthetic spectra.
\item $\da = -0.6 \times 10^{-5}$. Each transition in the synthetic
   spectrum was shifted according to this value of $\da$ which corresponds
   to the raw weighted mean value measured in the real QSO data.
\item Compression. The synthetic spectrum was compressed according to
   \begin{equation}\label{eq:comp}
   \lambda^\prime = \lambda - C(\lambda_c - \lambda)\,,
   \end{equation}
   where $\lambda$ is the initial wavelength of a given pixel and
   $\lambda^\prime$ is the new value obtained by compressing the spectrum
   about a central wavelength, $\lambda_c$, by a factor $C$. This mimics
   the compression effect of atmospheric dispersion, as described by
   equation \ref{eq:atdisp}. Thus, we chose $\lambda_c=5500{\rm \,\AA}$ as
   representative. In order to produce $\da \approx -0.6 \times 10^{-5}$ in
   the low-$z$ simulated spectra, we used $C=4\times 10^{-6}$.
\end{enumerate}
We ran 20 simulations of each absorption system, obtaining values of $\da$
for each of the 3 cases above. The results are compared with the real, raw
sample in Fig.~\ref{fig:cps_simplot_res}. Note that (i) we recover the
input values of $\da$ for the $\da = 0$ and $\da = -0.6\times 10^{-5}$
cases and (ii) the compression simulation shows a similar behaviour to that
of the atmospheric dispersion calculation in Fig.~\ref{fig:cps_ad_corr},
confirming that simple distortions of the wavelength scale should have had
an opposite overall effect on the high-$z$ systems compared to the low-$z$
systems.

We have applied the high-$z$ compression test to the simulations for each
of the above 3 cases. The results are shown in the lower three panels of
Fig.~\ref{fig:cps_comp_exp}, binned for clarity. The error bars on the
simulated pre- and post-removal values represent the rms from the 20
simulations of each absorption system. For the $\da = 0$ and $\da = -0.6
\times 10^{-5}$ simulations, note that $\da$ does not change systematically
after removing transitions to form the compression and expansion
samples. As expected, the compression of the synthetic spectra causes the
post-removal values of the compression sample (bottom panel of
Fig.~\ref{fig:cps_comp_exp}a) to shift towards more negative
values. However, the compression simulation for the expansion sample
(bottom panel of Fig.~\ref{fig:cps_comp_exp}b) reveals this sample to be
surprisingly insensitive to the artificial compression of the spectra.

\begin{table*}
\centering
\begin{minipage}{118mm}
\caption{Coarse results of the high-$z$ $q$-type removal test. Columns
  3 and 4 show the raw weighted and unweighted mean values of $\da$ (in
  units of $10^{-5}$) for the 26 relevant absorption systems before
  removing the transitions (the first three rows necessarily have the same
  value). Columns 5 and 6 give the weighted and unweighted mean after
  transitions of the specified $q$-type are removed. Note that the pre- and
  post-removal values of $\da$ and the 1\,$\sigma$ errors are not
  independent of each other. Also note the large errors in the post-removal
  unweighted means. A detailed comparison must refer to simulations of the
  QSO data (see Fig.~\ref{fig:cps_rmhizq}).}
\label{tab:hizq}
\begin{tabular}{lccccc}\hline
$q$-type removed &$N_{\rm abs}$&\multicolumn{2}{c}{Pre-removal}              &\multicolumn{2}{c}{Post-removal}             \\
                 &             &$\left<\da\right>_{\rm w}$&$\left<\da\right>$&$\left<\da\right>_{\rm w}$&$\left<\da\right>$\\\hline
Anchors          &26           &$-0.65 \pm 0.18$          &$-0.68 \pm 0.55$  &$-0.88 \pm 0.21$          &$-0.88 \pm 0.46$  \\
Positive-shifters&26           &$-0.65 \pm 0.18$          &$-0.68 \pm 0.55$  &$ 1.21 \pm 0.52$          &$ 0.25 \pm 1.39$  \\
Negative-shifters&26           &$-0.65 \pm 0.18$          &$-0.68 \pm 0.55$  &$-0.63 \pm 0.34$          &$-1.89 \pm 1.28$  \\
Mediocre-shifters&21           &$-0.48 \pm 0.22$          &$-0.39 \pm 0.59$  &$-0.41 \pm 0.22$          &$-0.26 \pm 0.72$  \\\hline
\end{tabular}
\end{minipage}
\end{table*}

\begin{figure*}
  \vbox{ \hbox{ \psfig{file=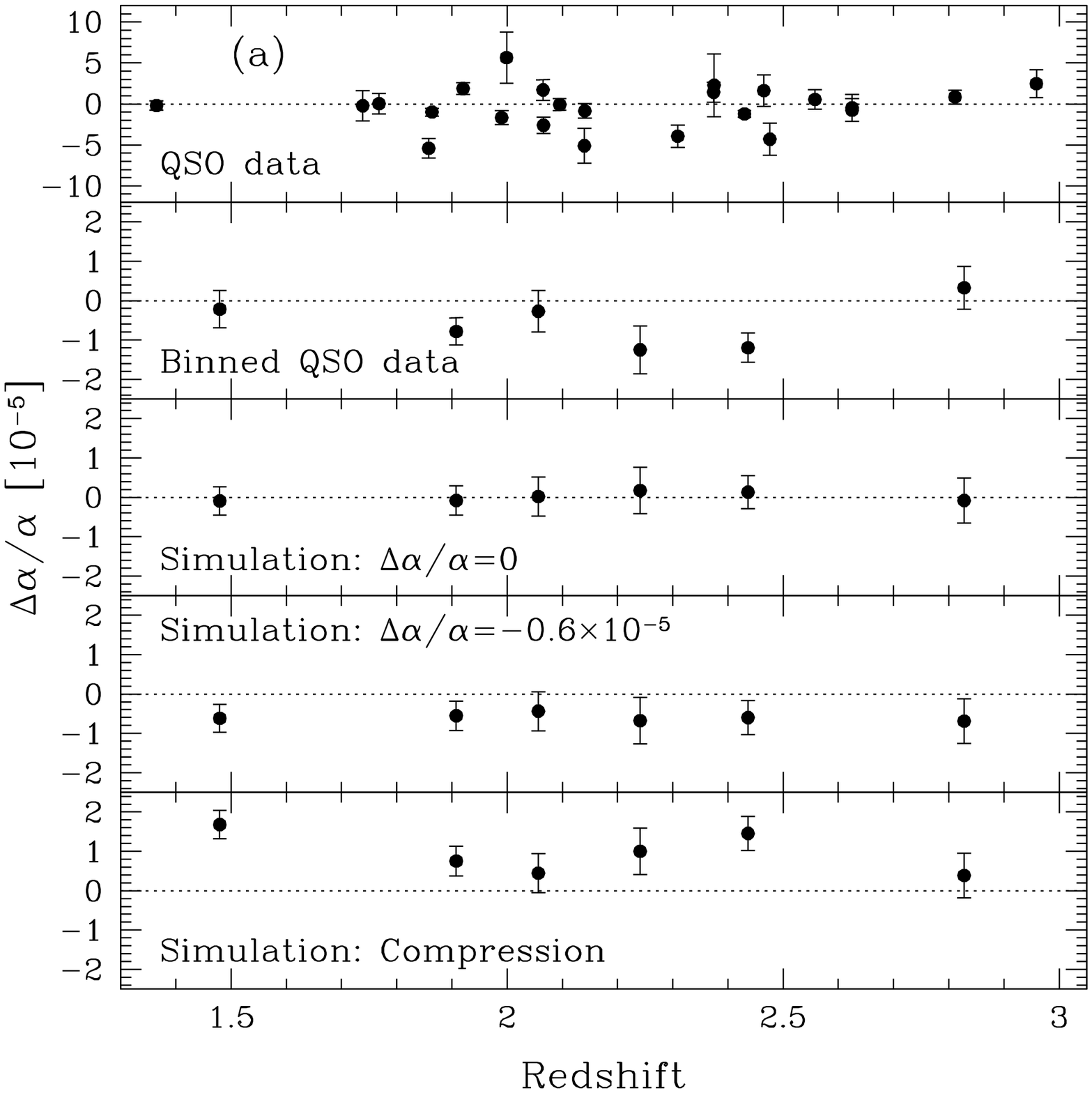,width=8.4cm}
    \hspace{0.5cm} \psfig{file=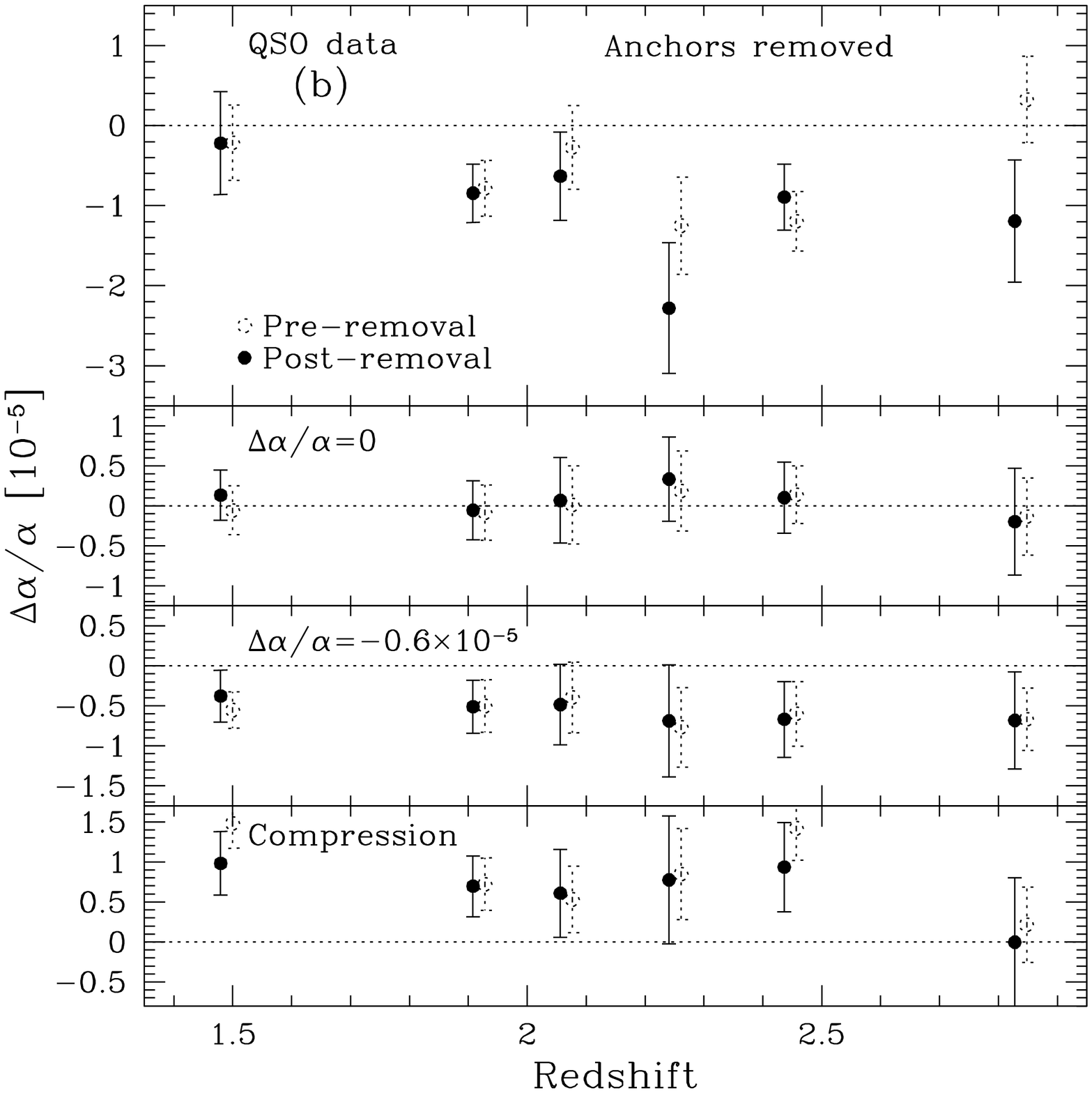,width=8.4cm} }
    \vspace{0.3cm} \hbox{ \psfig{file=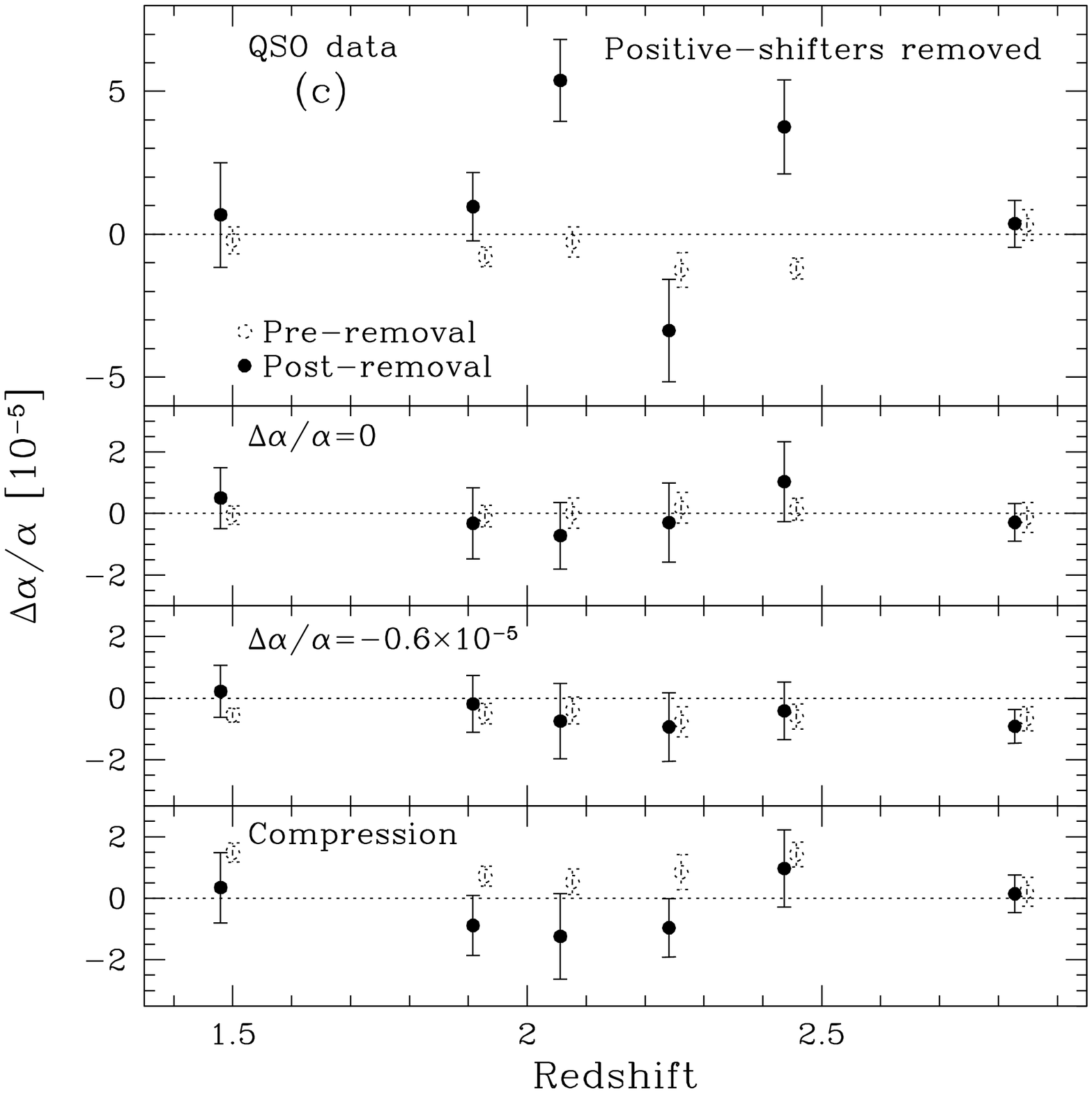,width=8.4cm}
    \hspace{0.5cm} \psfig{file=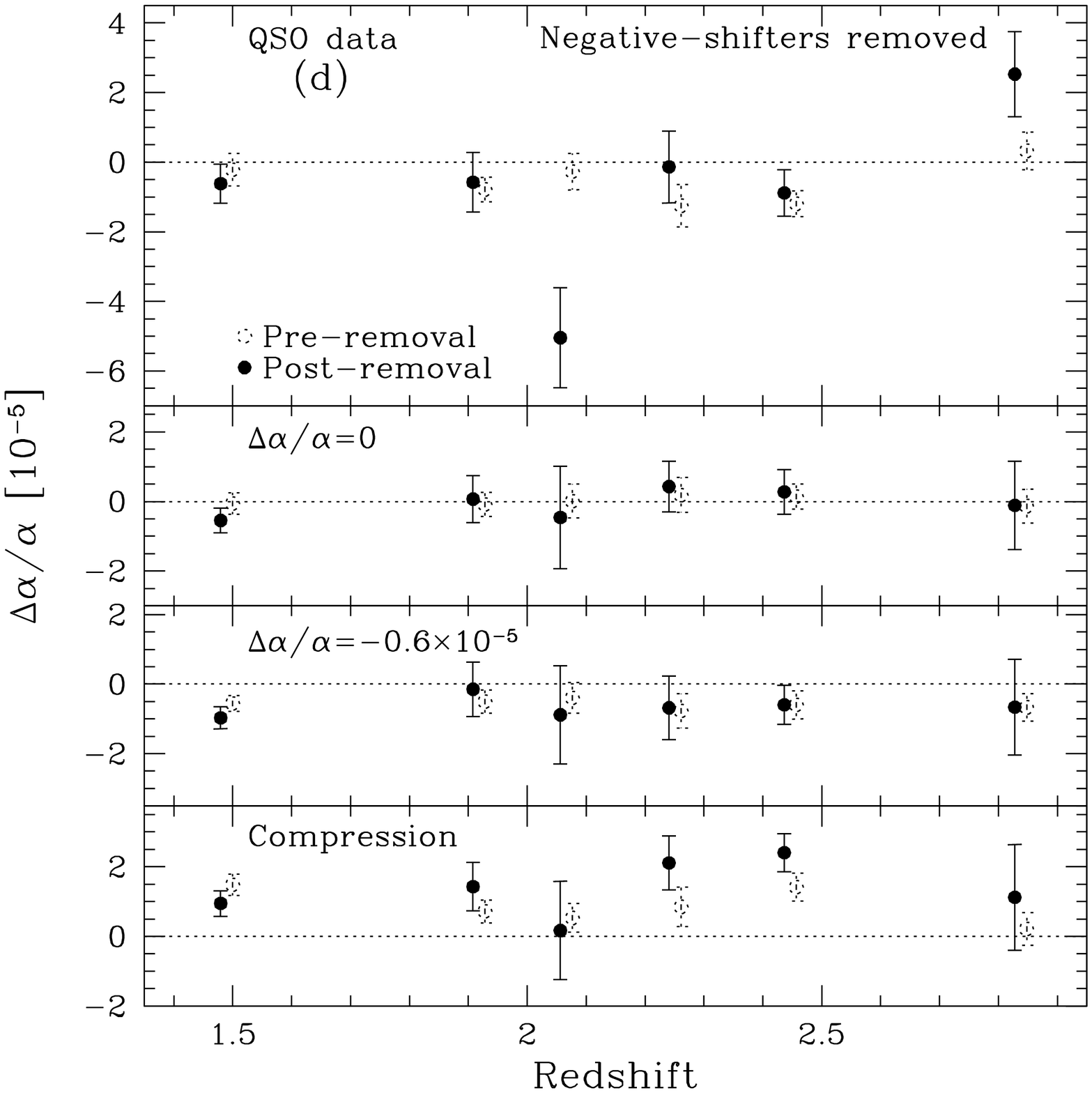,width=8.4cm} } }
    \caption{Detailed results of the high-$z$ $q$-type removal test for the
    QSO data and simulations (lower 3 panels). The error bars for the
    simulations represent the rms from 20 synthetic spectra. {\bf (a)} The
    upper panel shows the values of $\da$ for the 26 systems comprising the
    $q$-type sample, i.e.~those with at least one anchor, positive- and
    negative-shifter. These data are binned in the next panel. The lower
    three panels show the weighted mean values of $\da$ for the different
    simulations. These values and the binned QSO data are shown as dotted
    circles in (b), (c) and (d). {\bf (b)} Anchors removed. The synthetic
    spectra with artificial compression (lower panel) indicate that
    removing the anchors should not affect $\da$ significantly, even when
    distortions of the wavelength scale are present. The QSO values also
    show little deviation after line removal but some extra scatter may
    exist. {\bf (c)} and {\bf (d)} Positive- and negative-shifters
    removed. Although the compression simulation reveals a significant
    change in $\da$ upon line-removal, the extra scatter in the
    post-removal QSO data precludes a firm conclusion about the presence of
    systematic effects (see text for discussion).}
\label{fig:cps_rmhizq}
\end{figure*}

The top panels of Fig.~\ref{fig:cps_comp_exp} compare the pre- and
post-removal values of $\da$ for the real QSO data. The results are binned
in the same way as the simulations but the error bars represent the
1\,$\sigma$ error in the weighted mean of values within each bin. Of
particular note is the large scatter in the post-removal values compared to
the size of the rms error bars in the simulations. The scatter is also
large compared to the post-removal shift in $\da$ seen in the compression
simulations. We explained the origin of the extra scatter at high-$z$ in
Section \ref{ssec:scat} and formed the fiducial sample by adding an
additional random error to the systems likely to be most strongly
affected. We expect this scatter to significantly increase as transitions
are removed from the fit, as in the present case. The degree of extra
scatter is, to some extent, represented by the large uncertainties in the
unweighted means presented in Table \ref{tab:comp}. These values should
therefore be far more reliable in such circumstances.

It is therefore clear that the potential power of the compression
test (demonstrated in the bottom panel of Fig.~\ref{fig:cps_comp_exp}a) is
undermined by the extra scatter in the high-$z$ values. Unfortunately, one
can draw no conclusions about potential systematic errors from the
compression test without a significant increase in high-$z$ sample size.

\subsubsection{High-$z$ $q$-type removal}\label{sssec:rem_q}

If simple systematic errors are responsible for our measured non-zero
$\da$, it is surprising that $\da$ is so consistent between the low- and
high-$z$ samples. The lower panel of Fig.~\ref{fig:cps_simplot_res}
explicitly demonstrates this. It would also be surprising if we were to
take subsets of the QSO absorption lines, grouped according to the sign and
magnitude of $q$, and find consistent values of $\da$. We can apply such a
test to the high-$z$ sample only, since it contains transitions with
several distinct types of $q$ coefficient (see Fig.~\ref{fig:q_vs_wl}).

We have applied this high-$z$ $q$-type removal test to 26 absorption
systems. These contained at least one transition of each of the highly
delineated $q$-types, i.e.~anchors, positive- and negative-shifters. 21 of
these systems also contain at least one mediocre-shifter. If a low-order
distortion of the wavelength scale causes the observed non-zero $\da$,
removing all transitions of one $q$-type may result in a significant change
in $\da$. Table \ref{tab:hizq} compares the weighted and unweighted mean
values of $\da$ before and after $q$-type removal. Although removing the
anchors, mediocre-shifters and negative-shifters produces only small
changes in $\da$, removing the positive-shifters seems to cause to a
significant change. However, two important caveats should be noted: (i) the
pre- and post-removal values of $\da$ are not independent of each
other. This also applies to the 1\,$\sigma$ error bars. (ii) After removing
the positive- and negative-shifters, the errors in the unweighted mean
values are large, indicating a large scatter in $\da$. Both these points
make the results in Table \ref{tab:hizq} difficult to interpret. As in the
previous section, the unweighted mean values should be far more reliable
than the weighted means due to the extra scatter. This is confirmed below.

As for the high-$z$ compression test, comparison with simulations
facilitates interpretation of the above results. We have applied the
$q$-type removal test to the 3 simulations described in Section
\ref{sssec:rem_comp}. The results are shown in the lower three panels of
Fig.~\ref{fig:cps_rmhizq}, binned for clarity. The error bars on the
simulated pre- and post-removal values represent the rms from 20
simulations of each absorption system. Firstly, note the consistency
between the pre- and post-removal values for the $\da = 0$ and $\da = -0.6
\times 10^{-5}$ simulations in panels (b), (c) and (d). Also, in the
simulations of compressed spectra, we see that removing the anchor
transitions [panel (b)] does not significantly affect the values of
$\da$. However, the effect is more significant when the positive- and
negative-shifters are removed. Overall, we see $\da$ decrease when the
positive-shifters are removed whereas removing the negative-shifters
increases $\da$. This illustrates the potential power of the $q$-type
removal test in searching for low-order systematic distortions of the
wavelength scale.

The top panels of Fig.~\ref{fig:cps_rmhizq}b, c and d compare the pre- and
post-removal results for the real QSO data. The results are binned in the
same way as the simulations but the error bars represent the 1\,$\sigma$
error in the weighed mean of values within each bin. As for the high-$z$
compression test results in Fig.~\ref{fig:cps_comp_exp}, the results are
confused by the significant scatter in the post-removal QSO values. We even
observe extra scatter when removing the anchors from the real QSO
data. Again, this extra scatter is expected. We also expect it to be worse
here than for the compression test since (i) removing the positive- or
negative-shifters significantly reduces the sensitivity of the MM method to
$\da$, (ii) less absorption systems contain all three of the necessary
$q$-types than systems contributing to either the compression or expansion
samples and (iii) many combinations of transitions can be used for each
absorption system in the compression test. Also, by selecting systems which
contain anchors, positive- and negative-shifters, we tend to select systems
with transitions of very different line strengths (e.g.~transitions of
Al{\sc \,ii}, Cr{\sc \,ii} and Zn{\sc \,ii}; see
Fig.~\ref{fig:q_vs_wl}). We explained how this leads to extra scatter in
values of $\da$ in Section \ref{ssec:scat} and defined a `high-contrast'
sample of 22 systems at high-$z$. 18 of these systems are contained in the
sample of 26 systems where $q$-type removal is possible. The extra scatter
is clearly apparent in Fig.~\ref{fig:cps_rmhizq}a which shows the unbinned
pre-removal values of $\da$ for these 26 systems.

To summarize this section, simulations indicate that removing transitions,
grouped according to the magnitude and sign of $q$, is a potentially
powerful test for simple systematic errors in the high-$z$ data. However,
the extra scatter in the values of $\da$ for the relevant 26 absorption
systems precludes any conclusion about the existence or magnitude of such
systematic effects.

\section{Discussion and Conclusions}\label{sec:concs}

\subsection{Summary of new results}\label{ssec:summary}

We have used the MM method to analyse 3 samples of Keck/HIRES QSO spectra
containing a total of 128 absorption systems over the redshift range $0.2 <
z_{\rm abs} < 3.7$. All 3 samples independently yield consistent,
significantly non-zero values of $\da$ (Table \ref{tab:stats},
Fig.~\ref{fig:cps}). Combining the samples gives a weighted mean $\da =
(-0.574\pm 0.102)\times 10^{-5}$. For systems at $z_{\rm abs}<1.8$, the
predominant transitions arise from Mg{\sc \,ii} and Fe{\sc \,ii} whereas
the higher $z$ systems contain a variety of ions and transitions which have
a diverse dependence on $\alpha$. Therefore, if systematic effects were
responsible for the observed non-zero $\da$, one would expect significantly
different values at low- and high-$z$
(Fig.~\ref{fig:cps_simplot_res}). However, we find a similar $\da$ in both
cases, suggesting that our results are robust.

We identified a source of additional {\it random} scatter in the values of
$\da$ for 22 (out of 54) high-$z$ systems, i.e.~those for which the fitted
transitions have widely differing optical depths. In these systems, the
velocity structure is primarily constrained by the low optical depth
transitions. Any velocity components too weak to feature in low optical
depth transitions may, nevertheless, subtly affect the fitted line
positions of nearby velocity components in the high optical depth
transitions. We have therefore increased the errors on $\da$ for these
systems to match the observed scatter. Our most robust estimate for the
overall weighted mean becomes $\da = (-0.543\pm 0.116)\times 10^{-5}$,
representing 4.7\,$\sigma$ statistical evidence for a smaller $\alpha$ in
the QSO absorption systems (Fig.~\ref{fig:cps_fiducial}).

If one assumes that $\da=0$ at $z_{\rm abs}=0$ then a constant increase in
$\alpha$ with time is preferred by the data over a constant offset from the
laboratory value: $\dota = (6.40 \pm 1.35)\times 10^{-16}{\rm \,yr}^{-1}$
(Fig.~\ref{fig:cps_fits}). However, a bootstrap analysis demonstrates that
this preference has low significance. We find no evidence for dipolar
variations from the angular distribution of $\da$
(Fig.~\ref{fig:cps_aitoff}) and no evidence for spatial correlations in
$\alpha$ over 0.2--13\,Gpc comoving scales from the two point correlation
function (Fig.~\ref{fig:cps_2ptcorr}).

We have searched for possible instrumental and astrophysical systematic
effects which could mimic the above evidence for varying $\alpha$. As we
found in \citetalias{MurphyM_01b}, the two most important of these are due
to possible atmospheric dispersion effects and isotopic abundance
variations. Sixty per cent of the absorption systems may be affected by the
former. However, comparison of the affected and unaffected systems yields
no evidence for these effects (Fig.~\ref{fig:cps_ad}). Modelling of
atmospheric dispersion and its effect on the QSO spectra indicates that
$\da$ in the low-$z$ systems should be correlated with the zenith angle of
the QSO observations. We observe no such correlation
(Fig.~\ref{fig:cps_davxi}). Atmospheric dispersion effects clearly can not
explain our results. The effect of possible variations in the isotopic
ratios is more difficult to estimate. The isotopic abundance trends in low
metallicity stellar environments suggest that the $^{25,26}$Mg/$^{24}$Mg
and $^{29,30}$Si/$^{28}$Si ratios will be significantly lower in the QSO
absorption clouds. This systematic effect, if present, should have pushed
$\da$ to positive values. Thus, when it is incorporated into the analysis,
$\da$ becomes more negative (Fig.~\ref{fig:cps_noiso_comp}).

\subsection{Comparison with other QSO absorption lines constraints}\label{ssec:QSO}

The only other QSO absorption line method which directly constrains $\da$
is the AD method outlined in Section \ref{sssec:ad_method}. The MM method
constraints in the present work are more than an order of magnitude more
precise than the strongest AD method constraints \citep[obtained
in][]{MurphyM_01c}.

Comparison of H{\sc \,i} 21-cm and millimetre-band molecular rotational
absorption frequencies constrains variations in $y\equiv\alpha^2g_p$, where
$g_p$ is the proton $g$-factor \citep{DrinkwaterM_98a}. This technique
suffers from an important systematic error which is difficult to quantify:
the millimetre-band and radio continuum emission generally originate from
different regions of the background QSO, leading to possible line-of-sight
velocity differences between the millimetre-band and H{\sc \,i} 21-cm
absorption lines. \citet{DrinkwaterM_98a} compared Galactic H{\sc \,i}
21-cm and millimetre-band absorption lines to estimate the likely
line-of-sight velocity difference, $\left|\Delta v\right| = 1.2{\rm
\,km\,s}^{-1}$. This corresponds to a systematic effect for a single
measurement of $\left|\dy\right| = 0.4\times 10^{-5}$. This formed the
dominant (1\,$\sigma$) uncertainty in our recent measurements of $\dy$ for
two absorption systems \citep{MurphyM_01d}: $\dy = (-0.20 \pm 0.44)\times
10^{-5}$ for $z_{\rm abs}=0.2467$ towards PKS 1413+135 and $\dy = (-0.16
\pm 0.54)\times 10^{-5}$ for $z_{\rm abs}=0.6847$ towards TXS
0218+357. \citet{CarilliC_00a} argue that $\left|\Delta v\right|$ could be
as large as 10--100\,km\,s$^{-1}$ on the basis of typical subkiloparsec
motions of the ISM. They derive the upper limit $\left|\dy\right| <
1.7\times 10^{-5}$ for the same two absorption systems.

Comparison of optical and H{\sc \,i} 21-cm absorption frequencies
constrains variations in $x\equiv \alpha^2g_pm_e/m_p$ where $m_e/m_p$ is
the electron-to-proton mass ratio \citep{WolfeA_76a}. \citet{CowieL_95a}
obtained $\dx = (0.70 \pm 0.55)\times 10^{-5}$ from the $z_{\rm
abs}=1.7764$ system towards Q1331+170. However, the effect of line-of-sight
velocity differences has not been estimated for the optical/H{\sc \,i}
comparison but is likely to be at least $\left|\Delta v\right| \sim 1.2{\rm
\,km\,s}^{-1}$, as found for the millimetre/H{\sc \,i} technique. A
statistical sample of $\dx$ and $\dy$ measurements is required to provide
reliable constraints and to properly quantify the typical line-of-sight
velocity differences. This is currently limited by the small number of
systems known to exhibit both optical/millimetre-band and H{\sc \,i} 21-cm
absorption \citep[e.g.][]{CurranS_02b}.

The above constraints on $\dy$ and $\dx$ can only be compared with MM
constraints on $\da$ if one assumes the constancy of $g_p$ and $g_pm_e/m_p$
respectively. If one suspects variation in $\alpha$ then it seems wholly
unjustified to assume the constancy of other fundamental parameters like
$g_p$ and $m_e/m_p$. Some {\it model dependent} links between $\da$, $\dx$,
and $\dy$ and have been studied recently within the paradigm of grand
unification (e.g.~\citealt*{LangackerP_02a};
\citealt{DentT_03a,DmitrievV_03a}). \citet{LangackerP_02a} emphasize that
the observations should be used to constrain the theory and not {\it vice
versa}.

\subsection{Comparison with non-QSO absorption lines constraints}\label{ssec:nonQSO}

Several non-QSO absorption line constraints on $\da$ exist, falling into
two distinct classes, `local' and `early universe'. We direct the reader
to the recent review of \citet{UzanJ_03a} for a summary of these methods.

The local constraints include laboratory comparisons of atomic clocks
\citep{PrestageJ_95a} and fountains \citep{SortaisY_01a}, analysis of
$^{149}$Sm isotopic abundances from the Oklo natural fission reactor in
Gabon, Africa \citep{ShlyakhterA_76a,DamourT_96a}, and meteoritic
constraints on the long-lived $\beta$-decay rate of $^{187}$Re
\citep{PeeblesP_62a,DysonF_67a,DysonF_72a}. The strongest current
constraints are, respectively, $\left|\da\right| = (0.2 \pm 7.5) \times
10^{-15}$ \citep[$\tau = 4.7{\rm \,yr}$,][]{MarionH_03a}, $\da = (-0.04 \pm
0.15) \times 10^{-7}$ \citep[$\tau = 1.8{\rm \,Gyr}$,][]{FujiiY_00a} and
$\left|\da\right| < 3 \times 10^{-7}$ \citep[$\tau = 4.56{\rm
\,Gyr}$,][]{OliveK_02b}, where $\tau$ is the relevant look-back
time. However, none of these methods constrain $\alpha$ directly: the
laboratory measurements assume constant nuclear magnetic moments
\citep{MarionH_03a}, the Oklo bound can be weakened by allowing other
interaction strengths and mass ratios to vary
\citep{FlambaumV_02a,OliveK_02b} and \citet{UzanJ_03a} notes that the
meteoritic $\beta$-decay limit assumes a constant weak interaction
strength, $\alpha_{\rm w}$. Moreover, as discussed in Section
\ref{ssec:tempvar}, one can not reliably compare these local limits with MM
constraints on $\da$ without a detailed theory giving both temporal {\it
and} spatial variations of $\alpha$.

The early universe constraints come from analysis of the CMB anisotropies
(\citealt{HannestadS_99a}; \citealt*{KaplinghatM_99a}) and light-element
(e.g.~D, He, Li) big bang nucleosynthesis (BBN) abundances
\citep*{KolbE_86a}. The strongest constraints from the CMB are at the $\da
\sim 10^{-2}$ level if one considers the uncertainties in, and degeneracies
with, the usual cosmological parameters \citep[i.e.~$\Omega_{\rm m}$,
$\Omega_\Lambda$ etc.; e.g.][]{MartinsC_03b}. However, \citet{KujatJ_00a}
and \citet*{BattyeR_01a} note a crucial degeneracy between $\alpha$ and
$m_e$. Estimates based on the BBN abundance of $^4$He suffer from a large
uncertainty as to the electromagnetic contribution to the proton-neutron
mass difference. The least model-dependent limits are those of
\citet{NollettK_02a}, $\da = (3 \pm 7)\times 10^{-2}$, who, like
\citet*{BerstroemL_99a}, considered all light elements up to $^7$Li,
thereby avoiding this problem. However, \citet{UzanJ_03a} notes that
variations in $\alpha$ are degenerate with variations in $\alpha_{\rm w}$,
$\alpha_{\rm s}$ and $G$.

\subsection{Future MM method checks}\label{ssec:future}

All the alternative methods discussed in the previous two sections
constrain $\alpha$ in (sometimes model-dependent) combination with other
constants. Therefore, the only current avenue for confidently ruling out
the present evidence for varying $\alpha$ is to obtain {\it independent MM
constraints from QSO spectra}. The present work has shown that no known
systematic errors can mimic the effect of varying $\alpha$ in the
Keck/HIRES spectra. However, if subtle, unknown instrumental effects cause
the line-shifts we observe, the high quality spectra now available from the
VLT/UVES and Subaru/HDS will bear this out. If the present results are
confirmed, iodine cell calibration techniques, similar to those used to
identify extra-solar planets \citep[e.g.][]{MarcyG_92a}, could be applied
to selected absorption systems to confidently rule out instrumental
effects.

\section*{Acknowledgments}
We are indebted to Wallace Sargent for the new Keck/HIRES dataset presented
above. Without his extensive and careful observations, as well as the
support of numerous co-observers, the present work would not have been
possible. We are also grateful to Tom Barlow and Rob Simcoe for their
careful data reduction. We greatly appreciate the continued support of, and
fruitful discussions with, Chris Churchill, Jason Prochaska and Arthur
Wolfe who also provided the previous Keck/HIRES datasets. We also thank Tom
Bida and Steve Vogt for detailed information regarding Keck/HIRES,
particularly the latter for providing {\sc zemax} models of the
spectrograph. John Barrow, Charley Lineweaver and Jochen Liske contributed
valuable ideas and criticisms.

We are grateful to the John Templeton Foundation for supporting this
work. MTM received a Grant-in-Aid of Research from the National Academy of
Sciences, administered by SigmaXi, the Scientific Research Society. MTM is
also grateful to PPARC for support at the IoA under the observational
rolling grant (PPA/G/O/2000/00039). JKW thanks the IoA for hospitality
while carrying out some of this work. We thank the Australian Partnership
for Advanced Computing National Facility for numerically intensive
computing access.

Data presented herein were obtained at the W.M. Keck Observatory, which is
operated as a scientific partnership among the California Institute of
Technology, the University of California and the National Aeronautics and
Space Administration. The Observatory was made possible by the generous
financial support of the W.~M.~Keck Foundation.


\bsp_small

\label{lastpage}

\end{document}